\documentclass[a4paper,twocolumn,11pt,accepted=2023-02-13]{quantumarticle}

\pdfoutput=1

\usepackage[utf8]{inputenc}
\usepackage[english]{babel}
\usepackage[T1]{fontenc}
\usepackage{hyperref}

\usepackage{amsmath}
\usepackage{amsfonts}
\usepackage{amssymb}
\usepackage{amstext}
\usepackage[numbers]{natbib}
\usepackage{amsthm}
\usepackage{graphicx}
\usepackage{textcomp}
\usepackage{color}
\usepackage{mathtools}
\usepackage{here}
\usepackage{multirow}

\newtheorem{theorem}{Theorem}
\newtheorem{proposition}[theorem]{Proposition}
\newtheorem{lemma}[theorem]{Lemma}
\newtheorem{corollary}[theorem]{Corollary}

\theoremstyle{definition}

\theoremstyle{definition}
\newtheorem*{Proof}{Proof}

\newlength{\fighskip} \fighskip=2pt
\newlength{\figvskip} \figvskip=3pt

\newcommand*{\figbox}[2]{{
  \def\figscale{#1}
  \def\arraystretch{0.8}
  \arraycolsep=0pt
  \begin{array}{c}
    \vbox{\vskip\figscale\figvskip
      \hbox{\hskip\figscale\fighskip
        \includegraphics[scale=\figscale]{#2}}}
  \end{array}}}

\newcommand{\mc}[1]{\mathcal{#1}}
\newcommand{\ket}[1]{|#1\rangle} 
\newcommand{\ketbra}[2]{|#1\rangle\langle#2|} 

\DeclareMathOperator{\tr}{Tr}

\newcommand{\cP}{\mathcal{P}}

\newcommand{\cH}{{\mathcal H}}

\newcommand{\cT}{{\cal T}}
\newcommand{\cC}{{\cal C}}
\newcommand{\cK}{{\cal K}}

\newcommand{\cU}{\mathcal{U}}
\newcommand{\cL}{\mathcal{L}}

\newcommand{\cM}{{\mathcal M}}
\newcommand{\cJ}{{\mathcal J}}
\newcommand{\cD}{{\mathcal D}}

\def\>{{\rangle}}
\def\<{{\langle}}

\def\lsim{\mathrel{\rlap{\lower4pt\hbox{\hskip1pt$\sim$}}
		\raise1pt\hbox{$<$}}}                
\def\gsim{\mathrel{\rlap{\lower4pt\hbox{\hskip1pt$\sim$}}
		\raise1pt\hbox{$>$}}}                

\newcommand{\cS}{{\mathcal{S}}}

\newcommand{\mbb}[1]{\mathbb{#1}}

\begin{document}
	
\title{Black holes as clouded mirrors: the Hayden-Preskill protocol with symmetry}
	
\author[1,2,3]{Yoshifumi Nakata} \footnote{yoshifumi.nakata@yukawa.kyoto-u.ac.jp}
\author[4]{Eyuri Wakakuwa}
\author[2]{Masato Koashi}
\affiliation[1]{Yukawa Institute for Theoretical Physics, Kyoto university, Kitashirakawa Oiwakecho, Sakyo-ku, Kyoto, 606-8502, Japan}
\affiliation[2]{Photon Science Center, Graduate School of Engineering,~The University of Tokyo, Bunkyo-ku,~Tokyo 113-8656, Japan}
\affiliation[3]{JST, PRESTO, 4-1-8 Honcho, Kawaguchi, Saitama, 332-0012, Japan}
\affiliation[4]{Department of Communication Engineering and Informatics, Graduate School of Informatics and Engineering, The University of Electro-Communications, Tokyo 182-8585, Japan}
	
\begin{abstract}
The Hayden-Preskill protocol is a qubit-toy model of the black hole information paradox. Based on the assumption of scrambling, it was revealed that quantum information is instantly leaked out from the quantum many-body system that models a black hole.
In this paper, we extend the protocol to the case where the system has symmetry and investigate how the symmetry affects the leakage of information. We especially focus on the conservation of the number of up-spins.
Developing a partial decoupling approach, we first show that the symmetry induces a \emph{delay of leakage} and an \emph{information remnant}. 
We then clarify the physics behind them: the delay is characterized by thermodynamic properties of the system associated with the symmetry, and the information remnant is closely related to the symmetry-breaking of the initial state. These relations bridge the information leakage problem to macroscopic physics of quantum many-body systems and allow us to investigate the information leakage only in terms of physical properties of the system.
\end{abstract}

\maketitle

\section{Introduction}

Black holes are the most peculiar objects in the universe. While macroscopic properties of black holes can be fairly understood by general relativity, finding microscopic descriptions of black holes has been a central problem in fundamental physics. A significant step was made by the discovery of Hawking radiation~\cite{H1974, H1975}: due to a quantum effect, a quantum black hole emits thermal radiation. This discovery raises a question about whether the radiation carries away the information in the black hole. 
Although information leakage is unlikely in the classical case due to the no-hair theorem~\cite{I1967,I1968,C1971}, the holographic principle indicates that information should leak out as a black hole evaporates. This conflict is known as the information paradox. 

A novel approach to the information paradox was proposed from the theory of quantum information, known as the Hayden-Preskill (HP) protocol~\cite{HP2007}. In the protocol, a quantum black hole is modeled by a quantum many-body system consisting of a large number of qubits. After scrambling dynamics of the system, which is typically the case in quantum chaotic systems, the system is split into two random subsystems, one is a remaining black hole and the other is the `Hawking radiation'. Due to the scrambling dynamics, it was shown that information stored in the original many-body system can be recovered from the `radiation' even if the size of the corresponding subsystem is small. This phenomena is often expressed as that information is quickly leaked out from the black hole. 
This result spiked various research topics over quantum gravity, quantum chaos, and quantum information~\cite{SS2008,S2011,LSHOH2013,SS2014,SS2015,RS2015,RY2017,Y2019,L2020,SY1993,Kitaev1,Kitaev2,Jensen2016,MS2016,Sachdev2015,B2016,vKRPS2018,KVH2018,HQRY2016,PYHP2015,PEW2017,KC2019,HP2019},
including the trials of experimentally simulating quantum gravity in a lab~\cite{LFSLYYM2019,BGLLNSSSW2021,NLBGLSSSW2021}.

In this paper, we study the HP protocol when the quantum many-body system has symmetry. Since quantum many-body systems typically have symmetry, it is important to clarify how symmetry affects the protocol. 
We especially consider the system with conservation of the number of up-spins.
The conservation law makes it difficult to fully investigate the protocol by the technique used in~\cite{HP2007}. In~\cite{Y2019,L2020}, the difficulty was avoided by considering restricted cases with fixed conserved quantities, and possible connections to other conceptual puzzles of quantum black holes~\cite{BS2011,HO2019,HO2021,AHMNV2007} were argued.
Here, we develop a quantum information-theoretic technique to simultaneously deal with quantum information and symmetry, and investigate the HP protocol in general cases.

We first show that the symmetry changes \emph{when} and \emph{how} the information leaks out from the black hole. More specifically, we show the existence of a \emph{delay of information leakage} and an \emph{information remnant}, which are numerically shown to be macroscopically large for certain conditions of the initial many-body system.
The technique we have developed may be of independent interest in relation to covariant quantum error-correcting codes~\cite{WA2020,FNASPHP2020,HNPS2021,LK2022}. Therein, a lower bound on the recovery error, which corresponds to the information remnant in our terminology, was intensely studied for general types of symmetry.
We then clarify the physics behind these phenomena: the delay and the information remnant are, respectively, characterized by thermodynamic properties and symmetry-breaking of the many-body system. 
These relations unveil non-trivial connections between quantum information in a many-body system and the macroscopic physics associated with the symmetry.
Although we show the connections for the systems conserving the number of up-spins, we expect that they hold for a wider class of symmetries as far as they are global and abelian, such as $U(1)$ symmetry.

The paper is organized as follows. For clarity of the presentation, we provide a brief overview of our results in Sec.~\ref{S:OOR}.
We start the main analysis with introducing our notation in Sec.~\ref{S:Preliminaries}.
In Sec.~\ref{S:HPmodel}, we overview the original HP protocol, and propose a modified protocol with symmetry. The partial decoupling is explained and is applied to the HP protocol with symmetry in Sec.~\ref{S:IRPR}.
The physics behind the delay and the information remnant is discussed in Sec.~\ref{S:IRMP}.
We conclude this paper with discussions in Sec.~\ref{S:SD}. Technical statements are proven in Appendices.

\section{Overview of our results} \label{S:OOR}

\subsection{Brief introduction of the Hayden-Preskill protocol}

The HP protocol is a quantum information-theoretic toy model of a quantum black hole. In the model, a quantum black hole is represented by a closed quantum many-body system that undergoes unitary time evolution. 
After some time, a qubit is displaced from the system, which models an emission of a single qubit as the Hawking radiation from the black hole. The rest of the system again undergoes unitary time evolution, and another qubit is displaced. By repeating the unitary time evolution and displacement of a single qubit, the initial system is eventually split into two subsystems: one is considered to be a remaining black hole and the other is a number of qubits representing the Hawking radiation.

One may regard this model as the AdS/CFT correspondence of a quantum black hole, which predicts that a quantum black hole has a quantum mechanical description without gravity. This paper, however, does not argue the origin of the model and simply follows the original proposal of the protocol~\cite{HP2007}. By focusing on the minimal setting, we can investigate the effect of symmetry to the HP protocol from a purely information-theoretic perspective.

Following the original modelling, we refer to the initial many-body system as a \emph{BH} and to the random subsystems as a \emph{radiation} and a \emph{remaining BH}. The size of the radiation increases as time passes, and the size of a remaining BH decreases. Eventually, the whole BH becomes radiation.

A brief description of the HP protocol is as follows~\cite{HP2007}. 
The full details are given in Sec.~\ref{S:HPmodel}. See also Fig.~\ref{Fig:situ1} therein.
\begin{enumerate}
\item Quantum information source $A$ of $k$ qubits is thrown into an initial BH $B_{\rm in}$ composed of $N$ qubits, which has already emitted some radiation $B_{\rm rad}$. The quantum information in $A$ is kept track of by a reference system $R$. 
\item The new BH $S:=A B_{\rm in}$ of $N+k$ qubits undergoes scrambling dynamics and gradually emits new radiation, i.e., a random subsystem $S_{\rm rad}$ of $S$. The number of qubits in $S_{\rm rad}$ is denoted by $\ell$. The BH shrinks to the remaining BH $S_{\rm in}$ of $N+k-\ell$ qubits. As time goes on, $\ell$ increases.
\item The quantum information originally stored in $A$ is tried to be recovered from the radiation $B_{\rm rad}$ and $S_{\rm rad}$.
\end{enumerate}

It was shown in Ref.~\cite{HP2007} that, if the size $\ell$ of the new radiation $S_{\rm rad}$ exceeds a certain threshold, which is roughly given by
\begin{equation}
    \ell > k + \frac{N-H(B_{\rm in})}{2}, \label{Eq:nerw4}
\end{equation}
where $H(B_{\rm in})$ is the entropy of the initial BH $B_{\rm in}$, the remaining BH $S_{\rm in}$ is decoupled from the reference $R$.
As a result, the error in recovering the information from the radiations $B_{\rm rad}$ and $S_{\rm rad}$ decreases exponentially quickly as $\ell$ increases. 
The threshold depends on the properties of the initial BH $B_{\rm in}$, and can be the same order of the number $k$ of qubits in the quantum information source $A$.\\

\subsection{The HP protocol with symmetry and the main result} \label{SS:WR}
The main concern in this paper is to clarify what would occur if the system has symmetry. Since the dynamics cannot be fully scrambling in the presence of symmetry, information should be leaked out from the system in a different manner.
We particularly consider conservation of the number of up-spins, or equivalently, the $Z$-component of the angular momentum (\emph{$Z$-axis AM} for short). Since the corresponding symmetry is a $Z$-axial symmetry, we call the system the Kerr BH in analogy with actual black holes.
Throughout the paper, we denote by $L$ the mean of the $Z$-axis AM in the initial Kerr BH $B_{\rm in}$ and by $\delta L$ its standard deviation.

Assuming that the dynamics of the system is scrambling that respects the axial symmetry, we first show that partial decoupling occurs in the Kerr BH, rather than the full decoupling as in the case of BH without any symmetry.
This results in two substantial changes, \emph{the delay of information leakage} and \emph{the information remnant}.
The delay means that, compared to the BH without any symmetry, one needs to collect more radiation in order to recover quantum information of $A$. The information remnant means that a certain amount of the information of $A$ remains in the Kerr BH until it is all radiated.
Depending on the $Z$-axis AM $L$ and its fluctuation $\delta L$ of the initial Kerr BH $B_{\rm in}$, the delay ranges from $O(\sqrt{k})$ to macroscopically large values, and the information remnant varies from infinitesimal to constant. 

Our result implies that, with certain initial conditions, the symmetry radically changes the behavior of how radiation carries information away from the BH. In particular, when the Kerr BHs has extremely large $Z$-axis AM or fluctuation, there is a possibility that the Kerr BHs do keep nearly all information of $A$ until the last moment. This is in sharp contrast to the HP protocol without symmetry that predicts an instant leakage of information.
See Sec.~\ref{S:IRPR} for details.

\subsection{Physics behind the delay of information leakage and information remnant} \label{SS:AEC}

To physically understand the delay and the information remnant, we further investigate them from different perspectives. The details are given in Sec.~\ref{S:IRMP}.

We argue that the delay is ascribed to the structure of the entanglement generated by the scrambling dynamics with symmetry. To capture the information-theoretic consequence of such entanglement, we introduce a new concept that we call \emph{clipping of entanglement}, and propose the following condition for the information recovery from the radiation to be possible: for most of possible values of the $Z$-axis AM $n$ in the new radiation $S_{\rm rad}$, 
\begin{equation}
k < H(B_{\rm in}) + \log \biggl[ \frac{\dim \cH_n^{S_{\rm rad}}}{\dim \cH_{L-n}^{S_{\rm in}}} \biggr], \label{Eq:CoE}
\end{equation}
where $H(B_{\rm in})$ is the entropy of the initial Kerr BH $B_{\rm in}$, $\cH_n^{S_{\rm rad}}$ is the subspace of the Hilbert space of $S_{\rm rad}$ with the $Z$-axis AM $n$, and $\cH^{S_{\rm in}}_{L-n}$ is that of the remaining Kerr BH $S_{\rm in}$ with the $Z$-axis AM $L-n$.
This condition indeed reproduces the results based on the partial decoupling. Moreover, the results of the BH without symmetry~\cite{HP2007} can be also obtained: in this case, the condition simply reduces to
\begin{equation}
k < H(B_{\rm in}) + \log \biggl[ \frac{\dim \cH^{S_{\rm rad}}}{\dim \cH^{S_{\rm in}}} \biggr],
\end{equation}
due to the absence of conserved quantities.
From this, one obtains $\ell > k + (N-H(B_{\rm in}))/2$, which is the same as Eq.~\eqref{Eq:nerw4}.

A merit of the entanglement clipping is that the condition for the information leakage, Eq.~\eqref{Eq:CoE}, consists only of the entropy of the initial system and the dimension of the Hilbert spaces with respect to the conserved quantity. Hence, for quantum many-body systems with any extensive conserved quantity, we can easily compute the delay of information leakage.

Apart from the conciseness, the argument based on entanglement clipping has another merit that it unveils a non-trivial relation between the delay of information leakage and thermodynamic properties of the system.
We first show that the delay is characterized by up to the second-order derivatives of the entropy in $B_{\rm in}$ in terms of the $Z$-axis AM. The derivatives are then connected to thermodynamic properties of $B_{\rm in}$ as follows.
Let $\omega(T, \lambda)$ be the state function of $B_{\rm in}$ conjugate to the $Z$-axis AM, which equals to the angular velocity of the Kerr BH, and $\alpha(T, \lambda)$ be thermal sensitivity of the $Z$-axis AM.
We can show that, when the initial Kerr BH $B_{\rm in}$ has a small $Z$-axis AM,
\begin{equation}
{\rm delay} \propto \frac{L}{S}\sqrt{k_B | \omega(T, \lambda) \alpha(T, \lambda)|}, \label{a}
\end{equation}
where $k_B$ is the Boltzmann constant, and $S$ is the entropy of the initial Kerr BH. Note that both $\omega(T, \lambda)$ and $\alpha(T, \lambda)$ depend on the temperature $T$, but their product is temperature-independent.
The formula for a large $Z$-axis AM is also obtained. This allows us to understand the delay in terms of thermodynamic properties of the initial system.

The information remnant can also be intuitively understood. The fact that partial decoupling occurs instead of full decoupling implies that the $Z$-axis AM of the remaining Kerr BH $S_{\rm in}$ remains correlated to that with the reference $R$. In other words, the information originally stored in $A$ can be partially obtained by measuring the $Z$-axis AM in $S_{\rm in}$. From the viewpoint of the radiation $S_{\rm rad}$, this implies that the information about the coherence of the $Z$-axis AM in $A$ cannot be fully accessed, leading to the information remnant. The amount of the information remnant is closely related to the fluctuation of the $Z$-axis AM in the remaining Kerr BH $S_{\rm in}$: if the fluctuation is small, the change in the $Z$-axis AM caused by throwing $A$ into the Kerr BH is easy to detect.
We quantitatively confirm this expectation and show that 
\begin{align}
{\rm information} \ {\rm remnant} &\gtrsim \frac{1}{\sqrt{\langle \delta \nu^2 \rangle}}, \label{bb}
\end{align}
where $\sqrt{\langle \delta \nu^2 \rangle}$ is the standard deviation of the $Z$-axis AM in the remaining Kerr BH $S_{\rm in}$. We further show that the fluctuation $\sqrt{\langle \delta \nu^2 \rangle}$ is characterized by the degree $\zeta(S_{\rm in})$ of symmetry-breaking in $S_{\rm in}$, leading to
 \begin{align}
{\rm information} \ {\rm remnant} &\gtrsim \frac{1}{\sqrt{2 \zeta(S_{\rm in})}}. \label{b}
\end{align}
This reveals that the information remnant is characterized by the degree of symmetry-breaking of the Kerr BH.

The relations, Eqs.~\eqref{a} and~\eqref{b}, establish quantitative connections between the information leakage problem and macroscopic physics. We expect that they universally hold for any quantum chaotic systems with abelian symmetry. In those cases, the quantities related to the $Z$-axis AM in Eqs.~\eqref{a} and~\eqref{b} should be replaced by the corresponding conserved quantities.

A canonical instance is energy conservation, in which the relations similar to Eqs.~\eqref{a} and~\eqref{b} can be obtained.
The delay of information leakage when energy is conserved should be given in terms of the heat capacity $C_V$ of the quantum system as
\begin{equation}
{\rm delay} \propto \frac{E}{S}\sqrt{k_B |C_V|} \label{Eq:c444tt}
\end{equation}
where $E$ and $S$ are the energy and the entropy of the initial BH, respectively, and the amount of information remnant is determined by the energy fluctuation of the BH.

\subsection{Discussions}

Our results reveal, both in quantitative and physically intuitive manners, the presence of symmetry leads to non-negligible changes in the process of information leakage in the HP protocol. The changes were quantified by using the method based on partial decoupling as in Subsec.~\ref{SS:WR}, and were qualitatively estimated by physical quantities of the system as explained in Subsec.~\ref{SS:AEC}.
We are mainly concerned with rotational symmetry in our analysis and show that both the delay of information leakage and the information remnant can be large for certain initial conditions. Our approach, methods, and results offer a solid basis for studying quantum information in the many-body systems with symmetry.

In the context of quantum gravity, one of our main contributions is to show that more careful treatments of the symmetry, including the energy conservation, will be needed to fully understand the information recovery from the Hawking radiation. In fact, the presence of symmetry can result in the situation where information thrown into a quantum black hole cannot be recovered until it is completely evaporated. This conclusion is in sharp contrast to that of the original analysis of the HP protocol without symmetry and tells us that taking the black hole's symmetry, such as energy conservation, into account is crucial for understanding the quantum information perspective of quantum gravity through the HP protocol.

A couple of analyses were already carried out about the energy conservation in the HP protocol~\cite{Y2019,L2020}. Therein, the physical modes of the radiation responsible for the instant information recovery and the relations between the HP protocol and other puzzles in quantum gravity were discussed~\cite{BS2011,HO2019,HO2021,AHMNV2007}. Despite the fact that these studies shed novel light on quantum gravity, the arguments highly depend on information-theoretic assumptions about the details of the model, which are sometimes unimportant from the viewpoint of physics or even unphysical. Hence, the conclusions are strongly depending on information-theoretic details of the model: one may arrive at entirely different results depending of the assumptions.
To avoid such undesired situations, understanding the information perspectives of quantum gravity without referring to too much details about quantum information will be helpful.

Our approach provides a sophisticated method to this end. The simple formulas obtained in this paper can be used to understand information recovery only in terms of physical quantities. For instance, Eq.~\eqref{Eq:CoE} is a formula to investigate the information recovery only from the degeneracy of eigenspaces. This implies, in the case of energy conservation, that the information recovery can be assessed only from the energy spectrum of the Hamiltonian of a quantum black hole. If one accepts a controversial assumption that a quantum black hole has usual thermodynamic features, which may be justified by the AdS/CFT correspondence, Eq.~\eqref{Eq:c444tt} provides a more direct way of evaluating the information recovery only from macroscopic physical properties of the black hole. The information remnant induced by the energy conservation can be also discussed based on Eq.~\eqref{bb}, which may however be negligible in the thermodynamic limit if a black hole has a typical amount of energy fluctuation.

To summarize, our analysis provides 1. a caution not to literally accept the instant recovery of quantum information from the Hawking radiation in a realistic situation, and 2. convenient tools of investigating the information recovery only in terms of the quantities often studied in physics. Although our result itself does not provide conclusive answers to the most intriguing questions, such as whether the process of information leakage is changed by energy conservation in the leading order, our approach will be a stepping stone toward the full understanding of the information recovery from the Hawking radiation in a realistic situation with energy conservation. To this end, a couple of physical properties of a quantum black hole, such as the energy spectrum, its initial state, and the radiation process with energy conservation, should be clarified.

\section{Preliminaries} \label{S:Preliminaries}

Throughout the paper, we denote the sequence $\{i, i+1, \dots, j-1, j \}$ of integers between two natural numbers $i$ and $j$ ($i \leq j$) by $[i,j]$. The logarithm is always taken in base two.

The basic unit of quantum information is a qubit represented by a two-dimensional Hilbert space. The number of qubits in a system $S$ is denoted by $|S|$. We often write the relevant systems in the superscript, such as a Hilbert space $\mc{H}^S$ of a system $S$, an operator $X^{SR}$ on $SR$, and a superoperator $\mc{E}^{S\rightarrow B}$ from $S$ to $B$. For superoperators from $S$ to itself, we denote it by $\mc{E}^S$. 

The partial trace over a system $X$ is denoted by $\tr_X$. A reduced operator on $S$ of $\rho^{SR}$ is denoted simply by $\rho^S$, that is, $\rho^S = \tr_R [\rho^{SR}]$. Furthermore, we denote $(M^{S} \otimes I^R )\rho^{SR} (M^{S \dagger} \otimes I^R )$, where $I$ is the identity operator, by $M^{S}\rho^{SR} M^{S \dagger}$. The identity superoperator on $S$ is denoted by ${\rm id}^S$.

\subsection{Linear Operators and Superoperators}
We denote a set of linear operators from $\cH^A$ to $\cH^B$ by $\cL(\cH^A,\cH^B)$, and $\cL(\cH^A,\cH^A)$ by $\cL(\cH^A)$. We also use the following notation for the sets of positive semi-definite operators, quantum states, and sub-normalized states.
\begin{align}
&\mc{P}(\mc{H}) = \{ X \in \cL(\mc{H}) : X \geq 0 \}, \\ 
&\mc{S}(\mc{H})=\{\rho \in \mc{P}(\mc{H}) : \tr [\rho]=1 \}, \\
&\mc{S}_{\leq}(\mc{H})=\{\rho \in \mc{P}(\mc{H}) : \tr [\rho] \leq 1 \}.
\end{align}
A maximally entangled state between $S$ and $S'$, where $\cH^S \cong \cH^{S'}$, is denoted by $\Phi^{SS'}$. The completely mixed state on $\cH^S$ is denoted by $\pi^S = I^S/d_S$ ($d_S = {\rm dim} \cH^S$). 

A purification of $\rho^S \in \cS(\cH^S)$ by $R$ ($\cH^R \cong \cH^S$) is denoted by $\ket{\rho}^{SR}$, namely, $\tr_{R}[ \ketbra{\rho}{\rho}^{SR} ] = \rho^S$. For instance, since the marginal state of a maximally entangled state is the completely mixed state, we have $\Phi^S = \pi^S$.

The fundamental superoperators are the conjugations by a \emph{unitary}, an \emph{isometry}, and a \emph{partial isometry}. An isometry $V^{\cH \rightarrow \cK}$ is the linear operator such that
\begin{equation}
(V^{\cH \rightarrow \cK})^{\dagger} V^{\cH \rightarrow \cK} = I^{\cH}.
\end{equation}
When $\dim \cH = \dim \cK$, the isometry is called a unitary and also satisfies $V^{\cH \rightarrow \cK}  (V^{\cH \rightarrow \cK})^{\dagger} = I^{\cK}$. 
A partial isometry is a linear operator from $\cH$ to $\cK$ such that it is an isometry on its support. Projections, isometries, and unitaries are special classes of a partial isometry.

A \emph{quantum channel} $\mc{T}^{S \rightarrow B}$ is a completely-positive (CP) and trace-preserving (TP) map. A map is called CP if $({\rm id}^{S'} \otimes \mc{T}^{S \rightarrow B}) (\rho^{S'S}) \geq 0$ for any $\rho^{S'S} \geq 0$ and is TP if $\tr[\mc{T}^{S \rightarrow B} (\rho^{S})] = \tr[\rho^S]$. 
We also say a superoperator $\mc{T}^{S \rightarrow B}$ is sub-unital and unital if $\mc{T}^{S \rightarrow B}(I^S) \leq I^B$ and $\mc{T}^{S \rightarrow B}(I^S) =I^B$, respectively.

For a quantum channel $\cT^{S \rightarrow B}$, a \emph{complementary} channel is defined as follows. Let $V_{\cT}^{S \rightarrow BE}$ be a Steinspring dilation for a quantum channel $\mc{T}^{S \rightarrow B}$, i.e., $V^{S \rightarrow BE}_{\cT}$ is an isometry satisfying
\begin{equation}
\tr_{E} [V_{\cT}^{S \rightarrow BE} \rho^S (V_{\cT}^{S \rightarrow BE})^{\dagger}] = \mc{T}^{S \rightarrow B}(\rho^S),
\end{equation}
for any $\rho^S \in \mc{L}(\mc{H}^S)$.
The map $\bar{\mc{T}}^{S \rightarrow E}$ defined by
\begin{equation}
\bar{\mc{T}}^{S \rightarrow E}(\rho^S) :=\tr_{B} [V_{\cT}^{S \rightarrow BE} \rho^S (V^{S \rightarrow BE}_{\cT})^{\dagger}],
\end{equation}
for any $\rho^S \in \mc{L}(\mc{H}^S)$ is called a complementary channel.

Any superoperator $\mc{T}^{S \rightarrow B}$ from $S$ to $B$ has a \emph{Choi-Jamio\l kowski} representation defined by
\begin{equation}
\mathfrak{J}(\mc{T}^{S \rightarrow B}) := ({\rm id}^{S'} \otimes \mc{T}^{S \rightarrow B})(\Phi^{SS'}). \label{Eq:CJdef}
\end{equation}
This is sometimes called the channel-state duality.

\subsection{Norm and Entropy}
The Schatten $p$-norm for a linear operator $X \in \cL(\cH^A, \cH^B)$ is defined by $\| X \|_p := ( \tr [ (XX^{\dagger})^{p/2}] )^{1/p}$ ($p \in [1, \infty]$).
We particularly use the trace ($p=1$) and the Hilbert-Schmidt ($p=2$) norms. 
We define the fidelity between quantum states $\rho$ and $\sigma$ by 
\begin{equation}
F(\rho, \sigma) := \| \sqrt{\rho} \sqrt{\sigma} \|_1^2.
\end{equation}

The conditional min-entropy is defined by
\begin{multline}
H_{\rm min}(A|B)_{\rho} \\= \sup_{\sigma^B \in \mc{S}(\mc{H}^B)} \sup \{ \lambda \in \mathbb{R}| 2^{-\lambda} I^A \otimes \sigma^B - \rho^{AB} \geq 0 \},
\end{multline}
for $\rho^{AB} \in \cP(\cH)$. 
This is a variant of the conditional entropy $H(A|B)_{\rho}:=H(AB)-H(B)$, where $H(A)=-\tr [\rho^A \log \rho^A]$ is the von Neumann entropy. 
When ${\rm dim} \cH^B =1$, $H_{\rm min}(A|B)_{\rho}$ reduces to the min-entropy
\begin{equation}
H_{\rm min}(A)_{\rho} = \sup \{ \lambda \in \mathbb{R}| 2^{-\lambda} I^A - \rho^{A} \geq 0 \}.
\end{equation}

\subsection{Haar scrambling}
The Haar measure is often used to formulate the scrambling dynamics.
The Haar measure ${\sf H}$ on a unitary group ${\sf U}(d)$ of finite degree $d$ is the unique left- and right- unitarily invariant probability measure. Namely, it satisfies,
\begin{multline}
\text{for any $\mathcal{W} \subset {\sf U}(d)$ and $V \in {\sf U}(d)$}, \\ 
{\sf H}(V \mathcal{W}) = {\sf H} (\mathcal{W}V) = {\sf H}(\mathcal{W}).
\end{multline}
When a unitary $U$ is chosen with respect to the Haar measure ${\sf H}$, denoted by $U \sim {\sf H}$, it is called a Haar random unitary. 
Throughout the paper, the average of a function $f(U)$ over a random unitary $U \sim \mu$, where $\mu$ is a probability measure, is denoted by $\mathbb{E}_{U \sim \mu}[f(U)]$.

\section{Hayden-Preskill protocol} \label{S:HPmodel}

\begin{figure}[tb!]
\centering
\includegraphics[width=80mm,clip]{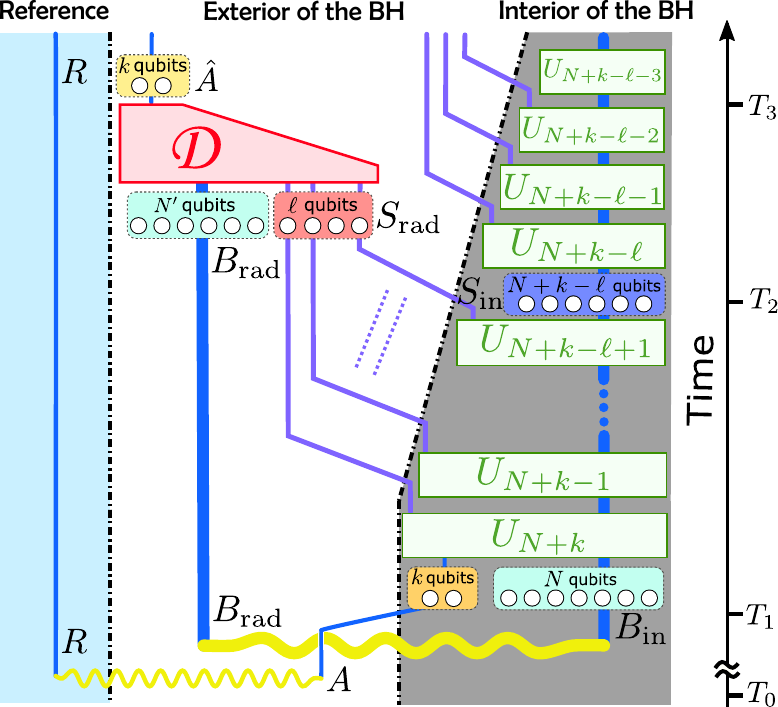}
\caption{Diagram of the HP protocol.
The blue lines represent trajectories of qubits, and the yellow wavy lines indicate that they may be entangled.  The black region in the right-hand side represents the BH system, and the blue region in the left-hand side is a reference system that is virtually introduced so as to keep track of the quantum information of $A$.
The green boxes in the black region represent the internal scrambling dynamics of the BH. Each $U_{N+k-i}$ is internal unitary time-evolution acting on $N+k-i$ qubits.
The purple lines represent the radiation of the BH.
The map $\cD$ is the recovery operation applied by Bob, aiming at recovering the information originally stored in $A$ from the past radiation $B_{\rm rad}$ and the new radiation $S_{\rm rad}$.
}
\label{Fig:situ1}
\end{figure}

We overview the general setting of the HP protocol in Subsec.~\ref{SS:HPoriginal1} and review the results in Ref.~\cite{HP2007} with a slight generalization in Subsec.~\ref{SS:HPoriginal}. We then explain how symmetry shall change the situation in Subsec.~\ref{SS:HPsymmetry}. The diagram of the protocol is given in Fig.~\ref{Fig:situ1}, and important quantities are summarized in Table~\ref{Tab:NotBH}.

\begin{table*}[tb!]
\centering
\caption{Notation in the investigation of the BH information problem} \label{Tab:NotBH}
\begin{tabular}{l|l}\hline
$B_{\rm in}$ ($N$ qubits) & Initial BH.  \\ \hline
$B_{\rm rad}$ ($N'$ qubits) & ``Past'' radiation that purifies the initial BH $B_{\rm in}$. \\ \hline
$A$ ($k$ qubits) & Quantum information source Alice throws into the BH $B_{\rm in}$. \\ \hline
$R$ ($k$ qubits) & Reference system that is maximally entangled with $A$. \\ \hline
$S$ ($N+k$ qubits) & The BH after the information source $A$ is thrown in ($S=A B_{\rm in}=S_{\rm in}S_{\rm rad}$). \\ \hline
$S_{\rm rad}$ ($\ell$ qubits) & New radiation from the BH $S$. \\ \hline
$S_{\rm in}$ ($N+k-\ell$ qubits) & Remaining BH after $S_{\rm rad}$ is evaporated.\\ \hline
\end{tabular}
\end{table*}

\subsection{General setting of the HP protocol} \label{SS:HPoriginal1}
In the HP protocol, we consider the situation depicted in Fig.~\ref{Fig:situ1}.
In the protocol, the BH $B_{\rm in}$ initially consists of $N$ qubits. The radiation emitted earlier, which we call the \emph{past radiation}, is denoted by a quantum system $B_{\rm rad}$.

We particularly consider two types of the BH. One is a ``pure'' BH, where $B_{\rm in}$ is in a pure state $\ket{\xi}^{B_{\rm in}}$, and the other is a ``mixed'' BH, where $B_{\rm in}$ is in a mixed state $\xi^{B_{\rm in}}$. The former models the quantum black hole that does not emit radiation yet or the one whose radiation has been measured by someone. The latter models a sufficiently old quantum black hole that has already emitted a large amount of radiation. For a mixed BH, the system $B_{\rm in} B_{\rm rad}$ is in a pure state $\ket{\xi}^{B_{\rm in} B_{\rm rad}}$, which is a purification of $\xi^{B_{\rm in}}$. 

The HP protocol is the following thought-experiment.
At time $T_1$, a person, Alice, throws a quantum information source $A$ of $k$ qubits into the BH $B_{\rm in}$. The quantum information source is defined by introducing a reference system $R$, and by setting the initial state between $A$ and $R$ to a maximally entangled state $\Phi^{AR}$.
The BH, now a composite system $S:=A B_{\rm in}$, undergoes the internal unitary dynamics and emits new radiation. 
We model this process by repeatedly applying an internal unitary dynamics followed by a displacement of a qubit in $S$ to the exterior of the BH. Since each radiation, i.e., the displacement of a qubit, shrinks the BH by one qubit, the unitary after $i$ qubits are radiated acts on $N+k-i$ qubits.
In Fig.~\ref{Fig:situ1}, the internal unitary dynamics acting on the $N+k-i$ qubits in the BH is denoted by $U_{N+k-i}$.

Suppose that the BH emits $\ell$ qubits of radiation, i.e., $|S_{\rm rad}| = \ell$, by time $T_2$. We call $S_{\rm rad}$ the \emph{new radiation}. The \emph{remaining BH} $S_{\rm in}$ is composed of $N+k-\ell$ qubits. Since the dynamics from time $T_1$ to $T_2$ is quantum mechanical, the dynamics should be represented by a quantum channel. From the construction, the dynamics $\cL^{S \rightarrow S_{\rm in} S_{\rm rad}}$ from $S$ to $S_{\rm in} S_{\rm rad}$ is simply a product of unitary dynamics, 
\begin{equation}
\cL^{S \rightarrow S_{\rm in} S_{\rm rad}} = \cU_{N+k-\ell+1} \circ \cU_{N+k-\ell+2} \circ \dots  \circ \cU_{N+k}, \label{Eq:dynamicsBH}
\end{equation}
where $\cU_{N+k-i}(\rho) := (I_i \otimes U_{N+k-i}) \rho (I_i \otimes U_{N+k-i})^{\dagger}$ and the identity $I_i$ acts on the radiated $i$ qubits.
The quantum channel from the BH $S$ to the remaining BH $S_{\rm in}$ and that to the new radiation $S_{\rm rad}$, denoted by $\cL^{S \rightarrow S_{\rm in}}$ and $\cL^{S \rightarrow S_{\rm rad}}$, are given by
\begin{align}
&\cL^{S \rightarrow S_{\rm in}}:= \tr_{S_{\rm rad}} \circ  \cL^{S \rightarrow S_{\rm in} S_{\rm rad}},\\
&\cL^{S \rightarrow S_{\rm rad}} := \tr_{S_{\rm in}} \circ  \cL^{S \rightarrow S_{\rm in} S_{\rm rad}},
\end{align}
respectively. Note that they are complementary to each other since $ \cL^{S \rightarrow S_{\rm in} S_{\rm rad}}$ is a unitary dynamics.

We then introduce another person Bob. He collects all the $\ell$ qubits of the new radiation $S_{\rm rad}$ and tries to recover the $k$-qubit information originally stored in $A$. He may additionally use the past radiation $B_{\rm rad}$.  We assume that Bob knows the initial state $\ket{\xi}^{B_{\rm in} B_{\rm rad}}$ and the whole dynamics $\cL^{S \rightarrow S_{\rm in} S_{\rm rad}}$. 
In the recovery process, Bob tries to reproduce the state $\Phi^{AR}$ maximally entangled with the reference $R$ by applying a quantum channel $\cD$ to the quantum systems he has, i.e., to the new and past radiations, $S_{\rm rad}$ and $B_{\rm rad}$.
If he succeeds, it implies that he can access the information source $A$ in the sense that, if the initial state of $A$ had been $\ket{\psi}$, Bob would be able to recover $\ket{\psi}$. 
Due to the no-cloning theorem~\cite{WZ1982}, this implies that the information stored in $A$ is not left in the interior of the BH. Thus, the information in $A$ has been carried away by the radiation $S_{\rm rad}$ of $\ell$ qubits. \\

To quantitatively address whether Bob succeeds in recovering information of $A$, we define and analyze a \emph{recovery error} $\Delta$. Let $\hat{\Phi}^{AR}$ be the final state of the whole process, i.e., 
\begin{equation}
\hat{\Phi}^{AR}_{\cD} := \cD^{S_{\rm rad} B_{\rm rad} \rightarrow A} \circ \cL^{AB_{\rm in} \rightarrow S_{\rm rad}} (\Phi^{AR} \otimes \xi^{B_{\rm in} B_{\rm rad}}). \label{Eq:StateD}
\end{equation}
Since the goal of Bob is to establish the state maximally entangled with the reference $R$, we define the recovery error by
\begin{equation}
\Delta(\xi, \cL)
:=
 \min_{\cD} \bigl[1-  F \bigl( \Phi^{AR}, \hat{\Phi}^{AR}_{\cD} \bigr) \bigr], \label{Eq:Delta19}
\end{equation}
where the minimum is taken over all possible quantum channels $\cD^{S_{\rm rad} B_{\rm rad} \rightarrow A}$ applied by Bob onto the new and the past radiation, $S_{\rm rad}$ and $B_{\rm rad}$. Note that $0 \leq \Delta(\xi, \cL) \leq 1$, and that the recovery error generally depends on the initial state $\xi$ of the initial BH $B_{\rm in}$ and its dynamics $\cL$.

\subsection{Review of the HP protocol without symmetry} \label{SS:HPoriginal}

When the BH has no symmetry, the dynamics $\cL^{S \rightarrow S_{\rm in} S_{\rm rad}}$ is considered to be fully scrambling. This is formulated by choosing each internal unitary dynamics $U_{N+k-i}$ in $\cU_{N+k-i}$ of the BH (see Eq.~\eqref{Eq:dynamicsBH}) at Haar random. When this is the case, we denote the dynamics by $\cL_{\rm Haar}$. The recovery error $\Delta$ in this case is characterized by the min-entropy $H_{\rm min}(B_{\rm in})_{\xi}$ of the initial state of the BH $B_{\rm in}$: for the scrambling dynamics $\cL_{\rm Haar}$ of the BH, the recovery error $\Delta$ satisfies~\cite{HP2007, DBWR2014}
\begin{multline}
\log_2[\Delta(\xi, \cL_{\rm Haar})] \\
\leq \min \bigl\{ 0, k+ \frac{N-H_{\rm min}(B_{\rm in})_{\xi}}{2}  -\ell \bigr \}, \label{Ineq:nosym}
\end{multline}
with high probability.
This clearly shows that, when $\ell \geq k + \frac{N-H_{\rm min}(B_{\rm in})_{\xi}}{2}$, the recovery error $\Delta(\xi, \cL_{\rm Haar})$ decreases exponentially quickly.
Note that the min-entropy $H_{\rm min}(B_{\rm in})_{\xi}$ is large if and only if the initial BH $B_{\rm in}$ and the past radiation $B_{\rm rad}$ are strongly entangled. Hence, in this framework, it is the entanglement between the initial BH $B_{\rm in}$ and the past radiation $B_{\rm rad}$ that determines the time scale for the information recovery to be possible.

For the pure BH, $H_{\rm min}(B_{\rm in})_{\xi}=0$, so that $\Delta(\xi, \mc{L}_{\rm Haar}) \leq 2^{N/2 + k-\ell}$. In contrast, for the mixed BH after the Page time~\cite{Page1993}, $H_{\rm min}(B_{\rm in})_{\xi} = N$, resulting in $\Delta_{tot}(\xi: \mc{L}_{\rm Haar}) \leq 2^{k-\ell}$. The latter case is particularly interesting since the recovery error does not depend on $N$. Thus, no matter how large the initial BH is, Bob can recover the $k$-qubit information when a little more than $k$ qubits are radiated. 

The mechanism behind Eq.~\eqref{Ineq:nosym} is \emph{decoupling}~\cite{HOW2005,HOW2007,HHWY2008,DBWR2014} in the sense that $\cL_{\rm Haar}^{S \rightarrow S_{\rm in}}(\Phi^{AR} \otimes \xi^{B_{\rm in}}) \approx \pi^{S_{\rm in}} \otimes \pi^R$, which is induced by the fully scrambling dynamics of the BH. Decoupling is a basic concept in quantum information theory, which guarantees that information is encoded into good codewords, enabling Bob to retrieve information from the new radiation $S_{\rm rad}$ easily.

\subsection{The HP protocol with symmetry} \label{SS:HPsymmetry}
Here, we describe how situation should be changed in the case of the Kerr BH with the $Z$-axial symmetry. 

\subsubsection{Symmetry, Hilbert spaces, and Quantum Information} \label{SS:BHsym}

When a quantum system has symmetry, the associated Hilbert space is accordingly decomposed. Since the axial symmetry is abelian, the Hilbert spaces $\cH^A$, $\cH^{B_{\rm in}}$, and $\cH^S$ are decomposed into the subspaces invariant under the $Z$-axial symmetry:
\begin{equation}
\cH^A = \bigoplus_{\kappa= 0}^{k} \mc{H}_{\kappa}^A,\  \cH^{B_{\rm in}} = \bigoplus_{\mu= 0}^{N} \mc{H}_{\mu}^{B_{\rm in}} \ \text{and}\ \cH^S= \bigoplus_{m= 0}^{N+k} \mc{H}_{m}^S,
\end{equation}
respectively. 
Note that the labels of the subspaces are based on the number of up-spins, but it can be readily transformed to the $Z$-axis AM. 
We denote the projection onto each subspace by $\Pi_{\kappa}^S$, $\Pi_{\mu}^{B_{\rm in}}$, and $\Pi_m^S$.

The symmetry also affects the quantum information stored in $A$. As mentioned, quantum information in $A$ is represented by the maximally entangled state $\Phi^{AR}$ with the reference $R$. 
The Hilbert space $\cH^R$ is also decomposed according to the decomposition of $\cH^A$, namely, $\cH^R = \bigoplus_{\kappa} \mc{H}^R_{\kappa}$, where $\mc{H}^R_{\kappa}$ is the range of $\tr_A[\Pi_{\kappa}^A \Phi^{AR} \Pi_{\kappa}^A]$.
Using the projection $\Pi^R_{\kappa}$ onto $\mc{H}^R_{\kappa}$, we define 
\begin{equation}
\Phi^{AR}_{\rm diag} = \sum_{\kappa} p_{\kappa} \Phi_{\kappa}^{AR},
\end{equation}
with $p_{\kappa}$ being $\tr[ (I^A \otimes \Pi^R_{\kappa}) \Phi^{AR}]$ and $\Phi_{\kappa}^{AR}$ being $(I^A \otimes \Pi^R_{\kappa}) \Phi^{AR} (I^A \otimes \Pi^R_{\kappa})/p_{\kappa}$.
Since $\Phi_{\rm diag}^{AR}$ is invariant under the axial rotation of $A$, this state contains the information of $A$ that is symmetry-invariant, leading us to call it \emph{symmetry-invariant} information.

Accordingly, we define two errors in recovering information when the BH has symmetry. One is $\Delta_{inv}$ for the symmetry-invariant information in $\Phi^{AR}_{\rm diag}$, and the other is $\Delta_{tot}$ for the total information in $\Phi^{AR}$. They are, respectively, defined by
\begin{align}
&\Delta_{inv}(\xi, \cL)
=
 \min_{\cD} \bigl[1-
 F \bigl( \Phi^{AR}_{\rm diag}, \hat{\Phi}^{AR}_{{\rm diag}, \cD} \bigr)\bigr], \label{Eq:Deltainv24}\\
 &\Delta_{tot}(\xi, \cL) 
=
 \min_{\cD} \bigl[1-
 F \bigl( \Phi^{AR}, \hat{\Phi}^{AR}_{\cD} \bigr) \bigr]. \label{Eq:Deltatotdef}
\end{align}
The recovery error $\Delta_{tot}$ for the total information is the same as $\Delta$ in Eq.~\eqref{Eq:Delta19}, but we below denote it by $\Delta_{tot}$ to clearly distinguish it from $\Delta_{inv}$.
Using a diagram is useful to represent the states $\hat{\Phi}^{AR}_{{\rm diag}, \cD}$ and $\hat{\Phi}^{AR}_{\cD}$ in Eqs.~\eqref{Eq:Deltainv24} and~\eqref{Eq:Deltatotdef}:
\begin{align}
&\hat{\Phi}^{AR}_{{\rm diag}, \cD} = \ \ \   \figbox{0.25}{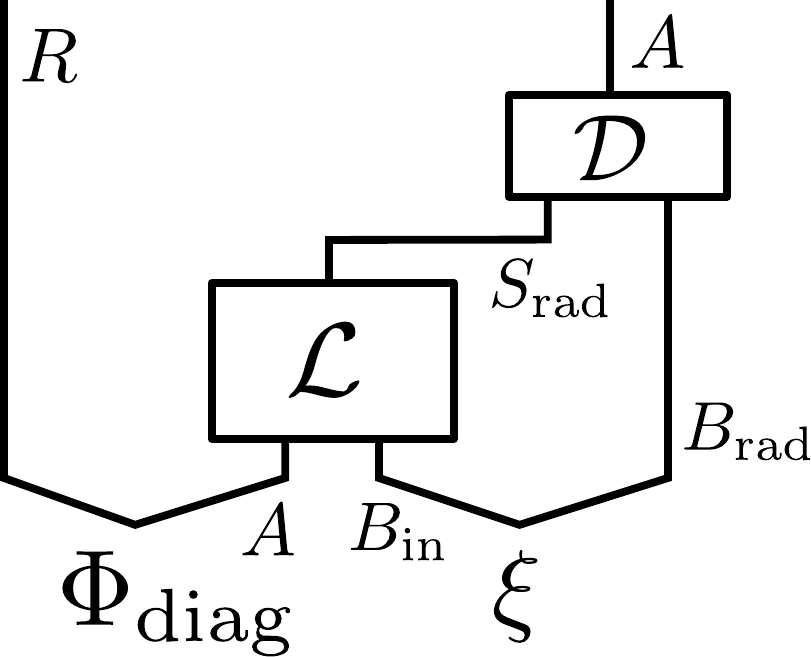}, \label{Eq:a}\\
&\hat{\Phi}^{AR}_{\cD} = \ \ \   \figbox{0.25}{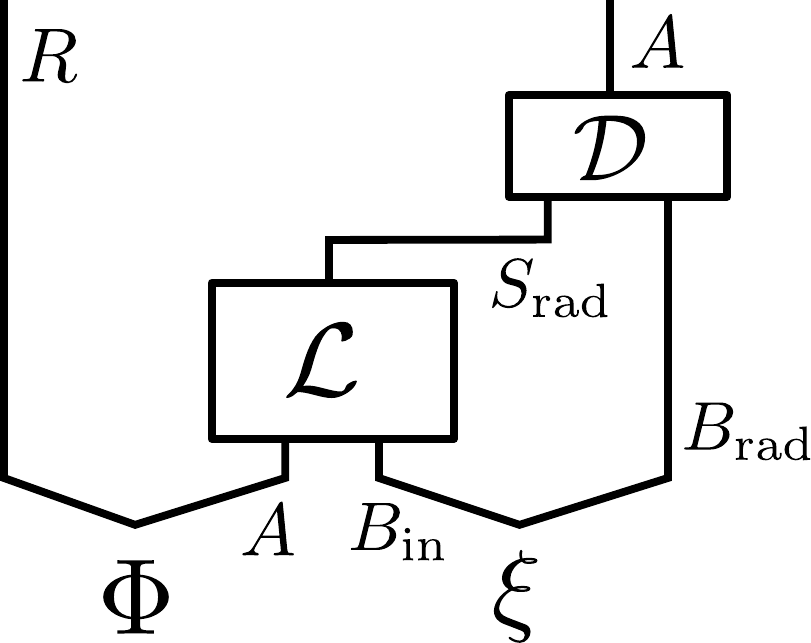}, \label{Eq:b}
\end{align}
respectively. Here, Eq.~\eqref{Eq:b} represents the state given by Eq.~\eqref{Eq:StateD}. The state defined by Eq.~\eqref{Eq:a} is defined similarly by replacing $\Phi$ in Eq.~\eqref{Eq:StateD} with $\Phi_{\rm diag}$.

The states $\hat{\Phi}^{AR}_{{\rm diag}, \cD}$ and $\hat{\Phi}^{AR}_{\cD}$ are related to each other by the pinching map $\cC^R$ on $R$, defined by $\cC^R(\rho):=\sum_{\kappa} \Pi_{\kappa}^R \rho \Pi_{\kappa}^R$ such as $\hat{\Phi}^{AR}_{{\rm diag}, \cD} = \cC^R (\hat{\Phi}^{AR}_{\cD})$. This is also represented as
\begin{align}
&\hat{\Phi}^{AR}_{{\rm diag}, \cD} = \ \ \   \figbox{0.25}{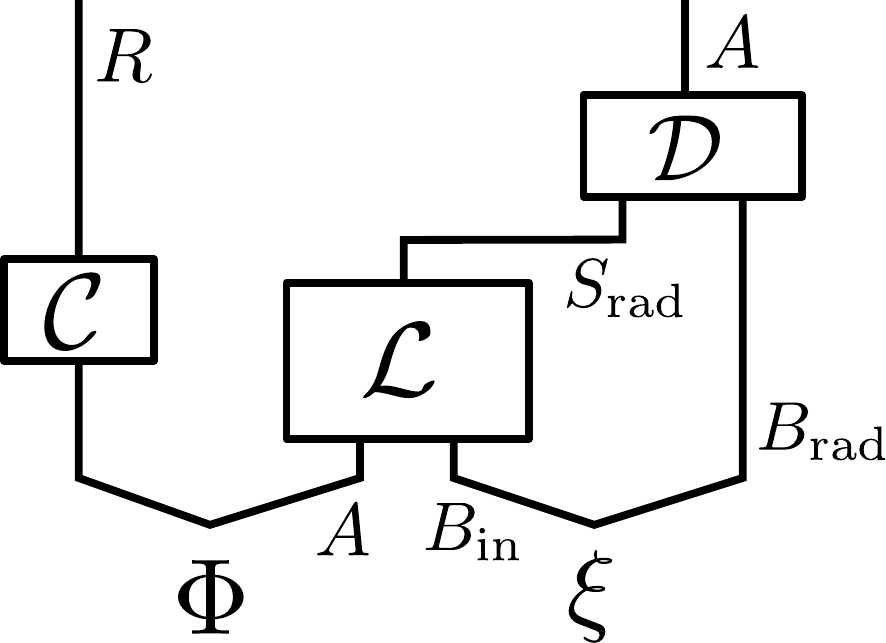}.
\end{align}

\subsubsection{Symmetry-preserving scrambling}

Since the Kerr BH has a $Z$-axial symmetry, its internal unitary dynamics, namely, each $U_{N+k-i}$ in Eq.~\eqref{Eq:dynamicsBH}, should be in the form of 
\begin{equation}
U_{N+k-i} = \bigoplus_{m =0 }^{N+k-i} U_{N+k-i}^{(m)}, \label{Eq:Ui}
\end{equation}
where $U_{N+k-i}^{(m)}$ acts on the subspace of $N+k-i$ spins spanned by the states with $m$ up-spins. These unitaries form a group. 
We assume that each  $U_{N+k-i}^{(m)}$ is Haar random on each subspace, namely, $U_{N+k-i}^{(m)}$ scrambles all quantum states with exactly $m$ up-spins. However, unlike the Haar scrambling, $U_{N+k-i}$ does not change the value of conserved quantity. We call such $U_{N+k-i}$  \emph{symmetry-preserving} scrambling.

We denote the corresponding dynamics of the Kerr BH by $\cL_{\rm Kerr}$. Using the unitarily invariant property of the Haar measure, it is straightforward to observe that the statistics induced by the dynamics $\cL_{\rm Kerr}^{S \rightarrow S_{\rm in} S_{\rm rad}}$ is the same as that induced by a single application of a symmetry-preserving scrambling $U^S_{\rm Kerr}$ on the Kerr BH $S$. Thus, in the following analysis, we consider $U^S_{\rm Kerr}$ instead of $\cL_{\rm Kerr}^{S \rightarrow S_{\rm in} S_{\rm rad}}$.

The symmetry-preserving scrambling unitary $U_{\rm Kerr}^S$ is formally defined by introducing a product of Haar measures: let ${\sf H}_m$ be the Haar measure on the unitary group acting on the subspace with $m$ up-spins. Then, we define the symmetry-preserving scrambling $U_{\rm Kerr}^S$ by 
\begin{equation}
U_{\rm Kerr}^S \sim {\sf H}_{\rm Kerr} := {\sf H}_0 \times {\sf H}_1 \times \dots \times {\sf H}_{N+k}.
\end{equation}

The fact that the symmetry-preserving scrambling cannot change the conserved quantity immediately implies the absence of full decoupling in the Kerr BH, which was also pointed out in Ref.~\cite{Y2019}. Since decoupling is the key technique used to derive the recovery error, Eq.~\eqref{Ineq:nosym}, it is unclear at all whether and how information retrieval can be done in the presence of symmetry. 
The main goal of this paper is to clarify the recovery errors $\Delta_{inv}$ and $\Delta_{tot}$ for the symmetry-invariant and total information when the dynamics is $\cL_{\rm Kerr}$, or equivalently $U_{\rm Kerr}$.

\section{Information recovery --partial decoupling approach--} \label{S:IRPR}
To investigate the recovery errors $\Delta_{inv}$ and $\Delta_{tot}$, we use the partial decoupling approach~\cite{WN2021}, which we briefly summarize in Subsec.~\ref{SS:PDThm}. We also provide a slight generalization. 
We then apply partial decoupling to the Kerr BH in Subsec.~\ref{SS:PDKerr}. We derive formulas for the upper bounds on the recover errors in Subsec.~\ref{SS:PDUB}. Based on the formulas, we numerically investigate the recovery errors in Subsec.~\ref{SS:PDNE}.

\subsection{Partial decoupling theorem} \label{SS:PDThm}

The partial decoupling theorem is proven by two of the authors in Ref.~\cite{WN2021}. 
We here explain its simplest form. In this subsection, the notation and the systems' labeling, such as $S$, $R$, and $E$, do not correspond to those in the HP protocol given in Sec.~\ref{S:HPmodel}.

The situation of partial decoupling is as follows. Let $S$ and $R$ be quantum systems. We assume that the Hilbert space $\cH^S$ has a direct-sum structure, such that 
\begin{equation}
\mc{H}^{S} = \bigoplus_{j \in \mc{J}} \mc{H}^{S}_j. \label{Eq:decompHS}
\end{equation}
The dimensions are denoted as $\dim \cH^S = D_S$ and $\dim \cH^S_j = d_j$.
Let $\varrho^{SR}$ be an initial state on $SR$. A random unitary $U^S$ in the form of $\bigoplus_{j \in \mc{J}} U^{S}_j$ is first applied to $S$, and then a given CPTP map $\cM^{S \rightarrow E}$ is applied. This leads to the state
\begin{equation}
\cM^{S \rightarrow E} (U^S \varrho^{SR} U^{S \dagger}). \label{Eq:resultingstate}
\end{equation}
Despite the fact that this state is dependent on $U^S$, the partial decoupling theorem states that, when $U^S \sim {\sf H}_{\times}:= {\sf H}_1 \times \dots \times {\sf H}_J$, where ${\sf H}_j$ is the Haar measure on the unitary group ${\sf U}(d_j)$ on $\cH_j^S$, the state is typically close to a fixed state independent of $U^S$ as far as a certain entropic condition is satisfied.
For a later purpose, we slightly extend the situation, where the initial state $\varrho^{SR}$ is subnormalized, i.e., $\varrho^{SR} \in \cS_{\leq}(\cH^{SR})$, and the map $\cM^{S \rightarrow E}$ is a trace-non-increasing CP map. 

The fixed (subnormalized) state is explicitly given by $\Gamma^{ER} \in \mc{S}_{\leq}(\mc{H}^{S^*\! RE})$, which is constructed as
\begin{equation}
\Gamma^{ER} = \sum_{j \in \mc{J}} \frac{D_S}{d_j}\zeta^{E}_{jj} \otimes \varrho^{R}_{jj}, \label{Eq:decompdeceomp}
\end{equation}
where 
\begin{align}
&\zeta^{E}_{jj} = \tr_{S'} \bigl[ \Pi^{S'}_j \zeta^{S'E} \bigr],\\
&\varrho^{R}_{jj} =\tr_{S} \bigl[ \Pi^S_j \varrho^{SR} \bigr],
\end{align}
with $\zeta^{S'E} = \mathfrak{J}(\cM^{S \rightarrow E})$ being the Choi-Jamio\l kowski representation of $\cM^{S \rightarrow E}$ (see Eq.~\eqref{Eq:CJdef}). 
Note that the state $\Gamma^{ER}$ has correlation between $E$ and $R$ through the classical value $j$ and has no quantum correlation. Hence, we call the state \emph{partially decoupled}.

The aforementioned entropic condition is given in terms of the conditional min-entropy of an extension $\Gamma^{S^*\!RE} \in \mc{S}_{\leq}(\mc{H}^{S^*\! RE})$ of $\Gamma^{ER}$ by a system $S^*=S'S''$, where $S'$ and $S''$ are replicas of the system $S$:
\begin{equation}
\Gamma^{S^*ER} := \sum_{j,j' \in \mc{J}} \frac{D_S}{\sqrt{d_j d_{j'}}}\zeta^{S'E}_{jj'} \otimes \varrho^{S'' R}_{jj'},
\end{equation}
where 
\begin{align}
&\zeta^{S'E}_{jj'} = \Pi^{S'}_j \zeta^{S'E} \Pi^{S'}_{j'},\\
&\varrho^{S''R}_{jj'} = \Pi^{S \rightarrow S''}_j \varrho^{SR} (\Pi^{S \rightarrow S''}_{j'})^{\dagger}.
\end{align}
Here, $\Pi_j^{S \rightarrow S''} = V^{S \rightarrow S''} \Pi_j^S$, with $V^{S \rightarrow S''}$ is a unitary that maps $S$ to $S''$.

Using these subnormalized states, the partial decoupling theorem is stated as follows.
\begin{theorem} \label{Thm:CoPDecoup}
For any $\delta>0$, it holds that
\begin{equation}
\bigl \| 
\cM^{S \rightarrow E}\bigl( U^S \varrho^{SR} U^{S \dagger} \bigr)  
-
\Gamma^{ER}
\bigr \|_1 
\leq
2^{-\frac{1}{2} H_{\rm min}(S^*|ER)_{\Gamma}} + \delta
\end{equation}
with probability at least
$1-\exp \bigl[- \frac{\delta^2 d_{\rm min}}{48 \|  \varrho^S \|_{\infty}} \bigr]$ in terms of the choice of the unitary $U^S \sim {\sf H}_{\times}$,
where $d_{\rm min} = \min_{j \in \cJ} \{d_j \}$. 
\end{theorem}

The sub-normalized states $\Gamma^{ER}$ and $\Gamma^{S^*\!RE}$ have relatively simple expressions if we use a diagram.
Let $\cC^{S \rightarrow S S^*}_{\cJ}$ be a CPTP map corresponding to a ``$\cJ$-correlator'' given by
\begin{equation}
\cC^{S \rightarrow S S^*}_{\cJ}(\rho^{S}) = C^{S \rightarrow S S^*} \rho^{S} (C^{S \rightarrow SS^*})^{\dagger},
\end{equation}
where $C^{S \rightarrow S S^*}$ is the isometry given by
\begin{equation}
C^{S \rightarrow S S^*}
=
\sum_{j \in \cJ} \Pi_j^{S \rightarrow S''} \otimes \ket{\Phi_j}^{S S'}.
\end{equation} 
Here, $\ket{\Phi_j}^{SS'}$ is the state maximally entangled only in the subspace $\cH_j^{S}\otimes \cH_j^{S'}$.
In other words, the $\cJ$-correlator generates the maximally entangled state in the subspace $\cH_j^{S''}\otimes \cH_j^{S'}$ depending on the value of $j \in \cJ$ in the system $S$, and swaps the system $S$ with $S''$.
Using this CPTP map, the subnormalized states $\Gamma^{ER}$ and $\Gamma^{S^*ER}$ are expressed as follows:
\begin{align}
&\Gamma^{ER} =\ \ \ \figbox{0.24}{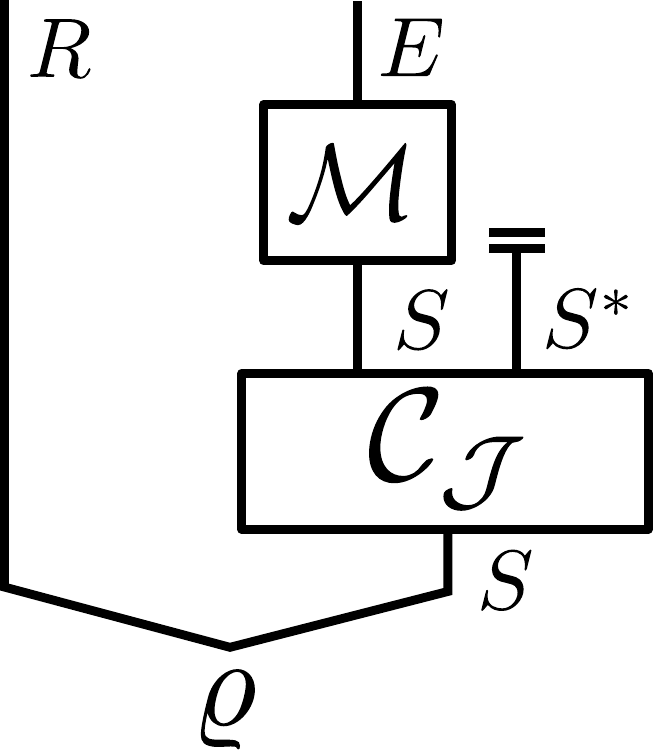}\ ,\\
&\Gamma^{S^*ER} =\ \ \ \figbox{0.24}{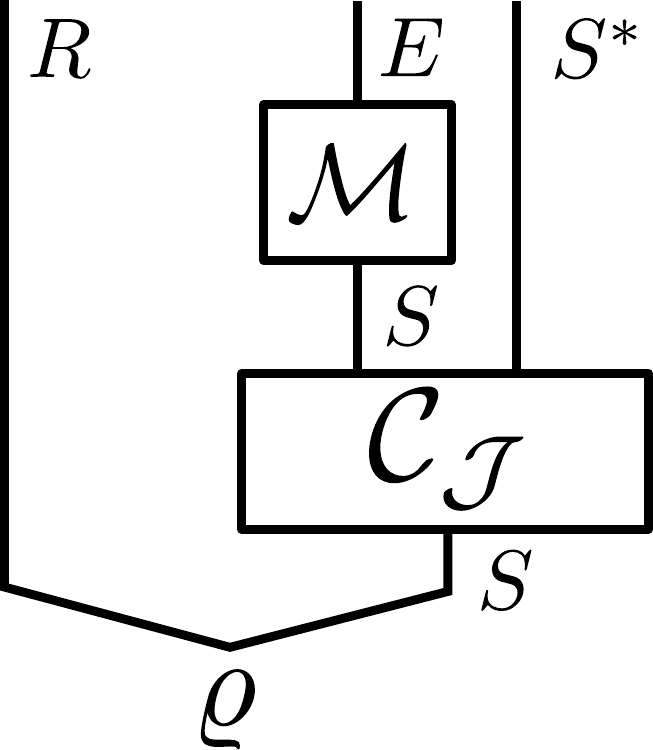}\ .
\end{align}
In the above diagram, the black horizontal double bars represent the trace over the corresponding system, namely, $S^*$ in this case.

Theorem~\ref{Thm:CoPDecoup} implies that, if $H_{\rm min}(S^*|ER)_{\Gamma}$ is sufficiently large, the state $\cM^{S \rightarrow E}\bigl( U^S \varrho^{SR} U^{S \dagger} \bigr)$ is close to $\Gamma^{ER} = \tr_{S^*}[\Gamma^{S^*ER}]$ with high probability.
In the diagram representation, this can be rewritten as
\begin{align}
&\figbox{0.27}{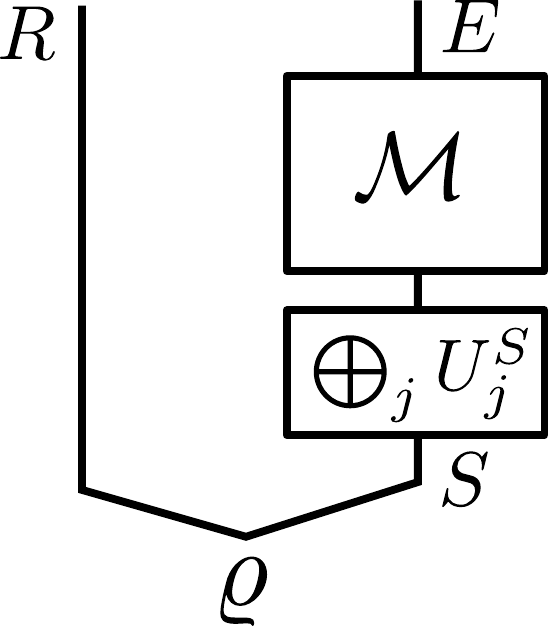}\ \ \  \approx\ \ \ \figbox{0.24}{TensorNetwork3}\ , \label{Eq:377777}
\end{align}
up to the approximation $2^{-H_{\rm min}(S^*|ER)_{\Gamma}/2}$ in the trace norm with high probability.\\

When $d_{\rm min}$ is small, such as $d_{\rm min}=1$, Theorem~\ref{Thm:CoPDecoup}  fails to provide a strong concentration since the probability in the statement becomes tiny for small $\delta$. When this is the case, we can set a ``threshold'' dimension $d_{\rm th}$.

\begin{corollary} \label{Co:CoPDGML}
Consider the same setting as in Theorem~\ref{Thm:CoPDecoup}. Let $d_{\rm th}$ be a positive integer, and $\Pi_{\geq}^S$ be given by $\Pi_{\geq}^S = \sum_{j \in \cJ_{\geq}} \Pi_j^S$, where $\cJ_{\geq} = \{ j \in \cJ : d_j \geq d_{\rm th} \}$. If $\tr[\varrho^{SR} \Pi_{\geq}^S] \geq 1- \delta$, then it holds for any $\delta>0$ that
\begin{multline}
\bigl \| 
\cM^{S \rightarrow E}\bigl( U^S \varrho^{SR} U^{S \dagger} \bigr)  
-
\Gamma^{ER}
\bigr \|_1\\
\leq
\frac{2^{-\frac{1}{2}H_{\rm min}(S^*|ER)_{\Gamma}} }{\sqrt{1-\delta}} + \delta + f(\delta),
\end{multline}
with probability at least $1- \exp \bigl[- \frac{\delta^2 d_{\rm th}}{48 C} \bigr]$,
where $C= \min \{1, \frac{\|  \varrho^S \|_{\infty}}{1-\delta} \}$, and $f(\delta)=2\sqrt{\delta} +\delta+ \frac{\delta}{1-\delta}$.
\end{corollary}

The proofs of Theorem~\ref{Thm:CoPDecoup} and Corollary~\ref{Co:CoPDGML} are given in Appendices~\ref{Proof:CoPDecoup} and~\ref{Pr:CoPDGML}, respectively.

\subsection{Partial decoupling in the Kerr BH} \label{SS:PDKerr}

We now apply the partial decoupling to the Kerr BH by identifying the dynamics $\cL_{\rm Kerr}^{S \rightarrow S_{\rm in} S_{\rm rad}}$ with the symmetry-preserving scrambling $U^S_{\rm Kerr} \sim {\sf H}_{\rm Kerr}$, which is in the form of
\begin{equation}
U^S_{\rm Kerr} = \bigoplus_{m =0}^{N+k} U_m^{S}.
\end{equation}
Since this is the same form as the random unitary used in the partial decoupling theorem, we can directly apply Theorem~\ref{Thm:CoPDecoup} as well as Corollary~\ref{Co:CoPDGML}.

We set $\varrho$ and $\cM$ in Theorem~\ref{Thm:CoPDecoup} to $\Phi^{AR} \otimes \xi^{B_{\rm in}}$ and $\tr_{S_{\rm rad}}$, respectively. The system $E$ in Theorem~\ref{Thm:CoPDecoup} corresponds to $S_{\rm in}$ in this case. 

Using the doubled replica $S^*=S'S''$ of the BH $S$, we have
\begin{align}
&\figbox{0.29}{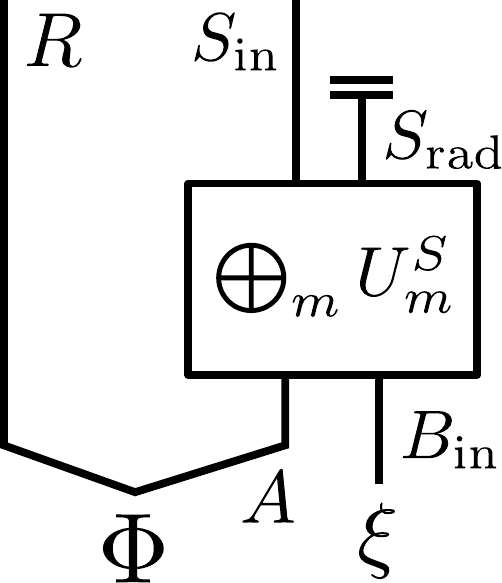} \ \ \approx \ \ \figbox{0.24}{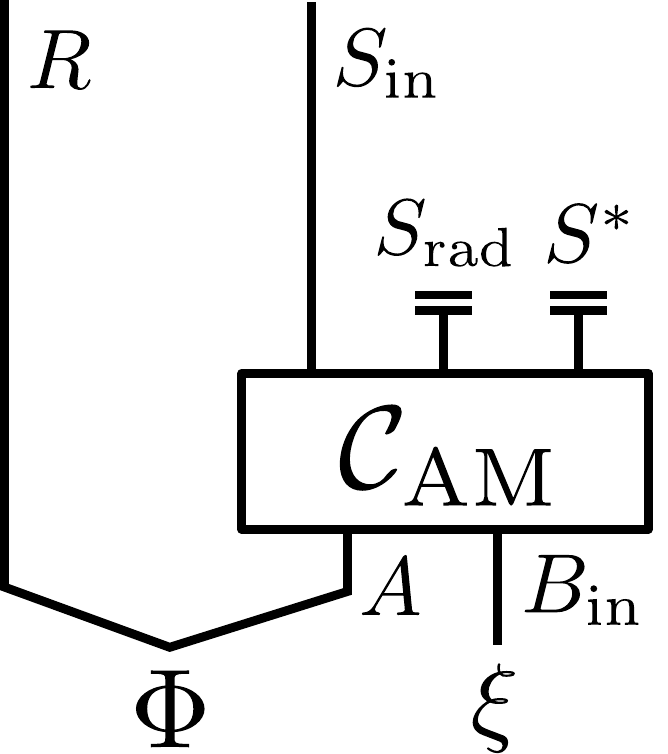}, \label{Eq:TNBH}
\end{align}
with high probability, where $\cC_{\rm AM}$ is the AM-correlator that generates the state maximally entangled in a subspace. More explicitly, $\cC_{\rm AM}$ is the conjugation map by the following isometry
\begin{equation}
C_{\rm AM}^{S \rightarrow S S^*} = \sum_{m=0}^{N+k}  \Pi_m^{S \rightarrow S''} \otimes \ket{\Phi_m}^{SS'}, \label{Eq:CAM}
\end{equation}
where $\ket{\Phi_m}^{SS'}$ is the state maximally entangled only in the subspace $\cH^{S}_m \otimes \cH^{S'}_m$.
According to Theorem~\ref{Thm:CoPDecoup}, Eq.~\eqref{Eq:TNBH} holds up to $\approx 2^{-H_{\rm min}(S^*|S_{\rm in}R)_{\Gamma}/2}$ in the trace norm, where
\begin{align}
\Gamma^{S^*S_{\rm in}R}&=\figbox{0.25}{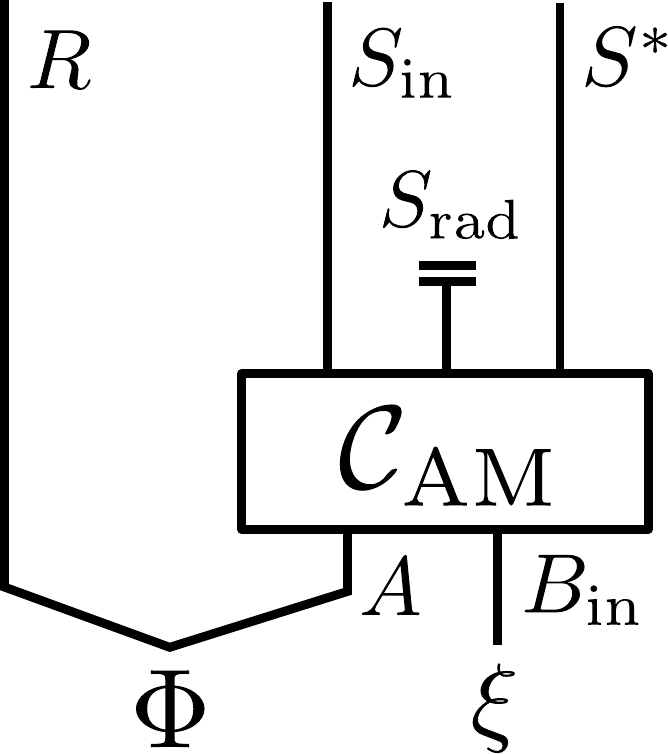}\ .
\end{align}
Hence, if $H_{\rm min}(S^*|S_{\rm in}R)_{\Gamma} \gg 1$, it is highly likely that the L.H.S. of Eq.~\eqref{Eq:TNBH} is partially decoupled, namely, the Kerr BH is partially decoupled from the reference $R$.



\subsection{Empirical smoothing of the conditional min-entropy} \label{SS:ES}

From Theorem~\ref{Thm:CoPDecoup}, the probability for Eq.~\eqref{Eq:TNBH} to hold up to the approximation $2^{-\frac{1}{2}H_{\rm min}(S^*|ER)_{\Gamma}} + \delta$ is at least $1-\exp [- \delta^2 2^k d_{\rm min}/48 ]$, where we have used $\| \pi^A \otimes \xi^{B_{\rm in}} \|_{\infty} \leq 2^{-k}$. Since $d_{\rm min} = 1$ for the axial symmetry, Theorem~\ref{Thm:CoPDecoup} fails to assure a high probability if $\delta < 2^{-k/2}$. This problem can be circumvented by using Corollary~\ref{Co:CoPDGML} instead of Theorem~\ref{Thm:CoPDecoup} and by setting a threshold dimension.
However, the degree of approximation $2^{-\frac{1}{2}H_{\rm min}(S^*|ER)_{\Gamma}}$ turns out not to be tight in general, which typically happens when the conditional min-entropy is used. A \emph{smoothing} is the technique that is exploited to obtain a better bound~\cite{RennerThesis}. We here explain how the bound can be improved by the smoothing technique.

The most general way of smoothing is to use the smooth conditional min-entropy instead of  the conditional min-entropy~\cite{T2016}. We here exploit a limited smoothing by ignoring ``less probable'' subspaces $\cH_n^{S_{\rm rad}}$ of the radiation $S_{\rm rad}$. This is because the smooth conditional min-entropy is computationally intractable. 

To make the idea more precise, we consider the subnormalized state in which the $Z$-axis AM in the new radiation $S_{\rm rad}$ is $n$. We can explicitly compute the average weight $p_n$, namely, the trace of the subnormalized state, by taking the average over $U_{\rm Kerr}^S$:
\begin{equation}
p_n = \frac{1}{2^k} \sum_{m=0}^{N+k} \sum_{\kappa =0}^k \chi_{m-\kappa} \frac{\binom{\ell}{n}\binom{N+k-\ell}{m-n} \binom{k}{\kappa}}{\binom{N+k}{m}}, \label{Eq:pnonon}
\end{equation}
where $\chi_{\mu} = \tr[\xi^{B_{\rm in}} \Pi^{B_{\rm in}}_{\mu}]$ (see Appendix~\ref{App:pn}).
Using $p_n$, we define a probable set $I_{\epsilon}$ by
\begin{equation}
I_{\epsilon} = \{ n \in [0, \ell] : p_n \geq \epsilon \},
\end{equation}
for $\epsilon \geq 0$.
We call the subspaces $\cH_n^{S_{\rm rad}}$ for $n \notin I_{\epsilon}$ rare events. Here, the parameter $\epsilon$ defines the range of the probable set and rare events.
We approximate, by using Theorem~\ref{Thm:CoPDecoup}, only the subnormalized state in the subspace corresponding to the probable events, and count the part corresponding to the rare events as an additional error.

This procedure can be formulated by introducing $\tilde{\Gamma}(\epsilon)$ corresponding to the probable events:
\begin{align}
&\tilde{\Gamma}^{S^*S_{\rm in}R}(\epsilon) :=\Pi^{S_{\rm rad}}_{\epsilon} \Gamma^{S^*S_{\rm in}R} \Pi^{S_{\rm rad}}_{\epsilon}, \label{Eq:Gammatilde}
\end{align}
where $\Pi^{S_{\rm rad}}_{\epsilon} := \sum_{n \in I_{\epsilon}} \Pi_n^{S_{\rm rad}}$.
The error from the rare events is then given by 
\begin{equation}
w(\epsilon) = 1 - \tr[ \tilde{\Gamma}^{S^*S_{\rm in}R}(\epsilon)]=\sum_{n \notin I_{\epsilon}} p_n.
\end{equation}
Using these, we have the following Proposition (see Appendix~\ref{App:Theta} for the proof).

\begin{proposition} \label{Prop:3}
Let $\epsilon \geq 0$. In the above setting, the probability that
\begin{align}
&\figbox{0.29}{TensorNetworkBH1}\ \  \approx \ \ \figbox{0.24}{TensorNetworkBH2-2}, \label{Eq:eer50}
\end{align}
holds up to approximation $\Theta_{\xi}^{\delta, \epsilon}$ in the trace norm is at least $1- \exp[-\frac{\delta ^2 d_{\rm min}(\epsilon)}{48}]$, where 
\begin{equation}
\Theta^{\delta, \epsilon}_{\xi}(N, k, \ell)
= 2^{1-\frac{1}{2}H_{\rm min}(S^*|S_{\rm in}R)_{\tilde{\Gamma}(\epsilon)}} + 2w(\epsilon) + 2\delta, 
\end{equation}
and
$d_{\rm min}(\epsilon) 
= \min \bigl\{ \binom{N+k}{m+n} : m \in [0,N+k-\ell], n \in I_{\epsilon} \bigr\}$.
\end{proposition}

In the definition of $\Theta^{\delta, \epsilon}_{\xi}$, there is a term $2^{1-\frac{1}{2}H_{\rm min}(S^*|S_{\rm in}R)_{\tilde{\Gamma}(\epsilon)}}$. This term does not recover the non-smoothing bound $2^{-\frac{1}{2}H_{\rm min}(S^*|S_{\rm in}R)_{\Gamma}}$ in the limit of $\epsilon \rightarrow 0$. This may imply that the term could be further improved by factor $2$.

\subsection{Upper bounds on recovery errors} \label{SS:PDUB}

Based on the partial decoupling of the Kerr BH, i.e., Proposition~\ref{Prop:3} and using the standard technique in quantum information theory~\cite{HOW2005,HOW2007,HHWY2008,DBWR2014}, we now provide upper bounds on the recovery errors $\Delta_{inv}(\xi, \cL_{\rm Kerr})$ and $\Delta_{tot}(\xi, \cL_{\rm Kerr})$.

The key technique is the Uhlmann's theorem, which states that, if any state $\rho^{AB}$ and a pure state $\ket{\sigma}^{AC}$ (both normalized) satisfy $\| \rho^A - \sigma^A\|_1 \leq \varepsilon$, there exists a CPTP map $\cD^{C \rightarrow B}$ such that 
\begin{equation}
1- \varepsilon \leq F\bigl(\cD^{C \rightarrow B}(\ketbra{\sigma}{\sigma}^{AC}), \rho^{AB} \bigr).
\end{equation}
We apply this to Eq.~\eqref{Eq:eer50}. Since $\xi^{B_{\rm in}} = \tr_{B_{\rm rad}}[ \ketbra{\xi}{\xi}^{B_{\rm in} B_{\rm rad}}]$, we have
\begin{align}
&\figbox{0.27}{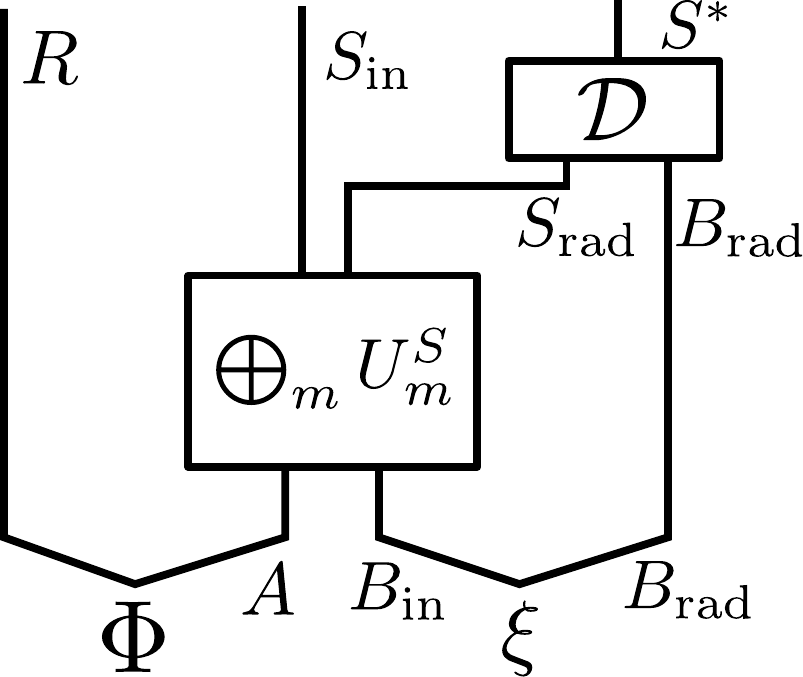} \approx \ \ \figbox{0.24}{TensorNetworkBH2full} \ ,
\end{align}
with approximation at least $1-\Theta_{\xi}^{\delta, \epsilon}$ in terms of the fidelity, where $\cD$ is a CPTP map from the radiations $S_{\rm rad} B_{\rm rad}$ to $S^* \cong S'S'' = S' A'' B''_{\rm in}$. 
Taking the partial trace over $S' B''_{\rm in} $ as well as the remaining BH $S_{\rm in}$, we obtain
\begin{align}
\figbox{0.27}{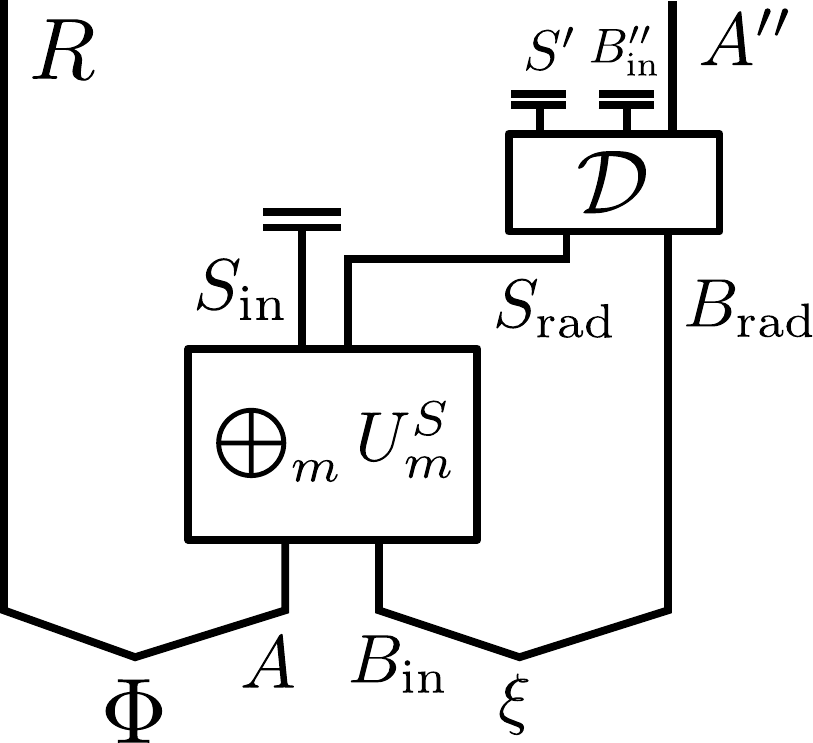} &\approx \ \ \figbox{0.24}{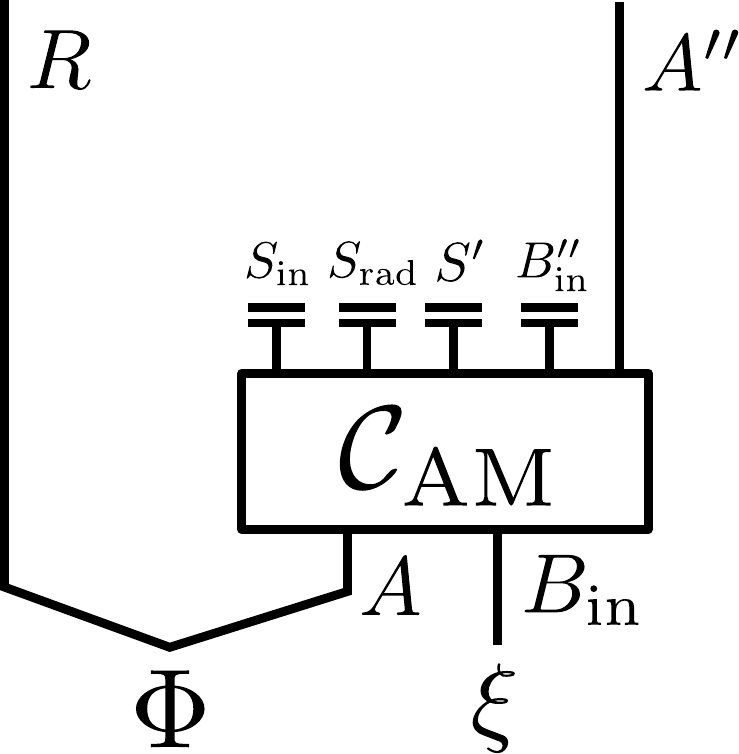}, \label{Eq:r4ccg34tgt}\\
&= \ \ \figbox{0.27}{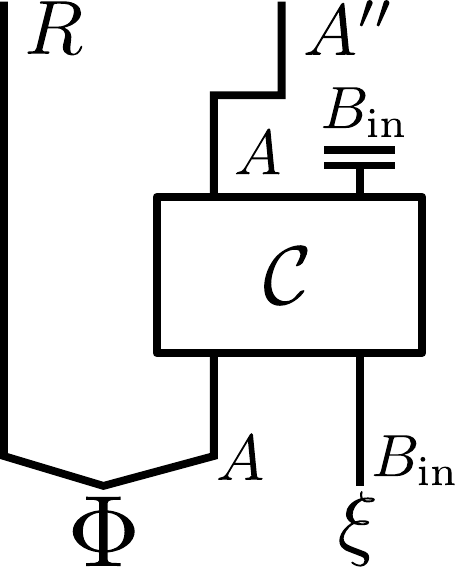}, \label{Eq:erc5555}
\end{align}
where $\cC$ is the pinching map with respect to the $Z$-axis AM defined by $\cC^{S}(\rho^{S}) = \sum_m \Pi_m^{S} \rho^{S} \Pi_m^{S}$. Note that $S=A B_{\rm in}$. The last line simply follows from the fact that the AM-correlator $\cC_{\rm AM}$ generates a normalized state depending on the value of the $Z$-axis AM in $S$ (see Eq.~\eqref{Eq:CAM}), but it is traced out in Eq.~\eqref{Eq:r4ccg34tgt}.

To obtain the recovery error $\Delta_{inv}$ for the symmetry-invariant information, we need to compute the fidelity to $\Phi_{\rm diag} = \cC^R(\Phi^{AR})= \sum_{\kappa =0}^k \Pi_{\kappa}^R \Phi^{AR} \Pi_{\kappa}^R$. Using Eq.~\eqref{Eq:erc5555}, we have
\begin{multline}
\figbox{0.25}{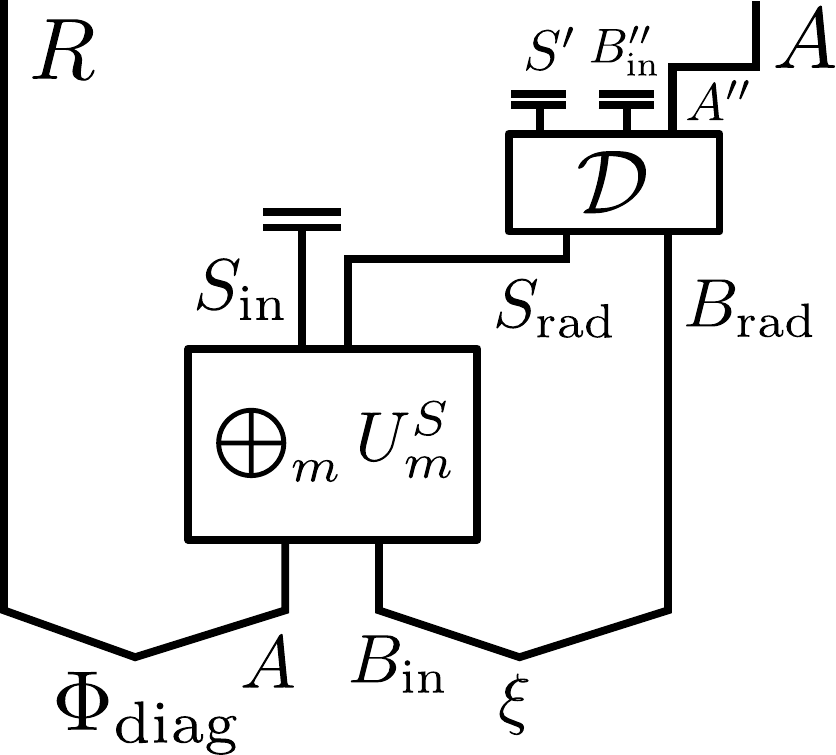} = \ \figbox{0.25}{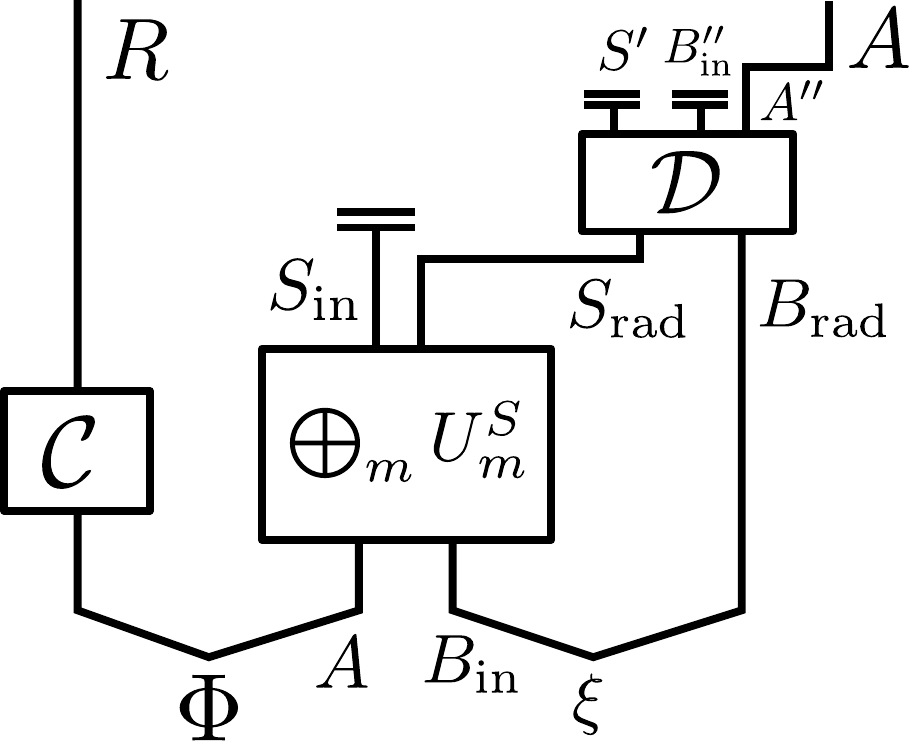},\\
\approx \ \ \figbox{0.27}{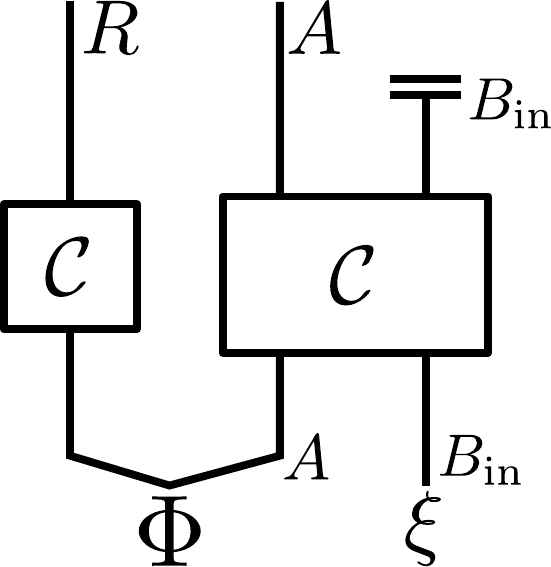} = \Phi_{\rm diag}^{AR}.
\end{multline}
Recalling that the approximation is at least $1-\Theta_{\xi}^{\delta, \epsilon}$ in the fidelity, we arrive at
\begin{equation}
\Delta_{inv}(\xi, \cL_{\rm Kerr})  
\leq 
1- F \bigl( \Phi^{AR}_{\rm diag}, \hat{\Phi}^{AR}_{{\rm diag}, \cD} \bigr)
\leq
\Theta_{\xi}^{\delta, \epsilon}. \label{Eq:rrc1}
\end{equation}

For the recovery error $\Delta_{tot}$ for the total information, we can similarly show that
\begin{align}
\Delta_{tot}(\xi, \cL_{\rm Kerr})
&\leq 
\bigl \| 
\cL_{\rm Kerr}^{S \rightarrow S_{\rm in}}(\Psi^{SR})  
- 
\Gamma^{S_{\rm in}} \otimes \pi^R
\bigr \|_1,\\
&\leq
\bigl \| 
\cL_{\rm Kerr}^{S \rightarrow S_{\rm in}}(\Psi^{SR})  
- 
\Gamma^{S_{\rm in}R}
\bigr \|_1 \notag \\
&\hspace{15mm}+
\bigl \| 
\Gamma^{S_{\rm in}R}
- 
\Gamma^{S_{\rm in}} \otimes \pi^R
\bigr \|_1,\\
&\leq
\Theta_{\xi}^{\delta, \epsilon}+ \eta_{\xi}, \label{Eq:rrc2}
\end{align}
where $\eta_{\xi}$ is given by
\begin{align}
\eta_{\xi} &= \| \Gamma^{S_{\rm in}R} - \Gamma^{S_{\rm in}} \otimes \pi^R \|_1 \label{Eq:44cg33456o]}\\
&=
\frac{1}{2^k}
\sum_{\nu=0}^{N+k-\ell} \sum_{\kappa=0}^{k} F_{\kappa, \nu}(\xi) \binom{N+k-\ell}{\nu}\binom{k}{\kappa}. \label{Eq:eta49}
\end{align}
Here,
\begin{multline}
F_{\kappa, \nu}(\xi) \\
=\biggl|
\sum_{m=0}^{N+k} \frac{\binom{\ell}{m-\nu}}{\binom{N+k}{m}}
\biggl(
\chi_{m-\kappa}
- 
\frac{1}{2^k}
\sum_{\kappa'=0}^{k}
\binom{k}{\kappa'}\chi_{m-\kappa'}
\biggr)
\biggr|,
\end{multline}
and $\chi_{\mu} = \tr[ \xi^{B_{\rm in}} \Pi^{B_{\rm in}}_{\mu}]$. See Appendix~\ref{App:eta} for the derivation.

In summary, we have derived upper bounds on the recovery errors as follows.

\begin{proposition} \label{Prop:4}
Let $\epsilon, \delta \geq 0$. The recovery errors $\Delta_{inv}$ and $\Delta_{tot}$ for symmetry-invariant and total information satisfy 
\begin{align}
&\Delta_{inv}(\xi, \cL_{\rm Kerr})  
\leq
\Theta_{\xi}^{\delta, \epsilon},\\
&\Delta_{tot}(\xi, \cL_{\rm Kerr})   \leq
\Theta_{\xi}^{\delta, \epsilon}+ \eta_{\xi}
\end{align}
with probability at least $1- \exp[-\frac{\delta ^2 d_{\rm min}(\epsilon)}{48}]$, where 
\begin{equation}
\Theta^{\delta, \epsilon}_{\xi}(N, k, \ell)
=2^{1-\frac{1}{2}H_{\rm min}(S^*|S_{\rm in}R)_{\tilde{\Gamma}(\epsilon)}} + 2w(\epsilon) + 2\delta,  \label{Eq:]rrtt}
\end{equation}
and
$d_{\rm min}(\epsilon) 
= \min \bigl\{ \binom{N+k}{m+n} : m \in [0,N+k-\ell], n \in I_{\epsilon} \bigr\}$.
\end{proposition}

Since Proposition~\ref{Prop:4} holds for any $\epsilon \geq 0$, we define $\Theta^{\delta}_{\xi}(N, k, \ell)$ by
\begin{equation}
\Theta^{\delta}_{\xi}(N, k, \ell) :=  \min_{\epsilon \geq 0} \Theta^{\delta, \epsilon}_{\xi}(N, k, \ell), \label{Eq:Theta46}
\end{equation}
and investigate it in the following analysis.
By definition, $\Theta^{\delta}_{\xi}(N, k, \ell)$ is the best possible upper bound of the recovery errors that can be obtained by the smoothing procedure described in Subsec.~\ref{SS:ES}.
We have numerically checked that this generally provides better bounds than those directly obtained from Theorem~\ref{Thm:CoPDecoup}, namely, those without smoothing the entropy.

\subsection{Numerical evaluation} \label{SS:PDNE}

The parameter $\delta$ in Proposition~\ref{Prop:4} controls the trade-off between the upper bounds on the recovery errors and the probability for the symmetry-preserving scrambling $\cL_{\rm Kerr}$ to result in that errors.
In the following numerical analysis, we consider only the term independent of $\delta$, that is,
\begin{align}
\Theta_{\xi}(N, k, \ell) = \min_{\epsilon \geq 0}
\bigl\{
2^{1-\frac{1}{2}H_{\rm min}(S^*|S_{\rm in}R)_{\tilde{\Gamma}(\epsilon)}} + 2w(\epsilon)
\bigr\}. \label{Eq:Theta52}
\end{align}
This suffices since, in the large-$N$ limit, we can typically choose $\delta$ such that $\Theta_{\xi}(N, k, \ell) \gg \delta$ and, at the same time, that the probability is arbitrarily close to one.
To see this, suppose that the $Z$-axis AM in $B_{\rm in}$ is a constant fraction of $N$. In this case, the dimension $d_{\rm min}(\epsilon)$ scales as $N^{c_0 \ell}$ ($c_0$ is some constant). Hence, we can choose $\delta$ to be values with a slightly greater scaling than $N^{-c_0 \ell/2}$, such as $N^{-c_1 \ell/2}$ with $c_1 < c_0$, and make the probability close to one in the large-$N$ limit.
On the other hand, as will be turned out later, $\Theta_{\xi}(N, k, \ell)$ typically scales as $2^{- c_2 \ell}$ ($c_2$ is some constant). Since $\Theta_{\xi}(N, k, \ell) \approx 2^{- c_2 \ell} \gg N^{-c_1 \ell/2} = \delta$ for sufficiently large $N$, $\delta$ is negligible and the probability is arbitrarily close to one.\\

To numerically compute $\Theta_{\xi}(N, k, \ell)$, we need to simplify $H_{\rm min}(S^*|S_{\rm in}R)_{\tilde{\Gamma}(\epsilon)}$ that appears in $\Theta_{\xi}$. This can be done by using the property of the conditional min-entropy for the special type of states.

Let $\{ \cH^A_j \}$ be mutually orthogonal subspaces of $\cH^A$, and $\pi_j^A$ be the completely mixed state on $\cH^A_j$. For any state $\Lambda^{ABC}$ in the form of $\sum_{j=0}^J p_j \pi_j^A \otimes \rho_j^{BC}$, where $\rho_j^{BC} \in \mc{S}(\mc{H}^{BC})$ and $\{ p_j \}$ is a probability distribution, it holds that
\begin{equation}
2^{-H_{\rm min}(AB|C)_{\Lambda}}
\leq
\sum_{j=0}^J \frac{p_j}{d_j} 2^{- H_{\rm min} (B|C)_{\rho_j}}, \label{Eq:eerrc}
\end{equation}
where $d_j = \dim \cH^A_j$.
This immediately follows from the definition of the conditional min-entropy. See Appendix~\ref{App:condmin} for the derivation.

We also use the fact that, for a pure state $\ket{\psi}^{AB}$, the conditional min-entropy is given by
\begin{align}
H_{\rm min}(A|B)_{\psi} &= -2 \log \bigl[\tr[\sqrt{\psi^A}] \bigr]\\
&= -2 \log \bigl[ \tr[\sqrt{\psi^B}] \bigr]. \label{Eq:purestate}
\end{align}

Below, we provide upper bounds on the recovery errors for the pure and mixed Kerr BH. In both cases, we characterize the state $\xi$ of the initial Kerr BH $B_{\rm in}$  by the expectation value $L$ of the $Z$-axis AM and the standard deviation $\delta L$, defined by
\begin{align}
& L =\tr [L_Z \xi] \label{Eq:LL}\\
& \delta L= \bigl( \tr [(L_Z - L)^2 \xi] \bigr)^{1/2},\label{Eq:deltaLL}
\end{align}
where $L_Z = \frac{\hbar}{2}\sum_{i = 1}^N Z_i$ with $\hbar$ being the Plank constant and $Z_i$ being the Pauli-$Z$ operator acting on the $i$th qubits. In the remaining of this paper, we set $\hbar =1$ except the last part of Subsec.~\ref{SS:DT}, where we discuss a physical origin of information leakage.

\begin{figure*}[th!]
\centering
\includegraphics[width=\textwidth, clip]{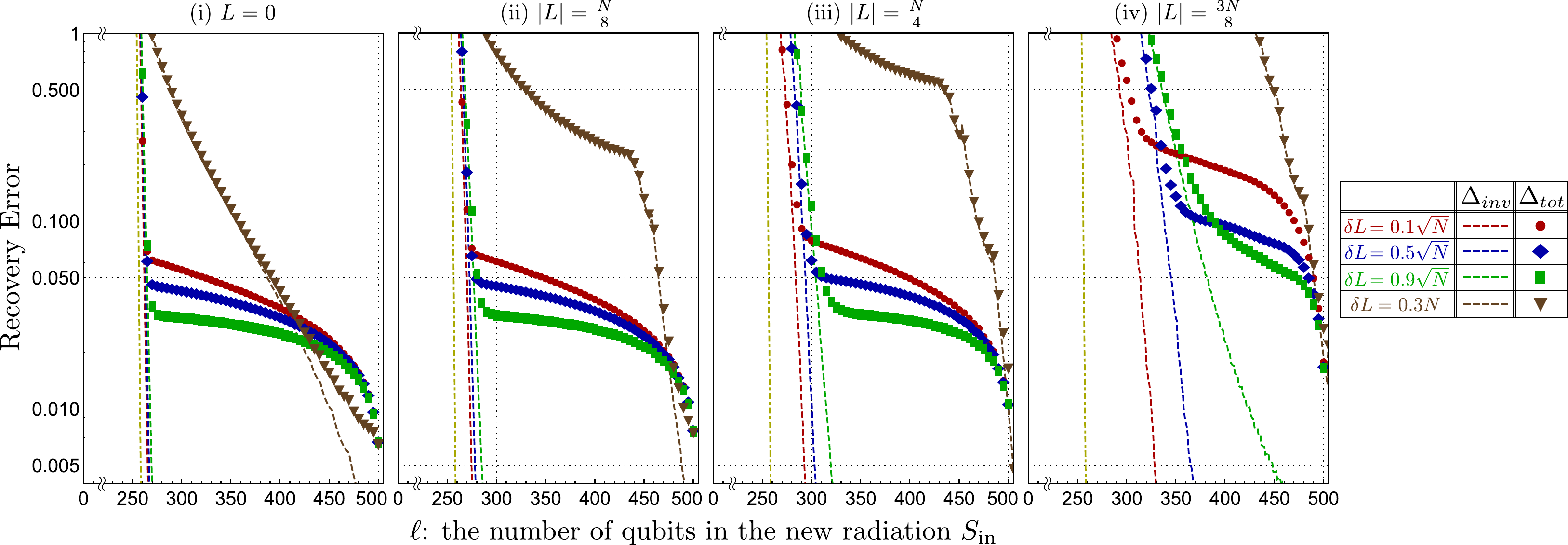}
\caption{Recovery Errors from the radiation emitted by the pure Kerr BH. Upper bounds on $\Delta_{inv}(\xi:\mc{L}_{\rm Kerr})$ (dashed lines) and those on $\Delta_{tot}(\xi:\mc{L}_{\rm Kerr})$ (filled markers) for the pure Kerr BH are plotted. We particularly consider the initial pure states $\ket{\xi}^{B_{\rm in}}$ with various $Z$-axis AM $L$ and fluctuation $\delta L$. 
The figures (i) - (iv) show the cases for $|L| = 0, N/8, N/4$, and $3N/8$, respectively, while $\delta L$ is chosen to be $0.1 \sqrt{N}$ (red), $0.5 \sqrt{N}$ (blue), $0.9 \sqrt{N}$ (green), and $0.3N$ (brown) for each $L$. 
For comparison, $\Delta_{inv}$ for $(L, \delta L)=(0, 0)$ is plotted by a yellow dash-dotted line in each figure.
The size $N$ of the initial Kerr BH and the size $k$ of the quantum information source are set to $500$ and $5$, respectively. We observe that the error $\Delta_{inv}$ for symmetry-invariant information decays exponentially quickly after $\ell = |S_{\rm in}|$ exceeds a certain number, while the error $\Delta_{tot}$ for the whole information typically has a plateau. The dependence of these features on $L$ and $\delta L$, as well as their physical mechanism, are discussed in the main text. Note however that, in any case, $\Delta_{inv}, \Delta_{tot} \approx 1$ unless $\ell > (N+k)/2$, which is a consequence of the no-cloning property of quantum information~\cite{WZ1982}. 
}
\label{Fig:noent}
\end{figure*}

\subsubsection{The pure Kerr BH}

Let us first consider the pure Kerr BH $B_{\rm in}$ with the initial state in the form of
\begin{equation}
\ket{\xi(L, \delta L)}^{B_{\rm in}} = \sum_{\mu=0}^N \sqrt{\chi_{\mu}(L, \delta L)} \ket{\varphi_{\mu}}^{B_{\rm in}}, \label{Eq:xipure}
\end{equation}
where $\mu$ stands for the number of up-spins in $B_{\rm in}$. We assume that $\{\chi_\mu(L, \delta L)\}_{\mu}$ are given by
\begin{equation}
\chi_\mu(L, \delta L) \propto \exp\biggl[ - \frac{(\mu - L-N/2)^2}{\delta L^2} \biggr], \label{Eq:chiGaussian}
\end{equation}
with a proper normalization such that $\sum_{\mu=0}^N \chi_{\mu}(L, \delta L) =1$, and $\ket{\varphi_{\mu}}^{B_{\rm in}}$ is an arbitrary pure state in $\cH^{B_{\rm in}}_{\mu}$.
Due to the symmetry-preserving scrambling, the recovery errors do not depend on the choice of $\ket{\varphi_{\mu}}^{B_{\rm in}}$.
Note that the expectation value of $L_Z$ and its standard deviation for the state $\ket{\xi(L, \delta L)}^{B_{\rm in}}$ deviates from $L$ and $\delta L$, respectively, due to the discretization, but the deviations are negligibly small.

In this case, the subnormalized state $\tilde{\Gamma}^{S^*S_{\rm in}R}(\epsilon)$ can be explicitly written as
\begin{equation}
\tilde{\Gamma}^{S^*S_{\rm in}R}_{L, \delta L}(\epsilon)
=
\sum_{n \in I_{\epsilon}} p_n \pi_n^{S'_{\rm rad}} \otimes \tilde{\Phi}^{S_{\rm in}S_{\rm in}'S''R}_n(L, \delta L), \label{Eq:amm4q1}
\end{equation}
where $\{ p_n \}$ is the probability distribution given by Eq.~\eqref{Eq:pnonon}, and 
\begin{widetext}
\begin{equation}
\ket{\tilde{\Phi}_n(L, \delta L)}^{S_{\rm in}S'_{\rm in}S''R}
\propto
\sum_{m=0}^{N+k}
\sqrt{\frac{1}{\binom{N+k}{m}}}
\bigl(\Pi_{m-n}^{S_{\rm in}} \ket{\Phi}^{S_{\rm in} S_{\rm in}'}\bigr) 
\otimes
\bigl(\Pi_{m}^{S \rightarrow S''} (\ket{\Phi}^{AR} \otimes \ket{\xi}^{B_{\rm in}} )\bigr),
\end{equation}    
\end{widetext}
with a proper normalization constant.
Here, we have used the notation that $\Pi_{a}$ is the zero operator if $a <0$.

Using Eqs.~\eqref{Eq:eerrc} and~\eqref{Eq:purestate}, we obtain 
\begin{equation}
H_{\rm min}(S^*|ER)_{\tilde{\Gamma}(\epsilon)} 
\geq 
k - \log [\gamma_{\rm pure}(N, k, \ell |\xi) ], \label{Eq:44f235554}
\end{equation}
where
\begin{multline}
\gamma_{\rm pure}(N,k,\ell|\xi) \\= \sum_{n \in I_{\epsilon}} \biggl( \sum_{m=0}^{N+k} \sum_{\kappa=0}^k  \sqrt{\frac{\chi_{m-\kappa}}{\binom{N+k}{m}}}\binom{N+k-\ell}{m-n} \binom{k}{\kappa} \biggr)^2, \label{Eq:gammapure}
\end{multline}
with $\chi_{m-\kappa} = \tr[ \xi^{B_{\rm in}} \Pi^{B_{\rm in}}_{m-\kappa}]$.
This bound does not depend on the sign of $L$. Namely, $\ket{\xi(\pm L, \delta L)}$ result in the same bound.
This is naturally expected since the rotation direction of the Kerr BH should not affect any features of the protocol.

From Eqs.~\eqref{Eq:Theta52} and~\eqref{Eq:44f235554}, we obtain
\begin{multline}
\Theta_{\xi}(N, k, \ell) \leq \\
\min_{\epsilon \geq 0}
\bigl\{
 2^{1-k/2} \sqrt{\gamma_{\rm pure}(N, k, \ell|\xi)} + 2w(\epsilon)
\bigr\}. \label{Ineq:174}
\end{multline}
Since $w(\epsilon) = \sum_{n \notin I_{\epsilon}} p_n$, we can numerically evaluate this bound, from which we obtain upper bounds on $\Delta_{inv}$ and $\Delta_{tot}$.\\

In Fig.~\ref{Fig:noent}, upper bounds on $\Delta_{inv}$ and $\Delta_{tot}$, i.e., 
\begin{align}
&\Delta_{inv}(\xi, \cL_{\rm Kerr})  
\leq
\Theta_{\xi},\\
&\Delta_{tot}(\xi, \cL_{\rm Kerr})   \leq
\Theta_{\xi}+ \eta_{\xi}
\end{align}
are provided as functions of the number $\ell$ of qubits in the new radiation $S_{\rm rad}$ for various $L$ and $\delta L$. 
We observe that $\Delta_{inv}$ starts decaying exponentially after a certain number of $\ell$. The number, as well as the decaying speed, strongly depends on both $|L|$ and $\delta L$. To characterize this, we introduce $\ell_{\Delta}(L, \delta L)$ for $\Delta>0$ by
\begin{equation}
\ell_{\Delta}(L, \delta L)
=
\min \bigl\{ \ell \in [0, N+k] | \Theta_{\xi} \leq \Delta \bigr\}. \label{Eq:ellDelta}
\end{equation}
Then, it is observed from Fig.~\ref{Fig:noent} that $\ell_{\Delta}(L, \delta L) \geq \ell_{\Delta}(0, 0)$ for any $\Delta \geq 0$. This implies that the Kerr BH has a \emph{delay} in the onset of releasing information compared to the trivial case ($L = \delta L =0$), which is given by the yellow dotted lines in Fig.~\ref{Fig:noent}. 
The delay is especially large when $L$ is large (see figure (iv)) or when $\delta L = \Theta(N)$ (see brown plots), indicating that the symmetry-invariant information cannot be recovered from the radiation if either $L$ or $\delta L$ is extremely large.

We also observe that $\Delta_{tot}$ behaves differently depending on whether $\delta L = O(\sqrt{N})$ or $O(N)$.
When $\delta L=O(\sqrt{N})$, $\Delta_{tot}$ first behaves similarly to $\Delta_{inv}$. 
However, soon after that, the decreasing of the error becomes slow, and $\Delta_{tot}$ remains at a non-negligible value until $\ell \approx N$. This indicates that a part of the information remains in the Kerr BH until the last moment. We call such residual information \emph{information remnant}. From the plots for different $L$ but same $\delta L$, we also observe that the amount of information remnant is independent of $L$.
When $\delta L=O(N)$, $\Delta_{tot} \approx \Delta_{inv}$ for any $\ell$, implying that there is no information remnant.

\subsubsection{The mixed Kerr BH}

\begin{figure*}[t!]
\centering
\includegraphics[width=\textwidth,clip]{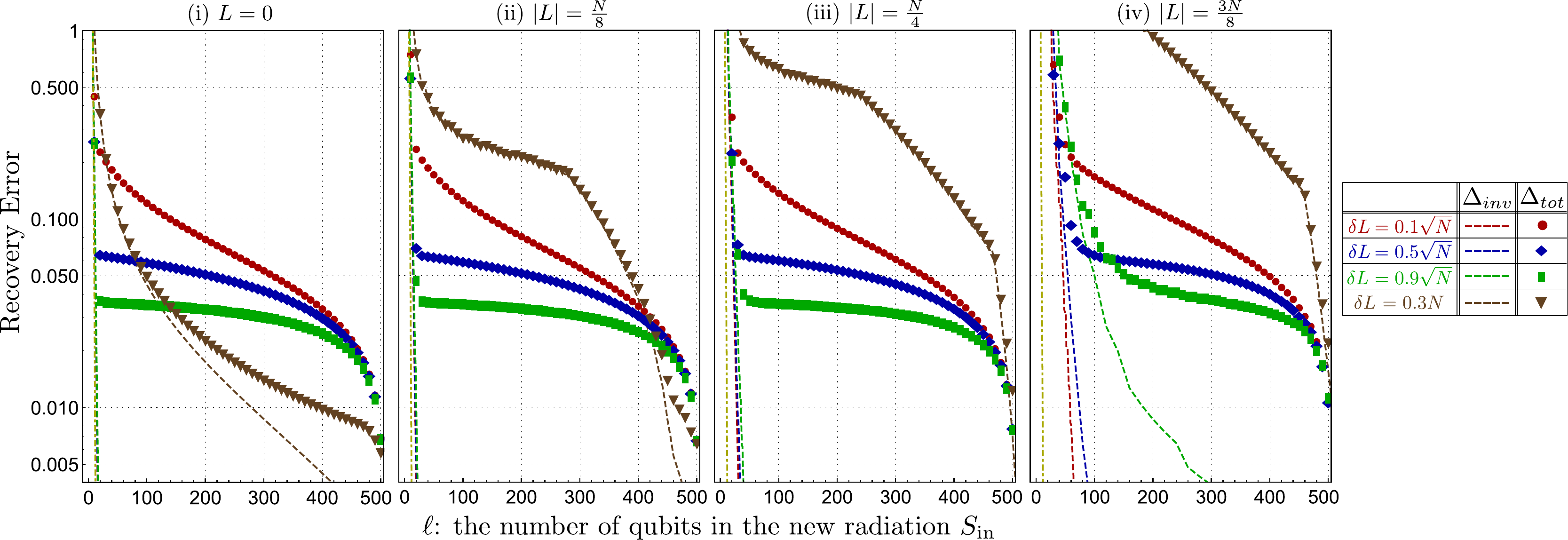}
\caption{Recovery Errors from the radiation emitted by the mixed Kerr BH. Upper bounds on $\Delta_{inv}(\xi:\mc{L}_{\rm Kerr})$ (dashed lines) and those on $\Delta_{tot}(\xi:\mc{L}_{\rm Kerr})$ (filled markers) for the mixed Kerr BH are plotted. The mixed Kerr BH $B_{\rm in}$ is entangled with the past radiation $B_{\rm rad}$. We consider a family of specific entangled states with various $Z$-axis AM $L$ and fluctuation $\delta L$.
The figures (i) - (iv) show the bounds for $|L| = 0, N/8, N/4$, and $3N/8$, respectively, where $\pm L$ leads to the same result as explained in the caption of Fig.~\ref{Fig:noent}, while $\delta L$ is chosen to be $0.1 \sqrt{N}$ (red), $0.5 \sqrt{N}$ (blue), $0.9 \sqrt{N}$ (green), and $0.3N$ (brown). The recovery error $\Delta_{inv}$ for $(L, \delta L)=(0, 0)$ is also plotted by a yellow dash-dotted line for comparison.
The size $N$ of the initial Kerr BH is set to $500$, and the size $k$ of the quantum information source to $5$.
Similarly to the pure Kerr BH, we observe that both $\Delta_{inv}(\xi:\mc{L}_{\rm Kerr})$ and $\Delta_{tot}(\xi:\mc{L}_{\rm Kerr})$ start decreasing at almost same timing, and that only $\Delta_{tot}$ stops decreasing soon after, except the one for $\delta L = 0.3N$.
}
\label{Fig:ent}
\end{figure*}

We next consider the mixed Kerr BH, which is initially entangled with the past radiation $B_{\rm rad}$. For simplicity, we only consider the following entangled states:
\begin{equation}
\ket{\xi(L, \delta L)}^{B_{\rm in} B_{\rm rad}} = \sum_{\mu=0}^N \sqrt{\chi_{\mu}(L, \delta L)} \ket{\Phi_{\mu}}^{B_{\rm in} B_{\rm rad}}, 
\end{equation}
where $\{ \chi_{\mu} \}_{\mu}$ are given by Eq.~\eqref{Eq:chiGaussian}, and $\ket{\Phi_{\mu}}^{B_{\rm in} B_{\rm rad}}$ is the maximally entangled state in $\cH^{B_{\rm in}}_{\mu} \otimes \cH^{B_{\rm rad}}_{\mu}$. 

In this case, the state $\Gamma^{S^*S_{\rm in}R}$ reduces to
\begin{equation}
\Gamma^{S^*S_{\rm in}R}
=
\sum_{n=0}^{\ell} \sum_{\mu=0}^{N} p_{n, \mu} \pi_{n}^{S'_{\rm rad}} \otimes \pi_{\mu}^{B''_{\rm in}} \otimes  \tilde{\Psi}_{n,\mu}^{S_{\rm in}S'_{\rm in}A''R}, \label{Eq:q243rd-0[pm:l1}
\end{equation}
where $p_{n,\mu}$ and the normalized pure state $\tilde{\Psi}_{n, \mu}^{S_{\rm in}S'_{\rm in}A''R}$ are given by
\begin{align}
&p_{n, \mu}
=
\frac{\chi_{\mu}(L, \delta L) \binom{\ell}{n}}{2^k} \sum_{m=0}^{N+k} \frac{\binom{N+k-\ell}{m-n} \binom{k}{m-\mu}}{\binom{N+k}{m}},
\end{align}
and
\begin{multline}
|\tilde{\Psi}_{n, \mu} \rangle ^{S_{\rm in}S'_{\rm in}A''R}
\propto \\
 \sum_{m=0}^{N+k}
\frac{1}{\sqrt{\binom{N+k}{m}}}
\bigl(\Pi^{S_{\rm in}}_{m-n} |\Phi \rangle^{S_{\rm in}S'_{\rm in}} \bigr) \otimes \bigl(\Pi^{A \rightarrow A''}_{m-\mu}|\Phi \rangle^{AR} \bigr),
\end{multline}
respectively. Note that $\sum_{\mu} p_{n, \mu} = p_n$, where $p_n$ is given by Eq.~\eqref{Eq:pnonon}.

Using Eqs.~\eqref{Eq:eerrc} and~\eqref{Eq:purestate}, we obtain
\begin{equation}
H_{\rm min}(S^*|ER)_{\tilde{\Gamma}(\epsilon)} 
\geq 
k - \log  \bigl[ \gamma_{\rm mixed}(N, k, \ell| \xi) \bigr]
\end{equation}
where 
\begin{multline}
\gamma_{\rm mixed}(N,k,\ell|\xi) := 
\sum_{\mu=0}^N \frac{\chi_\mu(L, \delta L)}{\binom{N}{\mu}}  \sum_{n \in I_{\epsilon}}\\
 \biggl( \sum_{m=0}^{N+k} \sqrt{\frac{1}{\binom{N+k}{m}}} \binom{N+k-\ell}{m-n} \binom{k}{m-\mu} \biggr)^2.
\end{multline}
This leads to
\begin{multline}
\Theta_{\xi}(N, k, \ell) \\
\leq 
\min_{\epsilon \geq 0}
\bigl\{
 2^{1-k/2} \sqrt{\gamma_{\rm mixed}(N, k, \ell|\xi)} + 2w(\epsilon)
\bigr\},
\end{multline}
and, we obtain upper bounds on $\Delta_{inv}$ and $\Delta_{tot}$. We again note that $\pm L$ provide the same bounds.

\begin{table*}[tb!]
  \centering
  \begin{tabular}{c|c||c|c}
  \multicolumn{2}{c||}{} &  \shortstack{\\  \ \ \  $\delta L = O(\sqrt{N})$  \ \ \ } &  \shortstack{\\ \ \ \  $\delta L = O(N)$  \ \ \ } \\ \hline \hline
\shortstack{\\ Delay for the pure Kerr BH} & for any $L$&  \shortstack{\\  $O(\sqrt{N})$}  &   \shortstack{\\$O(N)$} \\ \hline
     \multirow{2}{*}{  \begin{tabular}{c}
  \shortstack{\\ Delay for the mixed Kerr BH}  \end{tabular}}  & \shortstack{for small $L$} &  \shortstack{\\  $O(1)$} &  \shortstack{\\  $O(\sqrt{N})$?} \\ \cline{2-4}
 &    \shortstack{\\ for large $L$} &  \shortstack{\\  $O(1)$} &  \shortstack{\\ $O(N)$?}
  \end{tabular}
\caption{A summary of the delay of information leakage for the pure and mixed Kerr BHs, which initially have the $Z$-axis AM $L$ and the fluctuation $\delta L$.
The results are obtained from the numerical analysis based on partial decoupling, where we varies $N$ up to $500$. The order is only in terms of $N = |B_{\rm in}|$. In the case of the mixed Kerr BH with small $L$ and $\delta L = O(N)$, it was hard to decide the scaling from our numerics.}
\label{Tab:DelaySummary}
\end{table*}

The upper bounds on the recovery errors are numerically plotted in Fig.~\ref{Fig:ent} as functions of $\ell= |S_{\rm rad}|$. Similarly to the pure Kerr BH, the plots are for various $L$ and $\delta L$.
We observe that $\Delta_{inv}$ decays in a manner similar to the pure Kerr BH. That is, the recovery error starts decreasing after a certain number of $\ell$, as well as the onset of decreasing has a delay compared to the trivial case ($L = \delta L =0$, yellow dotted lines in Fig.~\ref{Fig:ent}).
While the delay is not so large for small $L$ and $\delta L$, it becomes substantially large when $|L|$ is large or $\delta L = \Theta(N)$.

The recovery error $\Delta_{tot}$ for the total information also behaves similarly to the pure Kerr BH, and it remains at a non-zero value until $\ell \approx N$. This implies that, even for the mixed Kerr BH, the information remnant exists. It turns out that the amount of information remnant depends only on $\delta L$ and is the same as that for the pure Kerr BH for the same $\delta L$, which will be elaborated on later.

The significant difference from the pure Kerr BH is that the timing at which the errors start decreasing can be much earlier. It indeed ranges widely from $O(k)$ to $O(N)$ depending on the $Z$-axis AM $L$ and on its fluctuation $\delta L$ of the initial Kerr BH.

\subsubsection{Scaling of the delay with respect to $N$}
We also provide numerical results about how the delay $\ell_{\Delta}(L, \delta L)$ scales with respect to the size $N$ of the initial Kerr BH $B_{\rm in}$ for various $Z$-axis AMs $L$ and their fluctuations $\delta L$. 
The result is summarized in Tab.~\ref{Tab:DelaySummary} for the pure and mixed Kerr BHs.
See Appendix~\ref{S:e} for the details.

\section{Information recovery --macroscopic physics approach--} \label{S:IRMP}

In Sec.~\ref{S:IRPR}, we have numerically shown that the delay and the information remnant are the consequences of the symmetry of the Kerr BH. 
In this section, we analyze them from different perspectives. This lead to great insights into their physical origins.

We consider the delay in Subsec.~\ref{SS:DT} and the information remnant in Subsec.~\ref{SS:IRSB}.

\subsection{Delay of information leakage and themodynamics} \label{SS:DT}

To interpret the delay of information leakage, we propose a new concept that we call \emph{clipping of entanglement} and show that it reproduces both the result for the BH without symmetry and the delay of information leakage for the Kerr BH.
As we will see, the clipping argument provides a concise condition for the information leakage to occur only in terms of dimensions of the Hilbert spaces. 

Before we start, we emphasize that the argument based on the entanglement clipping is applicable not only for the axial symmetry but for any symmetry with extensive conserved quantity.

\subsubsection{Clipping of Entanglement}

One of the significant features of the scrambling dynamics is that it generates nearly maximal entanglement between subsystems~\cite{Page1993,L1978,HLW2006}. We argue that such a maximal entanglement is responsible also for the quick information leakage.

For clarity, we first consider the BH without symmetry. Moreover, instead of directly considering the quantum information source $\Phi^{AR}$, i.e., the maximally entangled state between $A$ and the reference $R$, we consider the situation where the reference $R$ is measured in an arbitrary basis to obtain a $k$-bit outcome.
It is known that if the $k$-bit outcome can be inferred from the radiations $B_{\rm rad}$ and $S_{\rm rad}$, then the maximally entangled state $\Phi^{AR}$ can be reproduced~\cite{K2007}, that is, the quantum information in $A$ is recovered.

When the BH has no symmetry, its dynamics is fully scrambling and generates nearly maximal entanglement between $S_{\rm rad}$ and $S_{\rm in}$. This implies that the marginal state in the smaller subsystem should be nearly completely mixed.
In contrast, the support of the marginal state in the larger subsystem needs to be ``clipped'' at the dimension of the smaller one, because any bipartite pure state should have an equal size, known as the Schmidt rank, in the two subsystems~\cite{NC2000}. 

The information recovery can be understood from the entanglement clipping in the larger subsystem.
In the case of the pure BH, no information is obtained from the radiation when $|S_{\rm rad}| \leq |S_{\rm in}|$ since no clipping occurs in the new radiation $S_{\rm rad}$. Note that all the eigenvalues of the marginal state in $S_{\rm rad}$ are nearly the same, from which no information would be retrieved.
When $|S_{\rm rad}| \geq |S_{\rm in}|$, the clipping takes place in $S_{\rm rad}$, namely, the marginal state in $S_{\rm rad}$ is released from the full rank state. 
In the case of the mixed BH, the entanglement clipping in $S_{\rm rad}$ sets in much earlier since the BH is initially already on the verge of clipping due to the entanglement with the past radiation $B_{\rm rad}$, which is under control of the decoding person Bob.

To capture the information-theoretic consequence of the entanglement clipping in the new radiation $S_{\rm rad}$, we introduce a degree of clipping $C$ in $\cH^{S_{\rm rad}}$:
\begin{equation}
C(\cH^{S_{\rm rad}}) := H(B_{\rm in})_{\xi} + \log \biggl[ \frac{\dim \cH^{S_{\rm rad}}}{\dim \cH^{S_{\rm in}}} \biggr].
\end{equation}
Note that the degree of clipping $C(\cH^{S_{\rm rad}})$ can be negative. For instance, $C(\cH^{S_{\rm rad}}) = -(N+k)$ when $|S_{\rm rad}| = 0$ and $H(B_{\rm in})_{\xi} = 0$.
When $C(\cH^{S_{\rm rad}}) > k$, the marginal state in the new radiation $S_{\rm rad}$ is sufficiently clipped so that the $2^k$ states, corresponding to all the possible $2^k$ outcomes in $R$, can be fit in the marginal state without significant overlaps each other.
Thus, we propose that the condition for the information recovery is given by
\begin{equation}
C(\cH^{S_{\rm rad}}) > k.
\end{equation}
By explicitly writing down the dimensions, we obtain
\begin{equation}
\ell > k + \frac{N-H(B_{\rm in})_{\xi}}{2},
\end{equation}
as a condition for the information recovery to be possible. Note that this agrees well with the result of the HP protocol without symmetry, namely, Eq.~\eqref{Ineq:nosym}.\\

We can apply the same argument to the Kerr BH. 
In this case, it is of crucial importance that the symmetry-preserving scrambling in the form of $\bigoplus_m U^S_m$, where $U^S_m$ is the Haar scrambling in the subspace with $m$ up-spins, generates entanglement between $S_{\rm in}$ and $S_{\rm rad}$ only in the subspaces.
This introduces a special structure of entanglement between them.
To be more precise, we consider the initial Kerr BH $B_{\rm in}$ with a fixed $Z$-axis AM $L=\lambda N$ and $\delta L = 0$, where $\lambda \in [-1/2, 1/2]$. The number of up-spins in $B_{\rm in}$ is $\lambda N+N/2$. 
Due to the symmetry-preserving scrambling, the subspace $\cH_{n}^{S_{\rm rad}}$ with $n$ up-spins in $S_{\rm rad}$ is entangled only with $\cH_{n'}^{S_{\rm in}}$ satisfying $n + n' \approx \lambda N + (N+k)/2$. 

Taking this constraint into account, the entanglement clipping should be considered in each subspace $\cH_n^{S_{\rm rad}}$ separately. We thus define the degree of clipping in $\cH_n^{S_{\rm rad}}$ by
\begin{equation}
C(\cH_n^{S_{\rm rad}}) := H(B_{\rm in})_{\xi} + \log \biggl[ \frac{\dim \cH^{S_{\rm rad}}_{n}}{\dim \cH^{S_{\rm in}}_{\lambda N  + (N+k)/2 -n} } \biggr], \label{Eq:Cramp0}
\end{equation}
and require that
\begin{equation}
C(\cH_n^{S_{\rm rad}}) > k, \label{Eq:Cramp}
\end{equation}
for most $n$. 

We here emphasize that we have not used any property of the $Z$-axial symmetry of the Kerr BH except that it is additive. Thus, the above clipping condition should be valid for any BH with extensive conserved quantity.\\

If Eq.~\eqref{Eq:Cramp} is not satisfied for certain $n$, the information in the corresponding subspace should be counted as recovery error.
This point can be elaborated by considering the probability distribution $W(n)$ over the number $n$ of spins in $S_{\rm rad}$ induced by the random choice of $S_{\rm rad}$ from $S$. The probability is given by
\begin{equation}
W(n)  = \frac{ \dim \cH_{\lambda N+ (N+k)/2 - n}^{S_{\rm in}} \dim \cH_{n}^{S_{\rm rad}}}{\dim \cH_{\lambda N+ (N+k)/2}^S}. \label{Eq:DistS1}
\end{equation}
With respect to this probability distribution, we define probable values of $n$ and require for Eq.~\eqref{Eq:Cramp} to hold for any probable $n$. We count the total probability over non-probable values of $n$ as recovery errors.
This is similar to the empirical smoothing we exploited in Subsec.~\ref{SS:ES}.

\subsubsection{Comparison between entanglement clipping and partial decoupling}

To check the validity of the argument of entanglement clipping, let us compare the condition~\eqref{Eq:Cramp} with the partial decoupling result.
To this end, we need to specify the probable set of $n$. We define the probable set by the set of $n$ around its average $\langle n \rangle$ under the probability distribution of Eq.~\eqref{Eq:DistS1}.
The average is explicitly given by
\begin{equation}
\langle n \rangle= \biggl(\frac{L}{N+k} +1/2 \biggr) \ell.
\end{equation}
We require that Eq.~\eqref{Eq:Cramp} holds for all $n$ such that
\begin{equation}
|n - \langle n \rangle| \leq c  \sqrt{\langle \delta n^2 \rangle}, \label{Eq:deltaell}
\end{equation}
where $\delta n:=n - \langle n \rangle$.
Here, $c>0$ is a parameter related to the recovery error since taking a larger $c$ means that we require Eq.~\eqref{Eq:Cramp} for a wider range of $n$, resulting in less error.
We then define the number $\hat{\ell}_c(L)$ by the minimum of $\ell=|S_{\rm rad}|$ for the clipping condition $C(\cH_n^{S_{\rm rad}}) > k$ to hold for any $n$ satisfying Eq.~\eqref{Eq:deltaell}.

The quantity $\hat{\ell}_c(L)$ should correspond to the  $\ell_{\Delta}(L, 0)$, defined based on the partial decoupling by Eq.~\eqref{Eq:ellDelta}. In the following, we show that $\hat{\ell}_c(L) \approx \ell_{\Delta}(L, 0)$, which indicates that the clipping argument results in nearly the same prediction as the partial decoupling.

To this end, we provide an approximate expression of $\hat{\ell}_c(L)$. The dimensions in Eq.~\eqref{Eq:Cramp0} are given by
\begin{align}
&\dim \cH_{n}^{S_{\rm rad}} = \binom{\ell}{n},\\
&\dim \cH_{\lambda N+(N+k)/2 - n}^{S_{\rm in}} = \binom{N+k-\ell}{\lambda N+(N+k)/2 - n}.
\end{align}
These are well-approximated by a function $s(\lambda):=-(1/2 - \lambda) \log[1/2 - \lambda] -(1/2 + \lambda) \log[1/2 + \lambda]$. We then obtain
\begin{widetext}
\begin{equation}
\log \biggl[ \frac{\dim \cH_{n}^{S_{\rm rad}}}{\dim \cH_{\lambda N+ (N+k)/2 - n}^{S_{\rm in}}} \biggr]
=
\ell \ \! s \biggl( \frac{n}{\ell} - \frac{1}{2} \biggr)
-
(N+k-\ell) \ \! s \biggl( \frac{\lambda N+ (N+k)/2- n}{N+k-\ell} - \frac{1}{2} \biggr).
\end{equation}    
\end{widetext}

We can further simplify the expression by assuming $k \ll N$ and expanding $n$ around its average $\langle n \rangle  \approx (\lambda +1/2) \ell$ as
\begin{equation}
n \approx (\lambda +1/2) \ell + \delta n.
\end{equation}
It is straightforward to show
\begin{multline}
\log \biggl[ \frac{\dim \cH_{n}^{S_{\rm rad}}}{\dim \cH_{\lambda N+ (N+k)/2 - n}^{S_{\rm in}}} \biggr]\\
\approx
-(N+k- 2 \ell) \ \! s (\lambda) + 2  \delta n |s'(\lambda)|. \label{Eq:rrer81}
\end{multline}
We also introduce the initial degree of clipping $C_{\rm ini}$, i.e., the degree of clipping when $|S_{\rm rad}| = 0$. It is approximately given by 
\begin{equation}
C_{\rm ini} \approx H(B_{\rm in})_{\xi} - (N+k)s(\lambda) < 0.
\end{equation}
Using these, we arrive at
\begin{equation}
C(\cH^{S_{\rm rad}}_n)
\approx
C_{\rm ini} + 2 \ell s (\lambda)
+
2 \delta n
|s'(\lambda)|.
\end{equation}

As mentioned, we require $C(\cH^{S_{\rm rad}}_n) > k$ for all $\delta n$ such that $|\delta n| \leq c \sqrt{\langle \delta n^2 \rangle}$. The standard deviation $\sqrt{ \langle \delta n^2 \rangle}$ can be also rephrased in terms of $s(\lambda)$.
From Eq.~\eqref{Eq:DistS1}, we have
\begin{align}
\log[W(n)] 
& = -\frac{1}{2} \frac{(N+k)}{\ell (N+k-\ell)}  |s''(\lambda)| \delta n^2 + O(\delta n^3).
\end{align}
Since $W(n)$ can be approximated by a Gaussian distribution when $1 \ll k \ll N$, the variance $\langle \delta n^2 \rangle$ is approximately given by
\begin{equation}
\langle \delta n^2 \rangle \approx  \biggl( 1 - \frac{\ell}{N+k} \biggr) \frac{\ell}{|s''(\lambda)|}. \label{Eq:sss}
\end{equation}

Altogether, the number $\hat{\ell}_c(L)$ is given by the minimum $\ell$ that satisfies
\begin{equation}
C_{\rm ini} + 2 \ell s (\lambda)
+
2 c \sqrt{ \biggl( 1 - \frac{\ell}{N+k} \biggr) \frac{\ell}{|s''(\lambda)|} } 
|s'(\lambda)| > k.
\end{equation}
By solving this, we obtain an explicit form of $\hat{\ell}_c(L)$, which is
\begin{equation}
\hat{\ell}_c(L) \approx \ell_0(L) + c \ \! \ell_{\rm fl}(L), \label{Eq:lthCr} 
\end{equation}
where
\begin{align}
&\ell_0(L) = - \frac{C_{\rm ini}}{2 s(\lambda)} + \frac{k}{2 s(\lambda)},\\
&\ell_{\rm fl}(L)= \frac{|s’(\lambda)|}{s(\lambda)} \sqrt{\frac{\ell_0(L)}{|s''(\lambda)|} \biggl(1- \frac{\ell_0(L)}{N+k} \biggr) }. \label{Eq:nontri}
\end{align}

\begin{figure}[t!]
\centering
\includegraphics[width=0.45 \textwidth,clip]{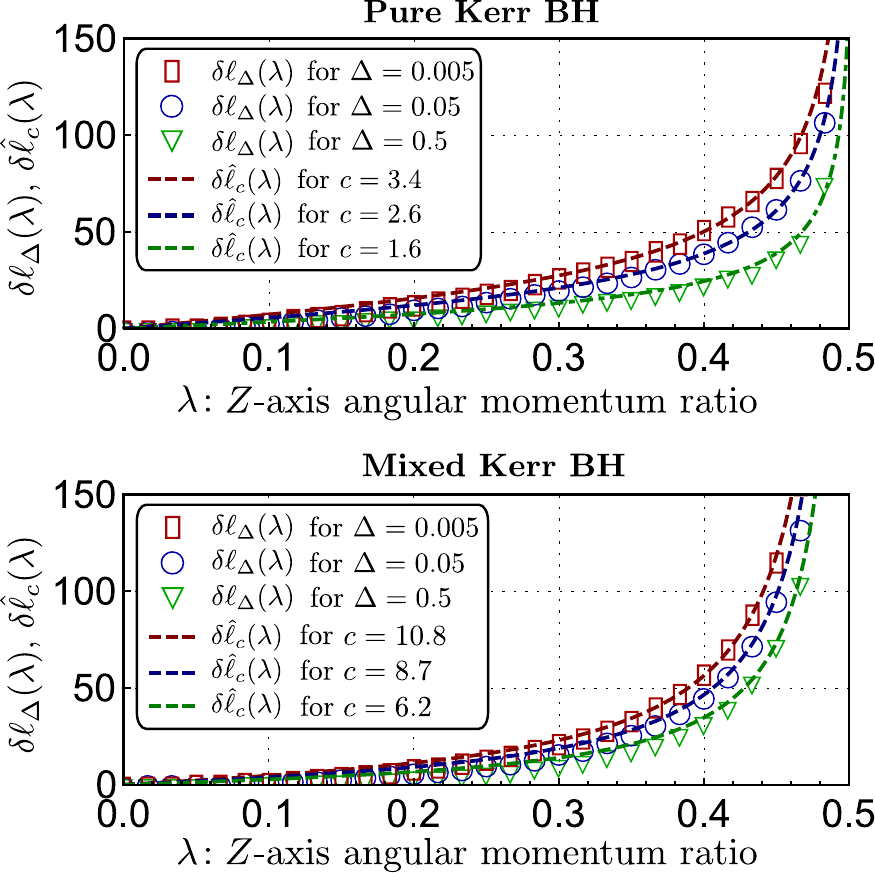}
\caption{Numerical comparison of the delays computed in two ways. One is $\delta \ell_{\Delta}(\lambda) := \ell_{\Delta}(\lambda N, 0) - \ell_{\Delta}(0,0)$, based on the partial decoupling (Eq.~\eqref{Eq:ellDelta}), and the other is $\delta \hat{\ell}_c(\lambda) := \hat{\ell}_c(\lambda N) - \hat{\ell}_c(0)$ obtained from the clipping of entanglement (Eq.~\eqref{Eq:lthCr}). 
They are both plotted as functions of the $Z$-axis AM ratio $\lambda =L/(\hbar N)$ of the initial Kerr BH $B_{\rm in}$. The upper and lower figures show the delays for the pure and mixed Kerr BHs, respectively. The size $N$ of the initial Kerr BH $B_{\rm in}$ is set to $300$, and the size $k$ of the quantum information source $A$ is fixed to $3$.
In both figures, $\delta \ell_{\Delta}(\lambda)$ is computed for $\Delta=0.005$ (red plots), $0.05$ (blue plots), and $0.5$ (green plots), while $\delta \hat{\ell}_c(\lambda)$ is computed for $c = 3.4$ (red dashed line), $2.6$ (blue dashed line), and $1.6$ (green dashed line) for the pure BH, and $10.8$ (red dashed line), $8.7$ (blue dashed line), and $6.2$ (green dashed line) for the mixed BH. 
In both pure and mixed Kerr BHs, they nearly coincide, indicating that the argument of entanglement clipping provides good estimations.}
\label{Fig:Delay}
\end{figure}

In Fig.~\ref{Fig:Delay}, we numerically compare $\hat{\ell}_c(L)$ and $\ell_{\Delta}(L, 0)$. 
We observe that, by choosing proper $c$ and $\Delta$, which both play the role of a smoothing parameter, they coincide very well. Hence, we conclude that the clipping argument provides a good estimate of the necessary number of qubits in the new radiation $S_{\rm rad}$ for the information to become recoverable from the radiations.\\

In Eq.~\eqref{Eq:lthCr}, we have decomposed $\hat{\ell}_c(L)$ into two terms. The first term, $\ell_0(L)$, represents a delay originated from an information theoretic reason: when the number of qubits in $S_{\rm rad}$ is increased by one, the degree of clipping is increased by $2s(\lambda)$ since a qubit has entropy $\approx s(\lambda)$. Thus, $|S_{\rm rad}|$ should be at least $\ell_0(L)$ so as to cancel the initial negative degree of clipping $C_{\rm ini}$, and more for $S_{\rm rad}$ to typically have a sufficiently large space to store $2^k$ states without significant overlaps. That is why we refer to $\ell_0(L)$ as an \emph{information-theoretic} delay.

It is the second term $\ell_{\rm fl}(L)$ that is a non-trivial consequence of the symmetry of the Kerr BH. It stems from the fluctuation of the $Z$-axis AM in the radiation, making the clipping condition harder to be fulfilled. We hence call $\ell_{\rm fl}(L)$ a \emph{fluctuational} delay.
Note also that, when the $Z$-axis AM $L$ of the initial Kerr BH is small, the delay of information leakage, $\delta \hat{\ell}_c(L) := \hat{\ell}_c(L)- \hat{\ell}_c(0)$, satisfies
\begin{equation}
\delta \hat{\ell}_c(L) \propto \ell_{\rm fl}(L),
\end{equation}
since $\ell_0(L) \approx \ell_0(0)$ and $\ell_{\rm fl}(0) = 0$ for small $L$.

\subsubsection{Delay and BH Thermodynamics}
The fluctuational delay $\ell_{\rm fl}(L)$ can be further rewritten in terms of thermodynamic quantities of the Kerr BH. In this context, it is more natural to quantify the delay by the amount of the $Z$-axis AM rather than the number of qubits. We hence introduce $L_{\rm fl}$ by
\begin{equation}
L_{\rm fl} := \hbar \lambda \ell_{\rm fl}(L), \label{Eq:Lell}
\end{equation}
and investigate it. In this Subsection, we explicitly write the Plank constant $\hbar$. Correspondingly, we also introduce $L_0 := \hbar \lambda \ell_0(L)$, which is the $Z$-axis AM corresponding to the information-theoretic delay $\ell_0(L)$.

In order to consider the thermodynamics of the Kerr BH, it is necessary to take its Hamiltonian into account. The Hamiltonian should be invariant under the rotation around the $Z$-axis and also sufficiently random for the dynamics to be symmetry-preserving scrambling.
Since we are interested in the thermodynamics properties that do not strongly rely on the randomness, we especially consider the Kerr BH with temperature $T$ sufficiently higher than the energy scale of the Hamiltonian, so that the thermodynamic entropy $S(\lambda, L, T)$ of the Kerr BH is independent of $T$, and the free energy $F(T, \lambda, L)$ is approximately given by $-T S(\lambda, L)$.

Let $\omega:= - \bigl(\frac{\partial F}{\partial L} \bigl)_{T, N}$ be an intensive state function conjugate to the $Z$-axis AM $L$, and $\alpha:=  \bigl(\frac{1}{L}\frac{\partial L}{\partial T}\bigl)_{\omega, N}$ be the sensitivity of the $Z$-axis AM to the temperature. 
More specifically, the state function $\omega(T, \lambda)$ is the angular velocity of the Kerr BH.
Since the thermodynamic entropy $S(\lambda, L)$ is given by 
\begin{equation}
S(\lambda, L) = k_B s(\lambda) N,
\end{equation}
where $k_B$ is the Boltzmann constant, we have 
\begin{align}
&\omega(T, \lambda) = \frac{k_B Ts'(\lambda)}{\hbar},\\
&\alpha(T,\lambda) = - \frac{s'(\lambda)}{ s''(\lambda) \lambda T},
\end{align}
in our particular model of the Kerr BH.
Hence, it follows that
\begin{equation}
\biggl| \frac{s'(\lambda)^2}{s''(\lambda)} \biggr|
=
\frac{\lambda \hbar}{k_B} | \omega(T,L) \alpha(T,L)|.
\end{equation}
Using these and assuming $k \ll N$, we obtain from Eqs.~\eqref{Eq:nontri} and~\eqref{Eq:Lell} that
\begin{align}
&L_{\rm fl}  \approx \frac{L}{S(\lambda, L)}  \sqrt{k_B \bigl|\omega(T,\lambda) \alpha(T, \lambda)  \bigr| \biggl( 1 - \frac{L_0}{L} \biggr) L_0}.\label{Eq:pert}
\end{align}
Note that $\omega(T,\lambda) \alpha(T, \lambda)$ is independent of the temperature $T$, and so is $L_{\rm fl}$.

After radiating the $Z$-axis AM by $L_0$ to fulfill the information-theoretic requirement, the Kerr BH must further radiate an extra amount $L_{\rm fl}$ of the $Z$-axis AM to release the information. 
It is clear from Eq.~\eqref{Eq:pert} that this amount $L_{\rm fl}$ is solely determined by intensive thermodynamic quantities of the initial Kerr BH, that is, the $Z$-axis AM per entropy $L/S(\lambda, L)$, the angular velocity $\omega(T, \lambda)$ of the Kerr BH, and the coefficient $\alpha(T, \lambda)$ of the thermal sensitivity of the $Z$-axis AM.\\

The delay of information leakage is, thus, closely related to the thermodynamic properties of the BH. This, in turn, implies that, if we understand the thermodynamics of the BH well, it is possible to predict how quickly the BH releases the quantum information therein.


We emphasize that the above argument is likely to be applicable to the system with any global abelian symmetry. A particularly important application is the case where energy is conserved during the unitary dynamics of the system. Assuming that the dynamics is energy-preserving scrambling in the sense that it scrambles only quantum states with the same energy scale, we can evaluate the delay of information leakage based on Eq.~\eqref{Eq:nontri}.
Suppose that the initial quantum many-body system BH is thermodynamical and has energy $E$ with negligibly small fluctuation. 
The thermodynamic entropy $S(E)$ of the system, where $E$ is the internal energy, and the heat capacity $C_V$ satisfy
\begin{equation}
|C_V| = k_B \biggl|\frac{S'(E)^2}{S''(E)} \biggr|.
\end{equation}
Thus, a similar calculation from Eq.~\eqref{Eq:nontri} leads to
\begin{equation}
E_{\rm fl}  \approx \frac{E}{S(E)}  \sqrt{k_B |C_V| \biggl( 1 - \frac{E_0}{E} \biggr) \frac{E_0}{E}},
\end{equation} 
where $E_0$ and $E_{\rm fl}$ are the energy corresponding to the information-theoretic and fluctuational delays, respectively.
As $E_0/E$ is between $0$ and $1$, the amount is basically determined by the ratio between the energy $E$ and entropy $S(E)$ of the initial black hole, and the heat capacity $C_V$, connecting the information leakage to thermodynamic quantities of the quantum many-body system.

\subsection{Information remnant and symmetry-breaking} \label{SS:IRSB}

We finally investigate the origin of the information remnant characterized by $\eta_{\xi}$ (see Eq.~\eqref{Eq:44cg33456o]}). As explained in Subsec.~\ref{SS:PDNE}, an important feature of $\eta_{\xi}$ is that it depends on the initial fluctuation $\delta L$ of the $Z$-axis AM in $B_{\rm in}$ but not on the $Z$-axis AM $L$ itself.

We start with a simple observation that the information remnant should exist when the unitary dynamics of the BH is symmetric for the following reason. First, throwing the system $A$ into the initial BH $B_{\rm in}$ changes the $Z$-axis AM of the BH, which remains unchanged by the subsequent unitary dynamics in $S=AB_{\rm in}$. When the system $S$ is eventually split to two random subsystems $S_{\rm in}$ and $S_{\rm rad}$, the change induced by throwing $A$ into the BH is inherited to both $S_{\rm in}$ and $S_{\rm rad}$. This implies that the original value of the $Z$-axis AM in $A$, which is kept stored in the reference system $R$, can be inferred to some extent from the remaining BH $S_{\rm in}$ by measuring its $Z$-axis AM. Hence, the information in $A$ about the coherence between difference values of the $Z$-axis AM cannot be fully accessed from the radiation $S_{\rm rad}$.

This observation also implies that the amount of information remnant should be related to that of the fluctuation of the $Z$-axis AM in the remaining BH $S_{\rm in}$: if the fluctuation overwhelms the change in the $Z$-axis AM caused by throwing $A$ into the BH, the information of $A$ shall not be obtained from the AM in the BH $S_{\rm in}$, making the information remnant negligible. 
To evaluate this, we observe that, after $S$ is split into $S_{\rm in}$ and $S_{\rm rad}$ of $N+k-\ell$ and $\ell$ qubits, respectively, the initial change caused by throwing of $A$ into the BH is also split. Since the system is split randomly, the change inherited to $S_{\rm in}$ is approximately given by
\begin{equation}
\frac{N+k-\ell}{N+k} O(k) = \biggl(1- \frac{\ell}{N+k-\ell} \biggr) O(k)
\end{equation}
on average. If this change is much smaller than the fluctuation in $S_{\rm in}$, then the information remnant is expected to be negligible. For instance, this is the case for any $\ell$ when the initial fluctuation $\delta L$ in $B_{\rm in}$ satisfies $\delta L \gg k$. In contrast, if $\delta L \ll k$, the information remnant remains non-negligible until sufficiently large amount of radiation is emitted such that the fluctuation in $S_{\rm in}$ becomes much larger than $\bigl(1- \frac{\ell}{N+k-\ell} \bigr) k$. As the $\ell$-qubit evaporation increases the fluctuation of the BH by $O(\sqrt{\ell})$, non-negligible information remnant must exist if $\bigl(1- \frac{\ell}{N+k-\ell} \bigr)k \leq \sqrt{\ell}$ as order estimation. 

In the following, we make this argument rigorous and provide a quantitative estimate of the amount of information remnant.
Since we are interested in the information remnant, we consider only the situation where the new radiation $S_{\rm rad}$ is sufficiently large so that the Kerr BH has already get partially decoupled and the state therein and the reference $R$ is $\Gamma^{S_{\rm in} R}$.
The state is explicitly given by (see Proposition~\ref{Prop:3})
\begin{multline}
\Gamma^{S_{\rm in} R}
=\sum_{m=0}^{N+k} \frac{2^{N+k}}{\binom{N+k}{m}} \tr_{S'}[\Pi_m^{S'} (\pi^{S'_{\rm rad}}  \otimes \Phi^{S_{\rm in} S'_{\rm in}}) ] \\ 
\otimes  \tr_S[\Pi_m^S (\Phi^{AR} \otimes \xi^{B_{\rm in}}) ].
\end{multline}

Let us now consider to what extent $S_{\rm in}$ has the information of the $Z$-axis AM originally in $A$. 
Since the value of the $Z$-axis AM in $A$ is stored in the reference $R$, this can be checked by measuring the $Z$-axis AM in $R$ and by evaluating how much one can estimate the measurement outcome of $R$ from $S_{\rm in}$. There should be various ways of estimating the outcome in $R$ from $S_{\rm in}$, but we particularly consider an estimation by measuring the $Z$-axis AM in $S_{\rm in}$. This provides a lower bound of the information remnant.

This situation is formulated by the projective measurement $\{ \Pi_{\kappa}^R \}$ on $R$ for its $Z$-axis AM and that $\{ \Pi_{\nu}^{S_{\rm in}} \}$ on $S_{\rm in}$. The labels $\kappa$ and $\nu$ represent each $Z$-axis AM, and $\Pi_{\kappa}$ and $\Pi_{\nu}$ are the projectors on the corresponding subspaces.
When the measured system $S_{\rm in}R$ is in state $\Gamma^{S_{\rm in} R}$, the marginal probabilities to obtain $\nu$ in $S_{\rm in}$ and $\kappa$ in $R$ are, respectively, given by
\begin{align}
q(\kappa) &:= \tr[\Pi_{\kappa}^{R} \Gamma^R] = \tr[\Pi_{\kappa}^{R} \pi^R],\\
P(\nu) &:= \tr[\Pi_{\nu}^{S_{\rm in}} \Gamma^{S_{\rm in}}],
\end{align}
where $\Gamma^{S_{\rm in}}_{\kappa} := \tr_R [\Pi_{\kappa}^R \Gamma^{S_{\rm in}R}]/q(\kappa)$.
We also use the conditional probability
\begin{align}
P(\nu | \kappa) &:= \tr\bigl[( \Pi_{\nu}^{S_{\rm in}} \otimes \Pi_{\kappa}^{R}) \Gamma^{S_{\rm in}R}]/q(\kappa),\\
&=\tr\bigl[\Pi_{\nu}^{S_{\rm in}} \Gamma^{S_{\rm in}}_{\kappa}],
\end{align}
which satisfies
\begin{equation}
P(\nu) = \sum_{\kappa} q(\kappa) P(\nu | \kappa). \label{Eq:w34}
\end{equation}

Using these probabilities, $\eta_{\xi}=\| \Gamma^{S_{\rm in} R} - \Gamma^{S_{\rm in}} \otimes \pi^R \|_1$ (see Eq.~\eqref{Eq:44cg33456o]}) is bounded from below by
\begin{equation}
\eta_{\xi} \geq \sum_{\kappa} q(\kappa) \sum_{\nu} \bigl| P(\nu|\kappa) - P(\nu) \bigr|. \label{Ineq4[c=gt5rt}
\end{equation}
This follows from the monotonicity of the trace distance under the CPTP map for transformation $\rho^{S_{\rm in} R} \mapsto \sum_{\nu, \kappa} \tr[(\Pi_{\nu}^{S_{\rm in}} \otimes \Pi_{\kappa}^R) \rho^{S_{\rm in} R}] \ketbra{\nu, \kappa}{\nu, \kappa}$ with $\ket{\nu, \kappa}$ being the mutually orthonormal states in an ancillary system.

For the sake of simplicity of the analysis, we now approximate the value of $\nu$ by continuous values and replace the above probabilities $P(\nu)$ and $P(\nu|\kappa)$ with probability density functions $p(\nu)$ and $p(\nu|\kappa)$, respectively. 
We also assume that the $\kappa$-dependence of $p(\nu|\kappa)$ is approximated by a shift without changing its common shape, which is given by a probability function $\bar{p}(\nu)$ with vanishing tails for a large deviation $| \nu - \langle \nu \rangle|$. More specifically, we assume that 
\begin{equation}
p(\nu | \kappa)  \approx \bar{p}(\nu - \alpha \delta \kappa), \label{Eq:eercccppp}
\end{equation}
where $\bar{p}(\mu) :=p(\mu | \kappa = \langle \kappa \rangle)$ with $\alpha = 1- \ell/(N+k)$, $\delta \kappa = \kappa - \langle \kappa \rangle$, and $\langle \cdot \rangle$ is the expectation over the probability distribution $q(\kappa)$.
Note that the rescaling by $\alpha$ in Eq.~\eqref{Eq:eercccppp} is needed since $S$ is composed of $N+k$ qubits while $S_{\rm in}$ is of $N+k - \ell$ qubits.
When $|\delta \kappa|$ is sufficiently small, we have
\begin{equation}
\bar{p}(\nu - \alpha \delta\kappa) = \bar{p}(\nu) - \alpha \delta \kappa \frac{d \bar{p}(\nu)}{d \nu} + O(\delta \kappa^2),
\end{equation}
which further implies that $p(\nu) = \bar{p}(\nu) + O(\delta \kappa^2)$ due to Eq.~\eqref{Eq:w34} and $\langle \delta \kappa \rangle = 0$.

\begin{figure*}[t!]
\centering
\includegraphics[width=\textwidth,clip]{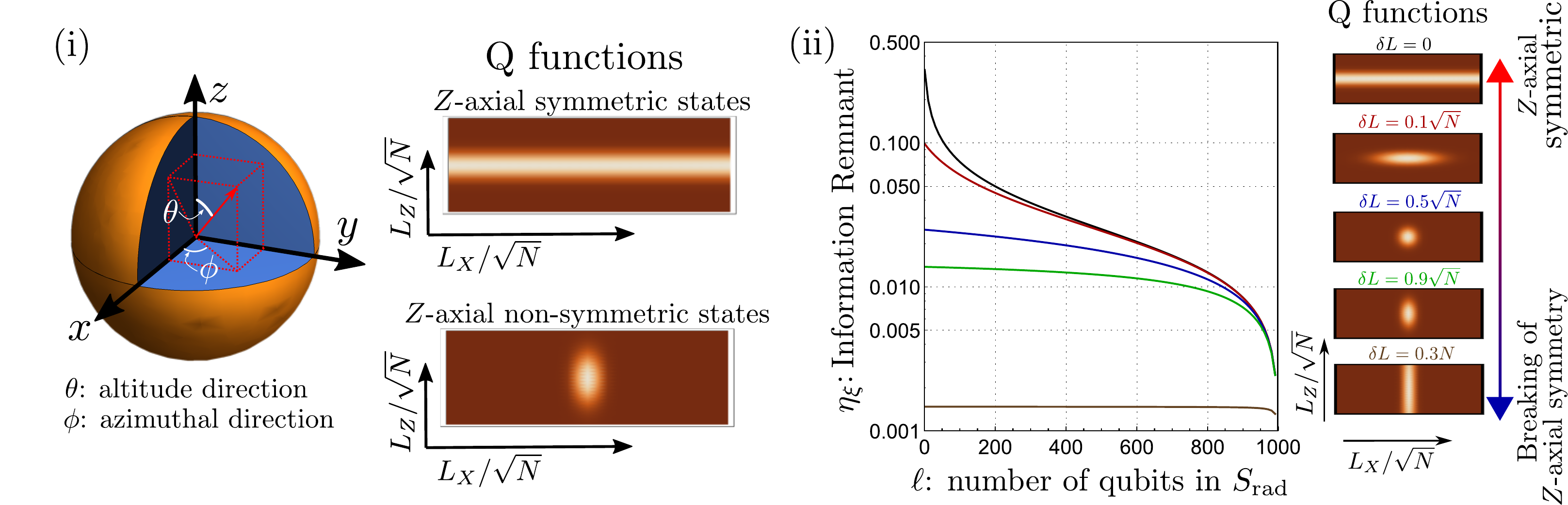}
\caption{Figure (i) explains the $Z$-axial symmetry of a pure state and the Q function. 
For a pure state $\ket{\xi}^{B_{\rm in}}$, the Q function is defined with variables $L_X$ and $L_Z$, which correspond to the $X$- and $Z$-axis AMs, respectively, by $Q_{\xi} \bigl((N/2)^{-1/2}L_X, (N/2)^{-1/2}L_Z \bigr) = \bigl| \langle \theta, \phi | \xi \rangle^{B_{\rm in}} \bigr|^2$, where $\ket{\theta, \phi} : = (\cos \frac{\theta}{2} \ket{0} + e^{i \phi} \sin \frac{\theta}{2} \ket{1})^{\otimes N}$ ($|\phi| \leq \pi/2$) and $(\theta, \phi)$ are those satisfying $\sin \theta \cos \phi = 2 L_X/N$ and $\cos \theta = 2 L_Z/N$. 
Note that the function is regarded as the Husimi Q function for $N \rightarrow \infty$ when $\langle L_Y \rangle \approx N/2$.
In Figure (i), we depict $| \langle \theta, \phi | \xi \rangle^{B_{\rm in}}|^2$: since it is a function of $(\theta, \phi)$, it can be visualized on a sphere.
The Q function is then depicted by projecting the sphere surface on $XZ$-plane, as shown in the right side of Figure (i) for two pure states as examples.
The states $\ket{\xi}$ are taken in the form of Eq.~\eqref{Eq:xipure} for $\delta L=0$ and $\delta L = 0.9 \sqrt{N}$, where $\ket{\varphi_n}$ is especially chosen as the state of equal superposition of all product basis states in $\cH_n^{B_{\rm in}}$. The state for $\delta L=0$ is apparently invariant under the rotation around the $Z$-axis: since the number of up spins is constant, the rotation changes only the global phase. This is visualized by the Q function, which is independent of $L_X$.
This is not the cases for $\delta L = 0.9 \sqrt{N}$: the state is variant and the figure is non-symmetric if they are rotated around the $Z$-axis. These indicate that the Q function visualizes how symmetric the states are.
Figure (ii) shows the semi-log plot of $\eta_{\xi}(N,k,\ell)$ for various pure states $\xi$ of the initial Kerr BH $B_{\rm in}$ with different fluctuations of the $Z$-axis AM, as well as the corresponding Q functions. The fluctuation $\delta L$ is taken as $\delta L =  0$ (black), $0.1 \sqrt{N}$ (red), $0.5 \sqrt{N}$ (blue), $0.9 \sqrt{N}$ (green), and $0.3N$ (brown). The size $N$ of the initial Kerr BH is fixed to $1000$, and $k$ is set to $1$. 
Note that the average $Z$-axis AM $L$ can be arbitrary since $\eta_{\xi}(N,k,\ell)$ does not depend on $L$. 
Comparing this with the Q functions for each $\xi(L, \delta L)$, we observe that $\eta_{\xi}(N,k,\ell)$ is small when the degree of symmetry-breaking is large. 
}
\label{Fig:IR}
\end{figure*}

Using these approximations and replacing the summation over $\nu$ in Eq.~\eqref{Ineq4[c=gt5rt} by integral, we obtain
\begin{align}
\eta_{\xi} &\geq \alpha \sum_{\kappa} q(\kappa) | \delta\kappa| \int d\nu \bigl| \frac{d \bar{p}(\nu)}{d \nu} \bigr| + O( \delta \kappa^2),\\
 &=  \alpha\langle | \delta \kappa | \rangle  \int d \nu \biggl| \frac{d \bar{p}(\nu)}{d \nu}  \biggr| + O( \delta \kappa^2) \label{intint}
\end{align}
where $\langle | \delta \kappa | \rangle := \sum_{\kappa} q(\kappa) |\delta \kappa|$ is the mean absolute deviation of the $Z$-axis AM in $R$. 
Recalling that we have assumed vanishing tails for $\bar{p}(\nu)$, the integral in Eq.~\eqref{intint} can be bounded from below by $2 \max_{\nu} \bar{p}(\nu)$. Thus, we have 
\begin{align}
\eta_{\xi} &\geq 2 \alpha \langle | \delta \kappa | \rangle \max_{\nu} \bar{p}(\nu) + O( \delta \kappa^2).
\end{align}

The value $\max_{\nu} \bar{p}(\nu)$ can be further bounded from below in terms of the variance $\langle \delta \nu^2 \rangle$ of the $Z$-axis AM in $S_{\rm in}$. 
This follows from the fact that the probability distribution that has the least standard deviation under the condition that the maximum probability is given is the rectangle function $r(x) = r_{\rm max}$ for $x \in [-1/(2 r_{\rm max}), 1/(2 r_{\rm max})]$ and $0$ otherwise. In that case, the variance $V$ is
\begin{equation}
V = \int_{-1/(2 r_{\rm max})}^{1/(2 r_{\rm max})} x^2 r_{\rm max} dx = \frac{1}{12\ \! r_{\rm max}^2}.
\end{equation}
Hence, for any probability density function $p(x)$, it holds that $\max p(x) \geq (2 \sqrt{3} \sqrt{V} )^{-1}$. Applying this to our case, we obtain $\max_{\nu} \bar{p}(\nu) \geq (2 \sqrt{3} \sqrt{\langle \delta \nu^2 \rangle} )^{-1}$.
We thus arrive at
\begin{equation}
\eta_{\xi} \geq \frac{1}{\sqrt{3}} \langle | \delta \kappa | \rangle \biggl( 1- \frac{\ell}{N+k} \biggr) \frac{1}{ \sqrt{\langle \delta \nu^2 \rangle}} + O(\delta \kappa^2).  \label{Eq:cierg}
\end{equation}
Since $\langle \delta \nu^2 \rangle$ is the variance of the $Z$-axis AM in the BH $S_{\rm in}$, this confirms the expectation that the information remnant is bounded from below by the fluctuation of the AM in the BH. In particular, when the fluctuation is small, the amount of information remnant should be necessarily large as expected.\\

We continue our analysis and show that the fluctuation is related to how much symmetric the BH state is. Due to Eq.~\eqref{Eq:cierg}, this further implies that the information remnant is also characterized by the symmetry of the state of the BH.
To start with, in Fig.~\ref{Fig:IR}, the degree of symmetry of the state $\xi^{B_{\rm in}}$ is visualized by using the so-called Q function. There, we consider only pure states for demonstration. By comparing the Q function of the state with the corresponding $\eta_{\xi}$, it is clear that the $Z$-axial symmetry in $\xi^{B_{\rm in}}$ is strongly broken when $\eta_{\xi}$ is small.

To analyze this relation in a quantitative manner, we introduce the degree of symmetry breaking by the sensitivity of the state under the symmetric action. 
Let $\ket{\Psi}^{S_{\rm in}S'_{\rm in}}$ be a purification of the state $\Psi^{S_{\rm in}}$, and $B_{\theta}(\Psi)$ denote
\begin{equation}
B_{\theta}(\Psi) := 1- \bigl| \langle \Psi | e^{i \theta L_Z} \otimes I  |\Psi \rangle \bigr|^2,
\end{equation}
where $L_Z$ is the $Z$-axis AM operator in $S_{\rm in}$ and the second identity acts on $S'_{\rm in}$. Then, we define the degree $\zeta(S_{\rm in})$ of symmetry-breaking by
\begin{equation}
\zeta(S_{\rm in}) := \frac{\partial^2}{\partial \theta^2} B_{\theta}(\Psi) \biggl|_{\theta = 0}.
\end{equation}
This quantifies how sensitive the state $\Psi$ is to the symmetric action, which is just a rotation around the $Z$-axis in our case. 
Note that we here considered $S_{\rm in}$, but the degree of symmetry-breaking can be similarly defined in any system, such as in the initial BH $B_{\rm in}$.

The degree $\zeta$ of symmetry-breaking is closely related to the variance $\langle \delta \nu^2 \rangle$ of the $Z$-axis AM $\nu$ in $\Psi^{S_{\rm in}}$.
Since we have
\begin{align}
&\bigl| \langle \Psi | e^{i \theta L_Z} \otimes I  |\Psi \rangle \bigr|^2 \notag \\
&=
\bigl| \tr [e^{i \theta L_Z} \Psi] \bigr|^2\\
&=
\bigl| 1 + i \theta \tr[L_Z \Psi] - \frac{\theta^2}{2} \tr[L_Z^2 \Psi] \bigr|^2 + O(\theta^3)\\
&=
1 + \theta^2 \bigl( (\tr[L_Z \Psi])^2 - \tr[L_Z^2 \Psi] \bigr) + O(\theta^3)\\
&=
1 - \theta^2 \langle \delta \nu^2 \rangle + O(\theta^3),
\end{align}
we obtain $\zeta(S_{\rm in}) = 2\langle \delta \nu^2 \rangle $.

Substituting this relation between the variance and the degree of symmetry-breaking into Eq.~\eqref{Eq:cierg}, it follows that
\begin{equation}
\eta_{\xi} \geq \sqrt{\frac{2}{3}}\langle | \delta \kappa | \rangle \biggl( 1- \frac{\ell}{N+k} \biggr) \frac{1}{\sqrt{ \zeta(S_{\rm in})}}. \label{Ineq:Etaeta}
\end{equation}
Thus, when information remnant is small, $\zeta(S_{\rm in})$ should be necessarily large, implying that the $Z$-axial symmetry in the remaining Kerr BH $S_{\rm in}$ should be broken strongly.
This clarifies yet another micro-macro correspondence of a quantum Kerr BH. \\

The symmetry-breaking in $S_{\rm in}$ has two origins: one is that in the initial Kerr BH $B_{\rm in}$, and the other is due to the new radiation, where $\ell$-qubit radiation increases the standard deviation of the $Z$-axis AM in the remaining Kerr BH $S_{\rm in}$ by $O(\sqrt{\ell})$, and so does the degree of symmetry breaking.
When $\sqrt{\ell}  \ll \delta L$ with $\delta L$ being the standard deviation of the $Z$-axis AM in the initial Kerr BH $B_{\rm in}$, the former origin is dominant. 
On the other hand, when $\sqrt{\ell}  \gg \delta L$, the symmetry-breaking in $S_{\rm in}$ is mostly originated from the radiation process.
We thus conclude that 
\begin{equation}
\eta_{\xi} \gtrsim 
\begin{cases}
\frac{a}{\sqrt{2}} \frac{1}{\delta L} & \text{when $\sqrt{\ell}  \ll \delta L$},\\
\frac{a}{\sqrt{\ell}} & \text{when $\sqrt{\ell}  \gg \delta L$},
\end{cases} \label{Eq:symbreB}
\end{equation}
where $a = \sqrt{2/3} \langle | \delta \kappa | \rangle ( 1- \ell/(N+k))$.
Here, we have used the fact that the degree of symmetry breaking in $B_{\rm in}$, $\zeta(B_{\rm in})$, is given by $2 (\delta L)^2$.

In the case of the pure Kerr BH, the information remnant is negligible in the thermodynamic limit ($N \rightarrow \infty$), assuming that $k$ remains constant. This is because the partial decoupling occurs in the pure Kerr BH only when $\ell \gtrsim N/2$. Thus, Eq.~\eqref{Eq:symbreB} implies that, regardless of $\delta L$, the information remnant vanishes when $N \rightarrow \infty$.
In contrast, for the mixed Kerr BH, there are cases where the partial decoupling sets in when $\ell = O(k)$. In this case, the fluctuation $\delta L$ of the $Z$-axis AM in the initial Kerr BH $B_{\rm in}$ determines the information remnant. In particular, if $\delta L$ is not so large that the lower case of Eq.~\eqref{Eq:symbreB} holds, a non-negligible amount of information remains in the Kerr BH even in the thermodynamic limit unless $\ell$ becomes sufficiently large.
This is in sharp contrast to the BH without symmetry, in which the information of $k$ qubits can be almost fully recovered from $\ell = O(k)$ radiation. Thus, the information remnant is a substantial feature induced by the symmetry when the initial Kerr BH is sufficiently mixed with small fluctuation $\delta L$ of the $Z$-axis AM.

Finally, we refer to the recent result~\cite{TS2021}, where a rigorous lower bound on the information remnant for any symmetry was derived in terms of the quantum Fisher information of the initial/remaining BH. Since the quantum Fisher information is another characterization of variance, our and their results are consistent with each other.

\section{Conclusion and Discussions} \label{S:SD}

In this paper, we have investigated the Hayden-Preskill protocol when the system conserves the number of up-spins, or equivalently the $Z$-component of the angular momentum.
Based on the partial decoupling approach, we have first provided general formulas for upper bounds on the errors in recovering the information. 
From the numerical evaluations of the recovery errors, we have shown that the symmetry induces two substantial deviations from the protocol without symmetry. One is the delay of information leakage and the other is the information remnant. Depending on the initial condition of the system, these phenomena can be negligible or macroscopically large. This is of particular interest since it implies that there are cases where symmetry substantially changes the process of information leakage from that without symmetry.

We have then investigated the delay and the information remnant from the macroscopic physics point of view. By introducing the clipping of entanglement, we have revealed that the delay is characterized by the thermodynamic properties of the initial system. The close relation between the information remnant and the symmetry-breaking of the system has been also revealed. These relations establish connections between the information leakage problem and macroscopic physics.

The results obtained by partial decoupling can be easily generalized to the systems with any abelian global symmetries, such as energy conservation. It is also possible to apply the partial decoupling approach to the systems with non-abelian global symmetry. See Ref.~\cite{WN2021} for the more general form of partial decoupling. We, however, leave the non-abelian case as an interesting future problem. In that case, the information cannot be simply divided into the symmetry-invariant and the other ones, and more careful analysis will be needed.

Another important open problem is about an approximate realization of the symmetry-preserving Haar scrambling. 
All of our analyses rely on the assumption that the internal unitary dynamics of the system is symmetry-preserving scrambling, which will not be rigorously satisfied since it takes exponentially long time to implement a Haar random unitary. Hence, it is of crucial importance to consider approximate implementations of symmetry-preserving scrambling.
Approximations of random unitary have been studied in terms of \emph{unitary designs}, and a couple of results are known about the time needed to achieve unitary designs~\cite{HL2009TPE,BHH2016,NHKW2017,HMHEGR2023}. It will be of great interest to investigate how symmetries affect such results~\cite{L2020,M2022}. More explicitly, it is interesting to address if we can construct \emph{symmetry-preserving} unitary designs by quantum circuits. 

It is also of interest to explicitly construct a decoder for the Hayden-Preskill protocol in the presence of symmetry. Without symmetry, a couple of decoders were proposed in the literature~\cite{YK2017, NMK2022}. It is not immediately clear if one can construct a decoder in the same way, which will be an interesting open problem.

\section{Acknowledgements}
Y.N. is supported by by JST, PRESTO Grant Number JPMJPR1865, Japan, and partially by JST CREST, Grant Number JPMJCR1671, Japan. Y.N. is partially supported by Grants-in-Aid for Transformative Research Areas (A) No.~JP21H05182 and No.~JP21H05183 from MEXT of Japan, as well as by JSPS KAKENHI Grant Number JP22K03464, Japan. Y.N. and M.K. are supported by JST Moonshot R\&D, Grant Number JPMJMS2061, Japan.

\bibliographystyle{quantum}
\bibliography{Bib3}

\begin{thebibliography}{10}

\bibitem{H1974}
Stephen~W. {Hawking}.
\newblock ``Black hole explosions?''.
\newblock \href{https://dx.doi.org/10.1038/248030a0}{Nature {\bf 248},
  30--31}~(1974).

\bibitem{H1975}
Stephen~W. Hawking.
\newblock ``Particle creation by black holes''.
\newblock \href{https://dx.doi.org/10.1007/BF02345020}{Communications in
  Mathematical Physics {\bf 43}, 199--220}~(1975).

\bibitem{I1967}
Werner Israel.
\newblock ``Event horizons in static vacuum space-times''.
\newblock \href{https://dx.doi.org/10.1103/PhysRev.164.1776}{Physical Review
  {\bf 164}, 1776--1779}~(1967).

\bibitem{I1968}
Werner Israel.
\newblock ``Event horizons in static electrovac space-times''.
\newblock \href{https://dx.doi.org/10.1007/BF01645859}{Communications in
  Mathematical Physics {\bf 8}, 245--260}~(1968).

\bibitem{C1971}
Brandon Carter.
\newblock ``Axisymmetric black hole has only two degrees of freedom''.
\newblock \href{https://dx.doi.org/10.1103/PhysRevLett.26.331}{Physical Review
  Letters {\bf 26}, 331--333}~(1971).

\bibitem{HP2007}
Patrick Hayden and John Preskill.
\newblock ``Black holes as mirrors: quantum information in random subsystems''.
\newblock \href{https://dx.doi.org/10.1088/1126-6708/2007/09/120}{Journal of
  High Energy Physics {\bf 2007}, 120}~(2007).

\bibitem{SS2008}
Yasuhiro Sekino and L~Susskind.
\newblock ``Fast scramblers''.
\newblock \href{https://dx.doi.org/10.1088/1126-6708/2008/10/065}{{Journal of
  High Energy Physics} {\bf 0810}, 065}~(2008).
\newblock  \href{http://arxiv.org/abs/0808.2096}{arXiv:0808.2096}.

\bibitem{S2011}
Leonard Susskind.
\newblock ``Addendum to fast scramblers''~(2011).
\newblock  \href{http://arxiv.org/abs/1101.6048}{arXiv:1101.6048}.

\bibitem{LSHOH2013}
Nima Lashkari, Douglas Stanford, Matthew Hastings, Tobias Osborne, and Patrick
  Hayden.
\newblock ``Towards the fast scrambling conjecture''.
\newblock \href{https://dx.doi.org/10.1007/jhep04(2013)022}{Journal of High
  Energy Physics {\bf 1304}, 022}~(2013).
\newblock  \href{http://arxiv.org/abs/1101.6048}{arXiv:1101.6048}.

\bibitem{SS2014}
Stephen~H. Shenker and Douglas Stanford.
\newblock ``Black holes and the butterfly effect''.
\newblock \href{https://dx.doi.org/10.1007/JHEP03(2014)067}{Journal of High
  Energy Physics {\bf 2014}, 67}~(2014).

\bibitem{SS2015}
Stephen~H. Shenker and Douglas Stanford.
\newblock ``Stringy effects in scrambling''.
\newblock \href{https://dx.doi.org/10.1007/JHEP05(2015)132}{Journal of High
  Energy Physics {\bf 2015}, 132}~(2015).

\bibitem{RS2015}
Daniel~A. Roberts and Douglas Stanford.
\newblock ``Diagnosing chaos using four-point functions in two-dimensional
  conformal field theory''.
\newblock \href{https://dx.doi.org/10.1103/PhysRevLett.115.131603}{Physical
  Review Letters {\bf 115}, 131603}~(2015).

\bibitem{RY2017}
Daniel~A. Roberts and Beni Yoshida.
\newblock ``Chaos and complexity by design''.
\newblock \href{https://dx.doi.org/10.1007/jhep04(2017)121}{{Journal of High
  Energy Physics} {\bf 1704}, 121}~(2017).
\newblock  \href{http://arxiv.org/abs/1610.04903}{arXiv:1610.04903}.

\bibitem{Y2019}
Beni Yoshida.
\newblock ``Soft mode and interior operator in the hayden-preskill thought
  experiment''.
\newblock \href{https://dx.doi.org/10.1103/PhysRevD.100.086001}{Physical Review
  D {\bf 100}, 086001}~(2019).

\bibitem{L2020}
Junyu Liu.
\newblock ``Scrambling and decoding the charged quantum information''.
\newblock \href{https://dx.doi.org/10.1103/PhysRevResearch.2.043164}{Physical
  Review Res. {\bf 2}, 043164}~(2020).

\bibitem{SY1993}
Subir Sachdev and Jinwu Ye.
\newblock ``Gapless spin-fluid ground state in a random quantum {Heisenberg}
  magnet''.
\newblock \href{https://dx.doi.org/10.1103/PhysRevLett.70.3339}{Physical Review
  Letters {\bf 70}, 3339--3342}~(1993).

\bibitem{Kitaev1}
Alexei Kitaev.
\newblock ``{`Hidden correlations in the Hawking radiation and thermal
  noise.'}, talk at {KITP}''~(2015).

\bibitem{Kitaev2}
Alexei Kitaev.
\newblock ``{`A simple model of quantum holography.'}, talks at
  {KITP}''~(2015).

\bibitem{Jensen2016}
Kristan Jensen.
\newblock ``Chaos in {${\mathrm{AdS}}_{2}$} holography''.
\newblock \href{https://dx.doi.org/10.1103/PhysRevLett.117.111601}{Physical
  Review Letters {\bf 117}, 111601}~(2016).

\bibitem{MS2016}
Juan Maldacena and Douglas Stanford.
\newblock ``Remarks on the sachdev-ye-kitaev model''.
\newblock \href{https://dx.doi.org/10.1103/PhysRevD.94.106002}{Physical Review
  D {\bf 94}, 106002}~(2016).

\bibitem{Sachdev2015}
Subir Sachdev.
\newblock ``Bekenstein-hawking entropy and strange metals''.
\newblock \href{https://dx.doi.org/10.1103/PhysRevX.5.041025}{Physical Review X
  {\bf 5}, 041025}~(2015).

\bibitem{B2016}
Mike Blake.
\newblock ``Universal charge diffusion and the butterfly effect in holographic
  theories''.
\newblock \href{https://dx.doi.org/10.1103/PhysRevLett.117.091601}{Physical
  Review Letters {\bf 117}, 091601}~(2016).

\bibitem{vKRPS2018}
Curt~W. von Keyserlingk, Tibor Rakovszky, Frank Pollmann, and Shivaji~L.
  Sondhi.
\newblock ``Operator hydrodynamics, otocs, and entanglement growth in systems
  without conservation laws''.
\newblock \href{https://dx.doi.org/10.1103/PhysRevX.8.021013}{Physical Review X
  {\bf 8}, 021013}~(2018).

\bibitem{KVH2018}
Vedika Khemani, Ashvin Vishwanath, and David~A. Huse.
\newblock ``Operator spreading and the emergence of dissipative hydrodynamics
  under unitary evolution with conservation laws''.
\newblock \href{https://dx.doi.org/10.1103/PhysRevX.8.031057}{Physical Review X
  {\bf 8}, 031057}~(2018).

\bibitem{HQRY2016}
Pavan Hosur, Xiao-Liang Qi, Daniel~A. Roberts, and Beni Yoshida.
\newblock ``Chaos in quantum channels''.
\newblock \href{https://dx.doi.org/10.1007/JHEP02(2016)004}{{Journal of High
  Energy Physics} {\bf 1602}, 004}~(2016).
\newblock  \href{http://arxiv.org/abs/1511.04021}{arXiv:1511.04021}.

\bibitem{PYHP2015}
Fernando Pastawski, Beni Yoshida, Daniel Harlow, and John Preskill.
\newblock ``Holographic quantum error-correcting codes: toy models for the
  bulk/boundary correspondence''.
\newblock \href{https://dx.doi.org/10.1007/JHEP06(2015)149}{Journal of High
  Energy Physics {\bf 2015}, 149}~(2015).

\bibitem{PEW2017}
Fernando Pastawski, Jens Eisert, and Henrik Wilming.
\newblock ``Towards holography via quantum source-channel codes''.
\newblock \href{https://dx.doi.org/10.1103/PhysRevLett.119.020501}{Physical
  Review Letters {\bf 119}, 020501}~(2017).

\bibitem{KC2019}
Tamara Kohler and Toby Cubitt.
\newblock ``Toy models of holographic duality between local hamiltonians''.
\newblock \href{https://dx.doi.org/10.1007/JHEP08(2019)017}{Journal of High
  Energy Physics {\bf 2019}, 17}~(2019).

\bibitem{HP2019}
Patrick Hayden and Geoffrey Penington.
\newblock ``Learning the alpha-bits of black holes''.
\newblock \href{https://dx.doi.org/10.1007/JHEP12(2019)007}{Journal of High
  Energy Physics {\bf 2019}, 7}~(2019).

\bibitem{LFSLYYM2019}
Kevin~A. Landsman, Caroline Figgatt, Thomas Schuster, Norbert~M. Linke, Beni
  Yoshida, Norman~Y. Yao, and Christopher Monroe.
\newblock ``Verified quantum information scrambling''.
\newblock \href{https://dx.doi.org/10.1038/s41586-019-0952-6}{Nature {\bf 567},
  61--65}~(2019).

\bibitem{BGLLNSSSW2021}
Adam~R. Brown, Hrant Gharibyan, Stefan Leichenauer, Henry~W. Lin, Sepehr
  Nezami, Grant Salton, Leonard Susskind, Brian Swingle, and Michael Walter.
\newblock ``Quantum gravity in the lab: Teleportation by size and traversable
  wormholes''~(2019).
\newblock  \href{http://arxiv.org/abs/1911.06314}{arXiv:1911.06314}.

\bibitem{NLBGLSSSW2021}
Sepehr Nezami, Henry~W. Lin, Adam~R. Brown, Hrant Gharibyan, Stefan
  Leichenauer, Grant Salton, Leonard Susskind, Brian Swingle, and Michael
  Walter.
\newblock ``Quantum gravity in the lab: Teleportation by size and traversable
  wormholes, part {II}''~(2021).
\newblock  \href{http://arxiv.org/abs/2102.01064}{arXiv:2102.01064}.

\bibitem{BS2011}
Tom Banks and Nathan Seiberg.
\newblock ``Symmetries and strings in field theory and gravity''.
\newblock \href{https://dx.doi.org/10.1103/PhysRevD.83.084019}{Physical Review
  D {\bf 83}, 084019}~(2011).

\bibitem{HO2019}
Daniel Harlow and Hirosi Ooguri.
\newblock ``Constraints on symmetries from holography''.
\newblock \href{https://dx.doi.org/10.1103/PhysRevLett.122.191601}{Physical
  Review Letters {\bf 122}, 191601}~(2019).

\bibitem{HO2021}
Daniel Harlow and Hirosi Ooguri.
\newblock ``Symmetries in quantum field theory and quantum gravity''.
\newblock \href{https://dx.doi.org/10.1007/s00220-021-04040-y}{Communications
  in Mathematical Physics {\bf 383}, 1669--1804}~(2021).

\bibitem{AHMNV2007}
Nima Arkani-Hamed, Luboš Motl, Alberto Nicolis, and Cumrun Vafa.
\newblock ``The string landscape, black holes and gravity as the weakest
  force''.
\newblock \href{https://dx.doi.org/10.1088/1126-6708/2007/06/060}{Journal of
  High Energy Physics {\bf 2007}, 060}~(2007).

\bibitem{WA2020}
Mischa~P. Woods and {\'{A}}lvaro~M. Alhambra.
\newblock ``Continuous groups of transversal gates for quantum error correcting
  codes from finite clock reference frames''.
\newblock \href{https://dx.doi.org/10.22331/q-2020-03-23-245}{{Quantum} {\bf
  4}, 245}~(2020).

\bibitem{FNASPHP2020}
Philippe Faist, Sepehr Nezami, Victor~V. Albert, Grant Salton, Fernando
  Pastawski, Patrick Hayden, and John Preskill.
\newblock ``Continuous symmetries and approximate quantum error correction''.
\newblock \href{https://dx.doi.org/10.1103/PhysRevX.10.041018}{Physical Review
  X {\bf 10}, 041018}~(2020).

\bibitem{HNPS2021}
Patrick Hayden, Sepehr Nezami, Sandu Popescu, and Grant Salton.
\newblock ``Error correction of quantum reference frame information''.
\newblock \href{https://dx.doi.org/10.1103/PRXQuantum.2.010326}{PRX Quantum
  {\bf 2}, 010326}~(2021).

\bibitem{LK2022}
Linghang Kong and Zi-Wen Liu.
\newblock ``Near-optimal covariant quantum error-correcting codes from random
  unitaries with symmetries''.
\newblock \href{https://dx.doi.org/10.1103/PRXQuantum.3.020314}{PRX Quantum
  {\bf 3}, 020314}~(2022).

\bibitem{WZ1982}
William~K. Wootters and Wojciech~H. Zurek.
\newblock ``A single quantum cannot be cloned''.
\newblock \href{https://dx.doi.org/10.1038/299802a0}{Nature {\bf 299},
  802--803}~(1982).

\bibitem{DBWR2014}
Frédéric Dupuis, Mario Berta, Jürg Wullschleger, and Renato Renner.
\newblock ``One-shot decoupling''.
\newblock \href{https://dx.doi.org/10.1007/s00220-014-1990-4}{Communications in
  Mathematical Physics {\bf 328}, 251--284}~(2014).

\bibitem{Page1993}
Don~N. Page.
\newblock ``Average entropy of a subsystem''.
\newblock \href{https://dx.doi.org/10.1103/PhysRevLett.71.1291}{Physical Review
  Letters {\bf 71}, 1291--1294}~(1993).

\bibitem{HOW2005}
Michał Horodecki, Jonathan Oppenheim, and Andreas Winter.
\newblock ``Partial quantum information''.
\newblock \href{https://dx.doi.org/10.1038/nature03909}{Nature {\bf 436},
  673--676}~(2005).

\bibitem{HOW2007}
Michał Horodecki, Jonathan Oppenheim, and Andreas Winter.
\newblock ``Quantum state merging and negative information''.
\newblock \href{https://dx.doi.org/10.1007/s00220-006-0118-x}{Communications in
  Mathematical Physics {\bf 269}, 107--136}~(2007).

\bibitem{HHWY2008}
Patrick Hayden, Micha\l{} Horodecki, Andreas Winter, and Jon Yard.
\newblock ``A decoupling approach to the quantum capacity''.
\newblock \href{https://dx.doi.org/10.1142/S1230161208000043}{Open Systems \&
  Information Dynamics {\bf 15}, 7--19}~(2008).

\bibitem{WN2021}
Eyuri Wakakuwa and Yoshifumi Nakata.
\newblock ``One-shot randomized and nonrandomized partial decoupling''.
\newblock \href{https://dx.doi.org/10.1007/s00220-021-04136-5}{Communications
  in Mathematical Physics {\bf 386}, 589--649}~(2021).

\bibitem{RennerThesis}
Renato Renner.
\newblock ``Security of quantum key distribution''.
\newblock \href{https://dx.doi.org/10.48550/ARXIV.QUANT-PH/0512258}{PhD
  thesis}.
\newblock {ETH Zurich}.
\newblock ~(2005).

\bibitem{T2016}
Marco Tomamichel.
\newblock ``Quantum information processing with finite resources''.
\newblock
  \href{https://dx.doi.org/https://doi.org/10.1007/978-3-319-21891-5}{SpringerBriefs
  in Mathematical Physics}. Springer Cham. ~(2016).

\bibitem{L1978}
Elihu Lubkin.
\newblock ``Entropy of an n‐system from its correlation with a
  {$k$}‐reservoir''.
\newblock \href{https://dx.doi.org/10.1063/1.523763}{Journal of Mathematical
  Physics {\bf 19}, 1028--1031}~(1978).

\bibitem{HLW2006}
Patrick Hayden, Debbie~W. Leung, and Andreas Winter.
\newblock ``Aspects of generic entanglement''.
\newblock \href{https://dx.doi.org/10.1007/s00220-006-1535-6}{Communications in
  Mathematical Physics {\bf 265}, 95--117}~(2006).

\bibitem{K2007}
Masato Koashi.
\newblock ``Complementarity, distillable secret key, and distillable
  entanglement''~(2007).
\newblock  \href{http://arxiv.org/abs/0704.3661}{arXiv:0704.3661}.

\bibitem{NC2000}
Michael~A. Nielsen and Isaac~L. Chuang.
\newblock ``Quantum computation and quantum information: 10th anniversary
  edition''.
\newblock \href{https://dx.doi.org/10.1017/CBO9780511976667}{Cambridge
  University Press}. ~(2010).

\bibitem{TS2021}
Hiroyasu Tajima and Keiji Saito.
\newblock ``Universal limitation of quantum information recovery: symmetry
  versus coherence''~(2021).
\newblock  \href{http://arxiv.org/abs/2103.01876}{arXiv:2103.01876}.

\bibitem{HL2009TPE}
Aram~W. Harrow and Richard~A. Low.
\newblock ``Efficient quantum tensor product expanders and k-designs''.
\newblock In Irit Dinur, Klaus Jansen, Joseph Naor, and Jos{\'e} Rolim,
  editors, Approximation, Randomization, and Combinatorial Optimization.
  Algorithms and Techniques.
\newblock
  \href{https://dx.doi.org/https://doi.org/10.1007/978-3-642-03685-9_41}{Pages
  548--561}.
\newblock Berlin, Heidelberg~(2009). Springer Berlin Heidelberg.

\bibitem{BHH2016}
Fernando G. S.~L. Brandão, Aram~W. Harrow, and Michał Horodecki.
\newblock ``Local random quantum circuits are approximate polynomial-designs''.
\newblock \href{https://dx.doi.org/10.1007/s00220-016-2706-8}{Communications in
  Mathematical Physics {\bf 346}, 397--434}~(2016).

\bibitem{NHKW2017}
Yoshifumi Nakata, Christoph Hirche, Masato Koashi, and Andreas Winter.
\newblock ``Efficient quantum pseudorandomness with nearly time-independent
  hamiltonian dynamics''.
\newblock \href{https://dx.doi.org/10.1103/PhysRevX.7.021006}{Physical Review X
  {\bf 7}, 021006}~(2017).

\bibitem{HMHEGR2023}
Jonas Haferkamp, Felipe Montealegre-Mora, Markus Heinrich, Jens Eisert, David
  Gross, and Ingo Roth.
\newblock ``Efficient unitary designs with a system-size independent number of
  non-clifford gates''.
\newblock \href{https://dx.doi.org/10.1007/s00220-022-04507-6}{Communications
  in Mathematical Physics {\bf 397}, 995--1041}~(2023).

\bibitem{M2022}
Iman Marvian.
\newblock ``Restrictions on realizable unitary operations imposed by symmetry
  and locality''.
\newblock \href{https://dx.doi.org/10.1038/s41567-021-01464-0}{Nature Physics
  {\bf 18}, 283--289}~(2022).

\bibitem{YK2017}
Beni Yoshida and Alexei Kitaev.
\newblock ``Efficient decoding for the hayden-preskill protocol''~(2017).
\newblock  \href{http://arxiv.org/abs/1710.03363}{arXiv:1710.03363}.

\bibitem{NMK2022}
Yoshifumi Nakata, Takaya Matsuura, and Masato Koashi.
\newblock ``Constructing quantum decoders based on complementarity
  principle''~(2022).
\newblock  \href{http://arxiv.org/abs/2210.06661}{arXiv:2210.06661}.

\bibitem{L2001}
Michel Ledoux.
\newblock ``{The Concentration of Measure Phenomenon}''.
\newblock
  \href{https://dx.doi.org/https://doi.org/http://dx.doi.org/10.1090/surv/089}{Mathematical
  Surveys and Monographs}. American Mathematical Society Providence, RI.
  ~(2001).

\bibitem{M2014}
Elizabeth Meckes.
\newblock ``Concentration of measure and the compact classical matrix groups''.
\newblock
  \url{https://www.math.ias.edu/files/wam/Haar_notes-revised.pdf}~(2014).

\bibitem{W1999}
Andreas Winter.
\newblock ``Coding theorem and strong converse for quantum channels''.
\newblock \href{https://dx.doi.org/10.1109/18.796385}{IEEE Transactions on
  Information Theory {\bf 45}, 2481--2485}~(1999).

\end{thebibliography}

\appendix
\section{Proof of Theorem~\ref{Thm:CoPDecoup}} \label{Proof:CoPDecoup}

To prove Theorem~\ref{Thm:CoPDecoup}, we use the theorem shown in Ref.~\cite{WN2021} and the concentration of measure phenomena for product measures.

\begin{theorem}[Non-smoothed partial decoupling theorem~\cite{WN2021}] \label{Thm:PDecoup}
In the situation described in Subsec.~\ref{SS:PDThm}, it holds that
\begin{multline}
\mbb{E}_{U^S \sim {\sf H}_{\times}} 
\bigl|\! \bigl| 
\mc{M}^{S \rightarrow E}\bigl( U^S \varrho^{SR} U^{S \dagger} \bigr)  
- 
\Gamma^{ER}
\bigr|\! \bigr|_1\\
\leq 
2^{-\frac{1}{2} H_{\rm min}(S^*|ER)_{\Gamma}},
\end{multline}
where $H_{\rm min}(S^*|ER)_{\Gamma}$ is the conditional min-entropy of $\Gamma^{S^*ER}$. 
\end{theorem}

It is well-known that a Haar random unitary has concentration properties when the degree of the group is large~\cite{L2001}, which  finds a number of applications in quantum information science.
In most cases, the Haar measure on the whole unitary group is considered. However, the concentration also happens for the product measure ${\sf H}_{\times} = {\sf H}_1 \times \cdots \times {\sf H}_J$.

To explain the concentration of measure on the product space, let us introduce the $L_2$-sum of the Hilbert-Schmidt norms. Let ${\sf U}_{\times}$ be a product of unitary groups ${\sf U}(d_1) \times \dots \times {\sf U}(d_J)$. For $U=(U_1, \cdots, U_J) \in {\sf U}_{\times}$ and $V=(V_1, \cdots, V_J) \in {\sf U}_{\times}$, the $L_2$-sum $D(U,V)$ of the Hilbert-Schmidt norms on ${\sf U}_{\times}$ is defined by
\begin{equation}
D(U,V) = \sqrt{ \sum_{j \in \cJ} \| U_j - V_j \|_2^2 }.
\end{equation}

We also use Lipschitz functions. A real-valued function $F$ on a metric space $(X ,d)$ is said to be \emph{Lipschitz} if
\begin{equation}
\| F \|_{\rm Lip} := \sup_{x \neq y \in X}\frac{|F(x) - F(y)|}{d(x,y)} < \infty.
\end{equation}
The quantity $\| F \|_{\rm Lip}$ is called an \emph{Lipschitz constant} of $F$. The function $F$ with a Lipschitz constant $L$ is called $L$-Lipschitz. 

The following provides the concentrating property on the product space.

\begin{theorem}[Concentration of measure on the product space~\cite{M2014}]\label{Lemma:CMporoduct}
Let ${\sf U}_{\times} = {\sf U}(d_1) \times \dots \times {\sf U}(d_J)$ equipped with the $L_2$-sum of Hilbert-Schmidt norms,
and ${\sf H}_{\times} = {\sf H}_1 \times \cdots \times {\sf H}_J$ be the product probability measure. Suppose that a function $F: {\sf U}_{\times} \rightarrow \mbb{R}$ is $L$-Lipschitz. Then, for every $\delta>0$,
\begin{equation}
F(U_1,\dots,U_J) \geq \mbb{E}_{(U_1,\cdots,U_J)\sim{\sf H}_{\times}}[F] + \delta,
\end{equation}
with probability at most $\exp[-\frac{\delta^2 d_{\rm min}}{12 L^2}]$,
where $d_{\rm min}=\min\{d_1,\dots, d_J\}$.
\end{theorem}

Based on Theorems~\ref{Thm:PDecoup} and~\ref{Lemma:CMporoduct}, we now prove Theorem~\ref{Thm:CoPDecoup} by identifying the unitary $U^S = \bigoplus_{j \in \cJ} U^S_j$ as a point $(U_1^S, \dots, U_J^S)$ on ${\sf U}_{\times} = {\sf U}(d_1) \times \cdots \times {\sf U}(d_J)$.

\begin{Proof}[Theorem~\ref{Thm:CoPDecoup}]
For a given $U^S = \bigoplus_{j \in \cJ} U^S_j$, let $(U_1^S, \dots, U_J^S)$ be the corresponding point on ${\sf U}_{\times}$ equipped with the $L_2$-sum of Hilbert-Schmidt norms $D$. By a slight abuse of notation, we also denote the point by $U^S$.

Let $F: {\sf U}_{\times} \rightarrow \mbb{R}$ be a function given by
\begin{multline}
F(U_1^S, \dots,U_J^S) \\
:= \bigl \| 
\cM^{S \rightarrow E}\bigl( U^S \varrho^{SR} U^{S \dagger} \bigr)  -
\tr_{S^*}[\Gamma^{S^*ER} ]
\bigr \|_1.
\end{multline}
We compute the Lipschitz constant of $F$. Using the triangle inequality, we obtain
\begin{align}
&|F(U_1^S, \dots,U_J^S) - F(V_1^S, \dots,V_J^S)| \notag \\
&\leq \| \cM^{S \rightarrow E} (U^S  \Psi^{SR} U^{S \dagger}) - \cM^{S \rightarrow E} ( V^S\varrho^{SR} V^{S \dagger}) \|_1\\
&\leq  \|U^S \varrho^{SR} U^{S \dagger}- V^S \Psi^{SR} V^{S \dagger}  \|_1\\
&\leq 2 \|(U^S -V^S) \ket{\varrho}^{SR}\|_2\\
&\leq 2  \| \varrho^S\|_{\infty}^{1/2} \|U^S -V^S\|_2.
\end{align}
Here, the second inequality follows from the monotonicity of the trace distance under any trace-non-increasing map, the third one from that $\Psi^{SR}$ is a pure state and the exact diagonalization with the use of the inequality $\sqrt{1-x^2} \leq \sqrt{2(1-x)}$ for any $x \in [0,1]$, and the last one from the fact that 
\begin{align}
\| M^X \ket{\phi}^{XY} \|_2^2 
&=\tr [M^X \phi^{XY} M^{X \dagger}]\\
&=\| \phi^{X} M^{X \dagger}M^X \|_1\\
&\leq \| \phi^{X}\|_{\infty} \| M^{X \dagger}M^X\|_1\\
&= \| \phi^{X}\|_{\infty} \| M^{X}\|_2^2,
\end{align}
due to the H{\"o}lder's inequality.

Explicitly writing  $U^S$ and $V^S$ as $\bigoplus_{j \in \mc{J}} U_j^{S}$ and $\bigoplus_{j \in \mc{J}} V_j^{S}$, respectively, $\|U^S -V^S\|_2$ is given by $\sum_{j \in \mc{J} } \bigl \|U_j^{S}-V_j^{S} \bigr|\! \bigr|_2^2$, i.e. $\|U^S -V^S\|_2 = D(U^S,V^S)$.
Hence, for any $U^S \in {\sf U}_{\times}$ and $V^S \in {\sf U}_{\times}$, we obtain
\begin{multline}
\frac{|F(U^S_{\rm rad}, \dots, U^S_J) - F(V^S_{\rm rad}, \dots, V^S_J)|}{D(U^S,V^S)} \\
\leq 2 \sqrt{\| \varrho^S \|_{\infty}},
\end{multline}
which implies that the Lipschitz constant of $F$ is bounded from above by $2 \sqrt{\| \varrho^S\|_{\infty}}$.

We then apply Theorem~\ref{Lemma:CMporoduct} and obtain
\begin{widetext}
\begin{align}
&{\rm Prob}_{U^S \sim {\sf H}_{\times}} \bigl[ F(U^S_{\rm rad},\dots, U^S_J)
\geq 2^{-\frac{1}{2} H_{\rm min}(S^*|ER)_{\Gamma}} + \delta\bigr] \notag \\
&\leq
{\rm Prob}_{U^S \sim {\sf H}_{\times}} \bigl[ F(U^S_{\rm rad},\dots, U^S_J) \geq \mbb{E}[F] + \delta\bigr]\\
&\leq
\exp\bigl[ -\frac{\delta^2 d_{\rm min}}{48 \| \varrho^S\|_{\infty}} \bigr],
\end{align}    
\end{widetext}
where $d_{\rm min} = \min_{j \cJ} \{ d_j\}$, and Theorem~\ref{Thm:PDecoup}, stating that $\mbb{E}[F] \leq 2^{-\frac{1}{2} H_{\rm min}(S^*|ER)_{\Gamma}}$, is used to obtain the first inequality. $\hfill \blacksquare$
\end{Proof}

\section{Proof of Corollary~\ref{Co:CoPDGML}} \label{Pr:CoPDGML}

To show Corollary~\ref{Co:CoPDGML}, we use the gentle measurement lemma:

\begin{lemma}[Gentle measurement lemma~\cite{W1999}] \label{Lem:GML}
Let $\Phi$ be in $\mc{S}(\mc{H})$ and $\Lambda$ be an Hermitian operator such that $0 \leq \Lambda \leq I$. If they satisfy $\tr[ \Lambda \Phi] \geq 1- \epsilon$, where $0 \leq \epsilon \leq 1$, then 
$\| \Phi - \Phi' \|_1 \leq  2 \sqrt{\epsilon}$, where
\begin{equation}
\Phi' = \frac{\sqrt{\Lambda} \Phi \sqrt{\Lambda} }{\tr[\Lambda \Phi]}.
\end{equation}
\end{lemma}

We also use a simple fact about the conditional min-entropy as given in Lemma~\ref{Lem:conmeas}.

\begin{lemma}[Conditional min-entropy after projective measurement] \label{Lem:conmeas}
Let $\Pi^A$ be a projection operator, $\Psi^{AB}$ be a quantum state.
A post-measured state $\tilde{\Psi}^{AB} := \Pi^A \Psi^{AB} \Pi^A/ \tr[\Pi^A \Psi^{AB}]$ satisfies
\begin{equation}
H_{\rm min}(A|B)_{\tilde{\Psi}} \geq H_{\rm min}(A|B)_{\Psi} + \log[\tr[\Pi^A \Psi^{AB}]].
\end{equation}
\end{lemma}

\begin{Proof}[Lemma~\ref{Lem:conmeas}]
Let $\sigma^B \in \cS(\cH^B)$ be the state such that $2^{-H_{\rm min}(A|B)_{\Psi}} I^A \otimes \sigma^B \geq \Psi^{AB}$. Then, we have
\begin{equation}
2^{-(H_{\rm min}(A|B)_{\Psi}+ \log[\tr[\Pi^A \Psi^{AB}]])} I^A \otimes \sigma^B \geq \tilde{\Psi}^{AB},
\end{equation}
which implies the desired result. $\hfill \blacksquare$
\end{Proof}

Using these lemmas, Corollary~\ref{Co:CoPDGML} can be shown as follows.

\begin{Proof}[Corollary~\ref{Co:CoPDGML}]
We first define the state $\tilde{\varrho}^{SR}$ by
\begin{equation}
\tilde{\varrho}^{SR} := \frac{\Pi_{\geq}^S \varrho_{SR} \Pi_{\geq}^S}{\tr [\Pi_{\geq}^S \varrho_{SR}] }
\end{equation}
We also use the fact that $\Gamma^{ER}:=\tr_{S^*}[\Gamma^{S^*ER} ] = \mbb{E}_{U^S \sim {\sf H}_{\times}} \bigl[\cM^{S \rightarrow E}\bigl( U^S\varrho^{SR} U^{S \dagger} \bigr) \bigr]$.
Using the triangle inequality, we obtain
\begin{widetext}
\begin{multline}
\bigl \| 
\cM^{S \rightarrow E}\bigl( U^S \varrho^{SR} U^{S \dagger} \bigr)  
-
\mbb{E}_{U^S \sim {\sf H}_{\times}}
\bigl[
\cM^{S \rightarrow E}\bigl( U^S \varrho^{SR} U^{S \dagger} \bigr) 
\bigr]
\bigr \|_1\\
\leq
\bigl \| 
\cM^{S \rightarrow E}\bigl( U^S \varrho^{SR} U^{S \dagger} -U^S \tilde{\varrho}^{SR} U^{S \dagger} \bigr)  
\bigr \|_1
+
\bigl \| 
\cM^{S \rightarrow E}\bigl( U^S \tilde{\varrho}^{SR} U^{S \dagger} \bigr)  
-
\mbb{E}_{U^S \sim {\sf H}_{\times}}
\bigl[
\cM^{S \rightarrow E}\bigl( U^S \tilde{\varrho}^{SR} U^{S \dagger} \bigr) 
\bigr]
\bigr \|_1 \\
+
\bigl \| 
\mbb{E}_{U^S \sim {\sf H}_{\times}}
\bigl[
\cM^{S \rightarrow E}\bigl( U^S \tilde{\varrho}^{SR} U^{S \dagger} \bigr) 
\bigr]
-
\mbb{E}_{U^S \sim {\sf H}_{\times}}
\bigl[
\cM^{S \rightarrow E}\bigl( U^S \varrho^{SR} U^{S \dagger} \bigr) 
\bigr]
\bigr \|_1. \label{Eq:kcowe}
\end{multline}
\end{widetext}
In the following, we evaluate each term in the right-hand side of Eq.~\eqref{Eq:kcowe} separately.

For the first term, noting that $\cC^{S\rightarrow E}$ is a trace-non-increasing map, and the trace norm is unitarily invariant, we have
\begin{align}
&\bigl \| 
\cM^{S \rightarrow E}\bigl( U^S \varrho^{SR} U^{S \dagger} -U^S \tilde{\varrho}^{SR} U^{S \dagger} \bigr)  
\bigr \|_1 \notag \\
&\leq 
\bigl \| 
\varrho^{SR}  -\tilde{\varrho}^{SR}
\bigr \|_1 \\
&\leq
2\sqrt{\epsilon},
\end{align}
where the last inequality follows from the gentle measurement lemma and the assumption that $\tr[\varrho^{SR} \Pi_{\geq}^S] \geq 1- \epsilon$.

To evaluate the second term, we use Theorem~\ref{Thm:CoPDecoup}. Recalling that $\tilde{\Psi}^{S}$ does not have support on the subspace $\cH_j^S$ with dimension being smaller than $d_{\rm th}$, it follows that, for any $\delta>0$,
\begin{widetext}
\begin{equation}
    \bigl \| 
\cM^{S \rightarrow E}\bigl( U^S \tilde{\varrho}^{SR} U^{S \dagger} \bigr)  
-
\mbb{E}_{U^S \sim {\sf H}_{\times}}
\bigl[
\cM^{S \rightarrow E}\bigl( U^S \tilde{\varrho}^{SR} U^{S \dagger} \bigr) 
\bigr]
\bigr \|_1
\leq
2^{-\frac{1}{2}H_{\rm min}(S^*|ER)_{\tilde{\Gamma}}} + \delta
\end{equation}    
\end{widetext}
with probability at least $1-\exp[- \frac{\delta^2 d_{\rm th}}{48 \| \tilde{\varrho}^S \|_{\infty}}]$.
Here, $\tilde{\Gamma}^{S^*ER}= \sum_{j,j' \in \mc{J}} \frac{D_S}{\sqrt{d_j d_{j'}}}\zeta^{SE}_{jj'} \otimes \tilde{\varrho}^{S'R}_{jj'}$.
Since $\Pi_{\geq}$ is commutable with $\Pi_j^S$ for any $j$, 
\begin{equation}
\tilde{\Gamma}^{S^*ER} = \frac{(I^S\otimes \Pi^{S'}_{\geq}) \Gamma^{S^*ER} (I^S \otimes \Pi^{S'}_{\geq})}{\tr[(I^S \otimes\Pi^{S'}_{\geq}) \Gamma^{S^*ER}]}.
\end{equation}
 Using Lemma~\ref{Lem:conmeas}, we have
\begin{align}
&H_{\rm min}(S^*|ER)_{\tilde{\Gamma}} \notag \\
&\geq H_{\rm min}(S^*|ER)_{\Gamma} + \log \bigl[\tr[\Pi^S_{\geq} \Gamma^{S^*ER}] \bigr] \\
&=H_{\rm min}(S^*|ER)_{\Gamma} + \log \bigl[\tr[\Pi^S_{\geq} \varrho^{SR}] \bigr] \\
&\geq H_{\rm min}(S^*|ER)_{\Gamma} + \log[ 1- \epsilon],
\end{align}
where the second line is obtained since $\frac{D_S}{\sqrt{d_j d_{j'}}} \tr[ \zeta^{SE}_{jj'}] = \delta_{jj'}$.
Furthermore, it holds that
\begin{align}
\| \tilde{\varrho}^S \|_{\infty} &\leq \min \bigl\{1, \frac{\| \Pi_{\geq}^S \|_{\infty}\| \varrho^S \|_{\infty}}{ \tr[\Pi^S_{\geq} \varrho^S]} \bigr\} \\
&\leq \min \bigl\{1, \frac{\| \varrho^S \|_{\infty}}{1-\epsilon} \bigr\} =:C,
\end{align}
where we have used the sub-multiplicativity of the operator norm, and the assumption that $\tr[\varrho^{SR} \Pi_{\geq}^S] \geq 1- \epsilon$.
Combining all of these together, the second term is bounded as
\begin{widetext}
    \begin{equation}
\bigl \| 
\cM^{S \rightarrow E}\bigl( U^S \tilde{\varrho}^{SR} U^{S \dagger} \bigr)  
-
\mbb{E}_{U^S \sim {\sf H}_{\times}}
\bigl[
\cM^{S \rightarrow E}\bigl( U^S \tilde{\varrho}^{SR} U^{S \dagger} \bigr) 
\bigr]
\bigr \|_1
\leq
\frac{2^{-\frac{1}{2}H_{\rm min}(S^*|ER)_{\Gamma}} }{\sqrt{1-\epsilon}} + \delta
\end{equation}
\end{widetext}
for any $\delta>0$ with probability at least $1- \exp[- \frac{\delta^2 d_{\rm th}}{48 C}]$.

To evaluate the third term in the right-hand side of Eq.~\eqref{Eq:kcowe}, we use the explicit form of the averaged operator
\begin{multline}
\mbb{E}_{U^S \sim {\sf H}_{\times}}
\bigl[
\cM^{S \rightarrow E}\bigl( U^S \Psi^{SR} U^{S \dagger} \bigr) 
\bigr]\\
=
\sum_{j \notin \mc{J}_{\geq}}
\frac{D_S}{d_j} \zeta^{E}_{jj} \otimes \varrho^{R}_{jj}.
\end{multline}
Further, using the relation $\Pi^S_{\geq} = \sum_{j \in \cJ_{\geq}} \Pi^S_j$, we obtain that
the third term $X$ in the right-hand side of Eq.~\eqref{Eq:kcowe} satisfy
\begin{align}
X
&=
\bigl \| 
\sum_{j \in \mc{J}} \frac{D_S}{d_j} \zeta^{E}_{jj} \otimes (\tilde{\varrho}^{R}_{jj} - \varrho^{R}_{jj})
\bigr \|_1\\
&\leq 
\sum_{j \notin \mc{J}_{\geq}} \bigl \| 
\frac{D_S}{d_j} \zeta^{E}_{jj} \otimes \varrho^{R}_{jj}
\bigr \|_1\notag \\
& \hspace{10mm}+
\sum_{j \in \mc{J}_{\geq}} \bigl \| 
\frac{D_S}{d_j} \zeta^{E}_{jj} \otimes (\tilde{\varrho}^{R}_{jj} - \varrho^{R}_{jj})
\bigr \|_1. \label{eq:34raefevdfvd|}
\end{align} 
Recalling that $\tr[\frac{D_S}{d_j} \zeta^{E}_{jj}] \leq 1$ and that $\tilde{\varrho}^{R}_{j} = \varrho^{R}_{j}/ \tr[\Pi_{\geq}^S \varrho^{SR}]$ for $j \in \cJ_{\geq}$, the first term in the right-hand side of Eq.~\eqref{eq:34raefevdfvd|} is bounded from above by
\begin{equation}
\sum_{j \notin \mc{J}_{\geq}} \tr[ \varrho^{R}_{jj}]
= 
\tr\bigl[ \varrho^{SR} ( I^S - \Pi^S_{\geq} ) \bigr] \leq \epsilon.
\end{equation}
An upper bound of the second term in the right-hand side of Eq.~\eqref{eq:34raefevdfvd|} is given by
\begin{align}
&\biggl|  \frac{1}{\tr[ \Pi^S_{\geq} \varrho^{SR}]} -1 \biggr| \sum_{j \in \mc{J}_{\geq}} \tr[\varrho^{SR}_{jj}] \notag \\
&\leq
\frac{\epsilon}{1-\epsilon} \sum_{j \in \mc{J}} \tr[\varrho^{SR}_{jj}]\\
&=
\frac{\epsilon}{1-\epsilon}.
\end{align}

Combining the upper bounds of all three terms in Eq.~\eqref{Eq:kcowe}, we obtain the desired result:
\begin{multline}
\bigl \| 
\cM^{S \rightarrow E}\bigl( U^S \varrho^{SR} U^{S \dagger} \bigr)  \\
-
\mbb{E}_{U^S \sim {\sf H}_{\times}}
\bigl[
\cM^{S \rightarrow E}\bigl( U^S \varrho^{SR} U^{S \dagger} \bigr) 
\bigr]
\bigr \|_1\\
\leq
\frac{2^{-\frac{1}{2}H_{\rm min}(S^*|ER)_{\Gamma}} }{\sqrt{1-\epsilon}} + \delta + f(\epsilon),
\end{multline}
with probability at least $1- \exp[- \frac{\delta^2 d_{\rm th}}{48 C}]$, where $f(\epsilon)=2\sqrt{\epsilon} + \epsilon + \frac{\epsilon}{1-\epsilon}$.  $\hfill \blacksquare$
\end{Proof}

\section{Computation of $p_n$} \label{App:pn}
We here show Eq.~\eqref{Eq:pnonon} by explicitly computing $p_n$, which is defined by
\begin{align}
&p_n := \mbb{E}_{U^S \sim {\sf H}_{\times}} \bigl[ p_n(U^S) \bigr], \\
&p_n(U^S) := \tr \bigl[ \Pi_n^{S_{\rm rad}} U^S (\Phi^{AR} \otimes \xi^{B_{\rm in}}) U^{S \dagger}   \bigr],\\
&\hspace{12mm} =\tr \bigl[ \Pi_n^{S_{\rm rad}} U^S (\pi^{A} \otimes \xi^{B_{\rm in}}) U^{S \dagger}   \bigr].
\end{align}
We first take the average of $U^S (\pi^{A} \otimes \xi^{B_{\rm in}}) U^{S \dagger}$ over $U^S \sim {\sf H}_{\times}$, which leads to
\begin{multline}
\mbb{E}_{U^S \sim {\sf H}_{\times}} \bigl[ U^S (\Pi^{A} \otimes \xi^{B_{\rm in}}) U^{S \dagger}   \bigr]\\
=
\sum_{m=0}^{N+k} \tr \bigl[ \Pi_m^S (\Pi^{A} \otimes \xi^{B_{\rm in}})  \bigr] \pi_m^S,
\end{multline}
due to the Schur's lemma.
By substituting this, we have
\begin{equation}
p_n =\sum_{m=0}^{N+k}
\tr \bigl[ \Pi_m^S (\pi^{A} \otimes \xi^{B_{\rm in}})  \bigr]
\tr \bigl[ \pi_m^S (I^{S_{\rm in}} \otimes \Pi^{S_{\rm rad}}_n)  \bigr].
\end{equation}

It is then straightforward to compute each term. The first trace is given by
\begin{align}
&\tr \bigl[ \Pi_m^S (\pi^{A} \otimes \xi^{B_{\rm in}})  \bigr]\\
&= 
\sum_{\kappa = 0}^k
\tr \bigl[ (\Pi_{\kappa}^{A} \otimes \Pi_{m-\kappa}^{B_{\rm in}} ) (\pi^{A} \otimes \xi^{B_{\rm in}})  \bigr],\\
&=
\frac{1}{2^k}
\sum_{\kappa = 0}^k
\binom{k}{\kappa} \chi_{m-\kappa},
\end{align}
where $\kappa_{m - \kappa} = \tr[\Pi_{m-\kappa}^{B_{\rm in}}\xi^{B_{\rm in}} ]$.
The second trace can be computed as
\begin{align}
&\tr \bigl[ \pi_m^S (I^{S_{\rm in}} \otimes \Pi^{S_{\rm rad}}_n)  \bigr] \notag\\
&=
\frac{1}{\binom{N+k}{m} } \tr \bigl[ \Pi_m^S (I^{S_{\rm in}} \otimes \Pi^{S_{\rm rad}}_n)  \bigr]\\
&=
\frac{1}{\binom{N+k}{m} } \tr \bigl[\Pi_{m-n}^{S_{\rm in}} \otimes \Pi^{S_{\rm rad}}_n  \bigr]\\
&=
\frac{1}{\binom{N+k}{m} }  \binom{N+k-\ell}{m-n} \binom{\ell}{n}.
\end{align}
In total, we have Eq.~\eqref{Eq:pnonon}.

\section{Empirical smoothing of the conditional entropy} \label{App:Theta}
We here show Proposition~\ref{Prop:3}. The statement is that, for any $\delta >0$, the dynamics $\cL^{S \rightarrow S_{\rm in}}_{\rm Kerr}$ leads to 
\begin{equation}
\bigl \|
\tr_{S_{\rm rad}} \bigl[ U^{S} \Psi^{SR} U^{S \dagger} \bigr]
- 
\Gamma^{S_{\rm in} R}
\bigr \|_1 
\leq 
\Theta_{\xi}^{\delta}(N,k,\ell), \label{Ineq:eeeeeee}
\end{equation}
with probability at least $1- \exp[-\delta^2 d_{\rm min}(\epsilon)/48]$, where $U^S:=\bigoplus_{m} U_m^{S}$ and $\Psi^{SR} := \Phi^{AR} \otimes \xi^{B_{\rm in}}$.
For the definitions of $\Theta^{\delta}_{\xi}(N, k, \ell)$, see Eq.~\eqref{Eq:Theta46}.

Using the triangle inequality, we have
\begin{multline}
\bigl \|
\tr_{S_{\rm rad}} \bigl[ U^{S} \Psi^{SR} U^{S \dagger} \bigr]
- 
\Gamma^{S_{\rm in} R}
\bigr \|_1  \\
\leq
\bigl \|
\tr_{S_{\rm rad}} [\Psi_U^{SR}]
- 
\tr_{S_{\rm rad}} [\Pi^{S_{\rm rad}}_{\epsilon} \Psi_U^{SR} \Pi^{S_{\rm rad}}_{\epsilon}]
\bigr \|_1\\
+
\bigl \|
\tr_{S_{\rm rad}} [\Pi^{S_{\rm rad}}_{\epsilon} \Psi_U^{SR} \Pi^{S_{\rm rad}}_{\epsilon}]
- 
\tilde{\Gamma}^{S_{\rm in} R}(\epsilon)
\bigr \|_1\\
+
\bigl \|
\tilde{\Gamma}^{S_{\rm in} R}(\epsilon)
- 
\Gamma^{S_{\rm in} R}
\bigr \|_1, \label{Eq:24nio}
\end{multline}
where $\Psi_U^{SR} = U^S \Psi^{SR} (U^S)^{\dagger}$.
Using Theorem~\ref{Thm:CoPDecoup} with the identification of $S_{\rm in}$ and $E$, the second term is bounded from above as
\begin{multline}
\bigl \|
\tr_{S_{\rm rad}} [\Pi^{S_{\rm rad}}_{\epsilon} \Psi_U^{SR} \Pi^{S_{\rm rad}}_{\epsilon}]
- 
\tilde{\Gamma}^{S_{\rm in} R}(\epsilon)
\bigr \|_1\\
\leq
2^{-\frac{1}{2} H_{\rm min}(S^*|S_{\rm in}R)_{\tilde{\Gamma}(\epsilon)}} + \delta, \label{Ineq:pppp}
\end{multline}
with probability at least $1- \exp[-\delta^2 d_{\rm min}(\epsilon)/48]$ for any $\delta >0$. Note that the minimum dimension is given by $d_{\rm min}(\epsilon)$ due to the application of $\Pi_{\epsilon}^{S_{\rm rad}}$.

When the second term of Eq.~\eqref{Eq:24nio} is small, the first term of Eq.~\eqref{Eq:24nio} should be also small. To observe this, we use the fact that $\Psi_U^{SR} \geq \Pi^{S_{\rm rad}}_{\epsilon} \Psi_U^{SR} \Pi^{S_{\rm rad}}_{\epsilon}$, which leads to
\begin{align}
&\bigl \|
\tr_{S_{\rm rad}} [\Psi_U^{SR}]
- 
\tr_{S_{\rm rad}} [\Pi^{S_{\rm rad}}_{\epsilon} \Psi_U^{SR} \Pi^{S_{\rm rad}}_{\epsilon}]
\bigr \|_1\\
&=
| \tr [\Psi_U^{SR}]
- 
\tr [\Pi^{S_{\rm rad}}_{\epsilon} \Psi_U^{SR} \Pi^{S_{\rm rad}}_{\epsilon}] |\\
&=
1 - \tr [\Pi^{S_{\rm rad}}_{\epsilon} \Psi_U^{SR} \Pi^{S_{\rm rad}}_{\epsilon}].
\end{align}
Using the monotonicity of the trace distance, we also obtain that
\begin{multline}
\bigl|
\tr [\Pi_{\epsilon}^{S_{\rm rad}}\Psi_U^{SR} \Pi_{\epsilon}^{S_{\rm rad}}]
- 
\tr[\tilde{\Gamma}^{S_{\rm in} R}(\epsilon)]
\bigr|\\
\leq
2^{-\frac{1}{2} H_{\rm min}(S^*|S_{\rm in}R)_{\tilde{\Gamma}(\epsilon)}} + \delta.
\end{multline}
Since $\tilde{\Gamma}^{S_{\rm in} R}(\epsilon) = \tr_{S^*}[\Gamma^{S^*ER}(\epsilon)]$, we have $\tr[\tilde{\Gamma}^{S_{\rm in} R}(\epsilon)] = \tr[\tilde{\Gamma}(\epsilon)] = 1-w(\epsilon)$ and so,
\begin{multline}
\tr [\Pi_{\epsilon}^{S_{\rm rad}}\Psi_U^{SR} \Pi_{\epsilon}^{S_{\rm rad}}]\\
\geq
1- w(\epsilon)-  2^{-\frac{1}{2} H_{\rm min}(S^*|S_{\rm in}R)_{\tilde{\Gamma}(\epsilon)}} - \delta.
\end{multline}
Thus, when Eq.~\eqref{Ineq:pppp} holds, it also holds that
\begin{multline}
\bigl \|
\tr_{S_{\rm rad}} [\Psi_U^{SR}]
- 
\tr_{S_{\rm rad}} [\Pi^{S_{\rm rad}}_{\epsilon} \Psi_U^{SR} \Pi^{S_{\rm rad}}_{\epsilon}]
\bigr \|_1\\
\leq 
 2^{-\frac{1}{2} H_{\rm min}(S^*|S_{\rm in}R)_{\tilde{\Gamma}(\epsilon)}} +w(\epsilon)  + \delta.
\end{multline}
Note however that this evaluation is not tight since, when $\epsilon=0$, the L.H.S. is trivially zero, but the R.H.S. is in general non-zero.

For the third term, note that $\Gamma^{S^*S_{\rm in}R} \geq \tilde{\Gamma}(\epsilon)^{S^*S_{\rm in}R}$, implying that  $\Gamma^{S_{\rm in} R} \geq \tilde{\Gamma}^{S_{\rm in} R}(\epsilon)$. Thus, we have 
\begin{align}
\| \tilde{\Gamma}^{S_{\rm in} R}(\epsilon)- \Gamma^{S_{\rm in} R} \|_1 &= \tr[ \Gamma^{S_{\rm in} R}- \tilde{\Gamma}^{S_{\rm in} R}(\epsilon)]\\
&= 1- \tr[\tilde{\Gamma}(\epsilon)]\\
&=w(\epsilon).
\end{align}

Altogether, we obtain that, for any $\delta >0$,
\begin{multline}
\bigl \|
\tr_{S_{\rm rad}} \bigl[ U^{S} \Psi^{SR} U^{S \dagger} \bigr]
- 
\Gamma^{S_{\rm in} R}
\bigr \|_1 \\
\leq
2^{1-\frac{1}{2} H_{\rm min}(S^*|S_{\rm in}R)_{\tilde{\Gamma}(\epsilon)}} + w(\epsilon) + 2 \delta
\end{multline}
with probability at least $1- \exp[-\delta^2 d_{\rm min}/48]$. Since the R.H.S. is equal to $\Theta_{\xi}^{\delta}(N,k,\ell)$, this completes the proof.


\section{Derivation of $\eta_{\xi}$} \label{App:eta}
Here, we derive Eq.~\eqref{Eq:eta49}, i.e.,
\begin{align}
&\eta_{\xi}(N,k,\ell) \notag\\
&:= \bigl \|
\Gamma^{S_{\rm in} R}- 
\Gamma^{S_{\rm in}} \otimes \pi^R
\bigr \|_1\\
&=
\frac{1}{2^k}
\sum_{\nu=0}^{N+k-\ell} \sum_{\kappa=0}^{k} F_{\kappa, \nu}(\xi) \binom{N+k-\ell}{\nu}\binom{k}{\kappa}, \label{Eq:D2}
\end{align}
where
\begin{multline}
F_{\kappa, \nu}(\xi) \\
=\biggl|
\sum_{m=0}^{N+k} \frac{\binom{\ell}{m-\nu}}{\binom{N+k}{m}}
\biggl(
\chi_{m-\kappa}
- 
\frac{1}{2^k}
\sum_{\kappa'=0}^{k}
\binom{k}{\kappa'}\chi_{m-\kappa'}
\biggr)
\biggr|,
\end{multline}
and $\chi_{\mu} = \tr[ \xi^{B_{\rm in}} \Pi^{B_{\rm in}}_{\mu}]$. 

It is straightforward to compute $\Gamma^{S_{\rm in} R}$ and $\Gamma^{S_{\rm in}}$ as
\begin{align}
&\Gamma^{S_{\rm in} R} = \notag \\
&\ \ 
\sum_{m=0}^{N+k} \sum_{n=0}^{\ell} \sum_{\kappa= 0}^{k} \frac{\binom{\ell}{n}\binom{k}{\kappa} \binom{N+k-\ell}{m-n}}{\binom{N+k}{m}}  \chi_{m-\kappa}  \pi^{S_{\rm in}}_{n-m} \otimes \pi_{\kappa}^{R},\\
&\Gamma^{S_{\rm in}}
= 
\frac{1}{2^k} \sum_{m=0}^{N+k} \sum_{n=0}^{\ell} \sum_{\kappa= 0}^{k}  \frac{ \binom{\ell}{n}  \binom{k}{\kappa}\binom{N+k-\ell}{m-n}}{\binom{N+k}{m}}  \chi_{m-\kappa} \pi_{m-n}^{S_{\rm in}},
\end{align}
where we have used $\Pi_m^{S'} = \sum_{n=0}^{\ell}\Pi_{m-n}^{S_{\rm in}'} \otimes  \Pi_{n}^{S_{\rm rad}'}$ and $\Pi_m^{S} = \sum_{\kappa=0}^{k}\Pi_{\kappa}^{A} \otimes  \Pi_{m-\kappa}^{R}$.
Since both are already diagonal, it is easy to compute their distance and obtain Eq.~\eqref{Eq:D2}.

\section{Derivation of Eq.~\eqref{Eq:eerrc}} \label{App:condmin}

We here derive Eq.~\eqref{Eq:eerrc}. To restate it, let $\{ \cH^A_j \}$ be mutually orthogonal subspaces of $\cH^A$, and $\pi_j^A$ be the completely mixed state on $\cH^A_j$. For any state $\Lambda^{ABC}$ in the form of $\sum_{j=0}^J p_j \pi_j^A \otimes \rho_j^{BC}$, where $\rho_j^{BC} \in \mc{S}(\mc{H}^{BC})$ and $\{ p_j \}$ is a probability distribution, it holds that
\begin{equation}
2^{-H_{\rm min}(AB|C)_{\Lambda}}
\leq
\sum_{j=0}^J \frac{p_j}{d_j} 2^{- H_{\rm min} (B|C)_{\rho_j}}, 
\end{equation}
where $d_j = \dim \cH^A_j$.

By definition of the conditional min-entropy, $\forall j \in [0,J]$, $\exists \sigma_j^C \in \mc{S}(\mc{H}^C)$ such that $2^{- H_{\rm min} (B|C)_{\rho_j}} I^B \otimes \sigma_j^C \geq \rho_j^{BC}$. Hence, we have
\begin{equation}
I^{AB} \otimes (2^{- H_{\rm min} (B|C)_{\rho_j}} \sigma_j^C) \geq \Pi_j^A \otimes \rho_j^{BC}.
\end{equation}
for all $j \in [1,J]$. This further implies that
\begin{equation}
\tr[\tilde{\sigma}^C] I^{AB} \otimes \frac{\tilde{\sigma}^C}{\tr[\tilde{\sigma}^C]} \geq \Lambda^{ABC},
\end{equation}
where $\tilde{\sigma}^C := \sum_{j =0}^J  p_j/d_j 2^{- H_{\rm min} (B|C)_{\rho_j}} \sigma_j^C$ is an un-normalized state. This concludes the proof.

\section{Comparison with the analysis based on the partial decoupling} \label{S:e}

\begin{figure*}[tb!]
\centering
\includegraphics[width=165mm,clip]{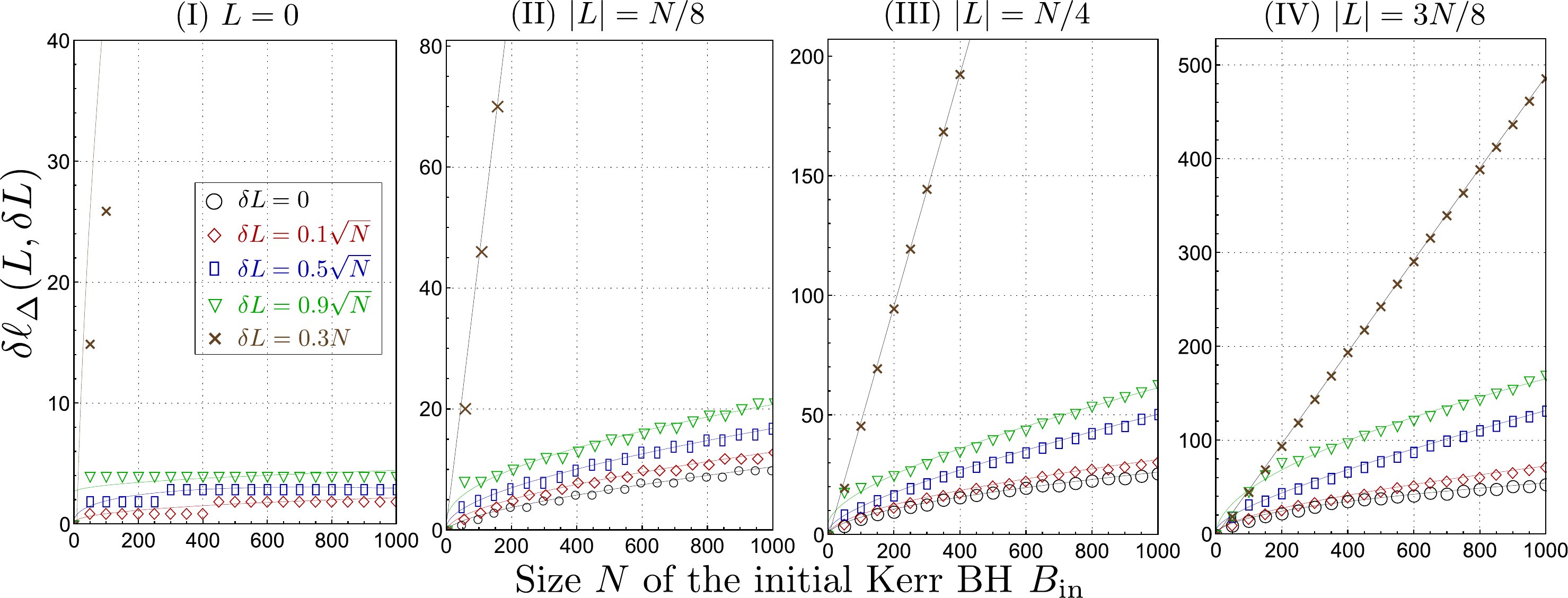}
\caption{
Figures of the delay $\delta \ell_{\Delta}(L, \delta L)$ for the pure Kerr BH for different $Z$-axis AM $L$ and different fluctuations $\delta L$. The size $k$ of the quantum information source $A$ is set to $1$. For simplicity, we set $\Delta =0.1$. In each plot, the solid line shows the fitting by the function $a x + b \sqrt{x} + c$, where $(a,b,c)$ is decided by the least squared fitting.
}
\label{Fig:LthNoEnt}
\end{figure*}

\begin{figure*}[tb!]
\centering
\includegraphics[width=165mm,clip]{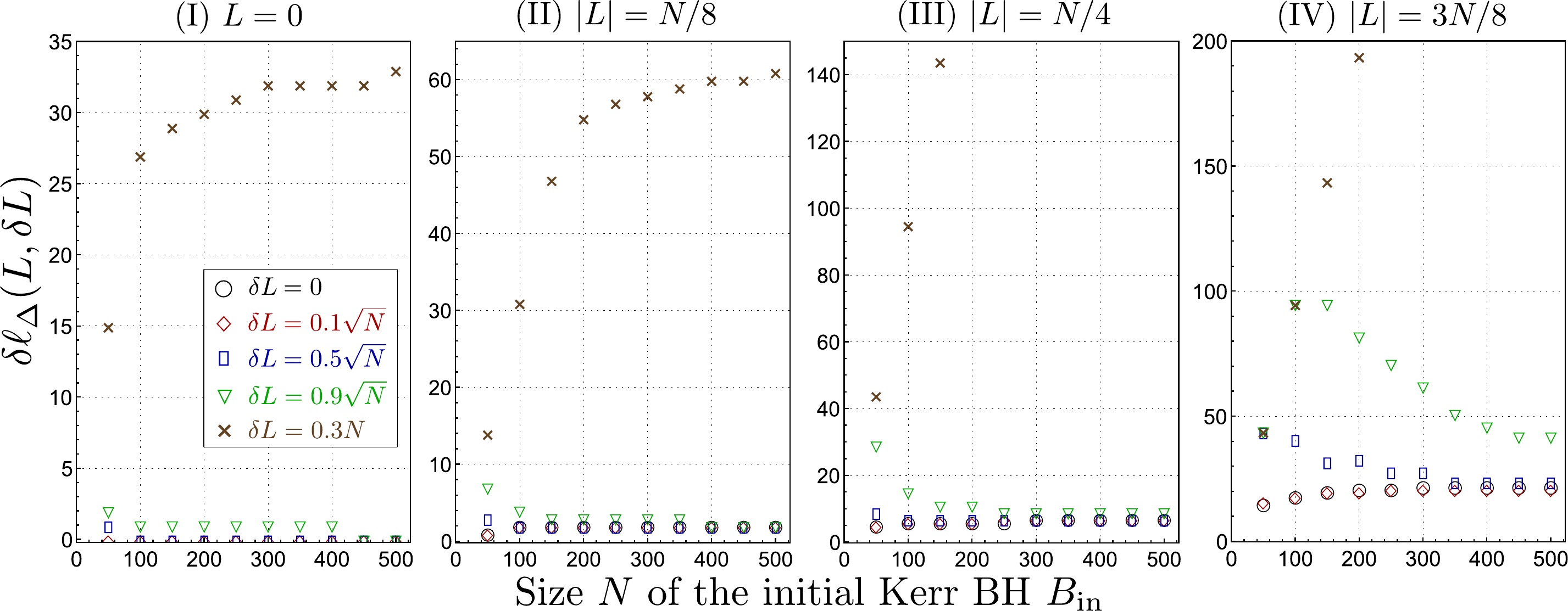}
\caption{
Figures for the delay $\delta \ell_{\Delta}(L, \delta L)$ for the mixed Kerr BH with different $Z$-axis AM $L$ and fluctuations $\delta L$ of the initial Kerr BH $B_{\rm in}$. They are plotted as a function of the number $N$ of qubits in the initial Kerr BH. The size $k$ of the quantum information source $A$ is set to $1$. The $\Delta$ is set to $0.1$ for simplicity.
}
\label{Fig:LthEnt}
\end{figure*}

Here, we provide numerical evaluations of the scaling of the delay $\delta \ell_{\Delta}(L, \delta L)$ with respect to the size $N$ of the initial Kerr BH $B_{\rm in}$ for various $Z$-axis AMs $L$ and their fluctuations $\delta L$.
The cases of pure and mixed Kerr BHs are depicted in Figs.~\ref{Fig:LthNoEnt} and~\ref{Fig:LthEnt}, respectively.

For the pure Kerr BH, we first observe from the case of $\delta L =0$ (black circles) that $\delta \ell_{\Delta}(L, 0) = O(\sqrt{N})$ for any $L$. 
Note that the delay is in comparison with the case of $L=\delta L = 0$. Hence, the delay for $\delta L=0$ is not plotted in Fig.~\ref{Fig:LthNoEnt} (I).
When $\delta L \neq 0$, the scaling seems to depend on if $\delta L = O(\sqrt{N})$ or $O(N)$. For $\delta = O(\sqrt{N})$, it appears that $\delta \ell_{\Delta} (L, \delta L) = O(\sqrt{N})$, while $\delta \ell_{\Delta} (L, \delta L) = O(N)$ for $\delta L=O(N)$. 
Thus, we conclude that, for any $L$, the delay $\delta \ell_{\Delta}(L, \delta L)$ gradually changes from $O(\sqrt{N})$ to $O(N)$ as $\delta L$ increases from $O(\sqrt{N})$ to $O(N)$.

For the mixed BH, the delay $\delta \ell_{\Delta}(L, 0)$ is independent of $N$. This seems to be also the case for $\delta L = O(\sqrt{N})$. When $\delta L = O(N)$, $\delta \ell_{\Delta}(L, \delta L)$ appears to increase as $N$ increases. It also seems that the scaling of $\delta \ell_{\Delta}(L, \delta L)$ with respect to $N$ depends on $L$: when $L = 0$, the delay may grow very slowly as $N$ becomes large, while it seems to scale linearly with $N$ when $|L| \geq N/4$. 

\end{document}